\documentclass[prx,twocolumn,10pt,aps,longbibliography,nofootinbib]{revtex4-1}

 \usepackage[utf8]{inputenc}
 \usepackage[greek,english]{babel}
 \usepackage{bm}
 \usepackage{verbatim}
 \usepackage{amsmath}
 \usepackage{amssymb}
 \usepackage{amsthm}
 \usepackage{latexsym}
 \usepackage{amsfonts}
 \usepackage{epsfig}
 \usepackage{epstopdf}
 \usepackage{wasysym} 
 \usepackage{subfigure}
 \usepackage{color}
 \definecolor{darkblue}{rgb}{0,0,.5}
 \definecolor{BLUE}{rgb}{0,0,1}
 \definecolor{BLACK}{rgb}{0,0,0}
 \usepackage[linktocpage, colorlinks=true, linkcolor=darkblue, citecolor=darkblue]{hyperref}
 \usepackage[all]{hypcap}
 \usepackage[makeroom]{cancel}

\newcommand{\C}[1]{{\cal{#1}}}
\newcommand{\bb}[1]{\textbf{#1}}
\newcommand{\lr}[1]{{\langle {#1}\rangle}}

\newcommand{\rl}[0]{{\rangle\langle}}

\begin{document}

\title{Classicality, Markovianity and local detailed balance from pure state dynamics}

\author{Philipp Strasberg$^1$}
\author{ Andreas Winter$^{1,2,3}$}
\author{Jochen Gemmer$^4$}
\author{Jiaozi Wang$^4$}
\affiliation{$^1$F\'isica Te\`orica: Informaci\'o i Fen\`omens Qu\`antics, Departament de F\'isica, Universitat Aut\`onoma de Barcelona, 08193 Bellaterra (Barcelona), Spain}
\affiliation{$^2$ICREA -- Instituci\'o Catalana de Recerca i Estudis Avan\c{c}ats, Pg.~Lluis Companys, 23, 08010 Barcelona, Spain}
\affiliation{$^3$Institute for Advanced Study, Technische Universit\"at M\"unchen, Lichtenbergstra{\ss}e 2a, D-85748 
Garching, Germany}
\affiliation{$^4$Department of Physics, University of Osnabr\"uck, D-49076 Osnabr\"uck, Germany}

\date{\today}

\begin{abstract}
 When describing the effective dynamics of an observable in a many-body system, the \emph{repeated randomness 
 assumption}, which states that the system returns in a short time to a maximum entropy state, is a crucial hypothesis 
 to guarantee that the effective dynamics is classical, Markovian and obeys local detailed balance. While the latter 
 behaviour is frequently observed in naturally occurring processes, the repeated randomness assumption is in blatant 
 contradiction to the microscopic reversibility of the system. Here, we show that the use of the repeated randomness 
 assumption can be justified in the description of the effective dynamics of an observable that is both \emph{slow} 
 and \emph{coarse}, two properties we will define rigorously. Then, our derivation will invoke essentially only the 
 eigenstate thermalization hypothesis and typicality arguments. 
 While the assumption of a slow observable is subtle, as it provides only a necessary but not sufficient condition, 
 it also offers a unifying perspective applicable to, e.g., open systems as well as collective observables of many-body 
 systems. All our ideas are numerically verified by studying density waves in spin chains. 
\end{abstract}

\maketitle

\newtheorem{lemma}{Lemma}[section]

\section{Introduction}
\label{sec intro}

The equal-a-priori-probability postulate and the related maximum entropy principle are central axioms of statistical 
mechanics. For instance, for an isolated system observed to have an energy $E$ these principles imply that the correct 
ensemble to describe the situation is 
\begin{equation}\label{eq microcanonical ensemble}
 \Pi_E/V_E,
\end{equation}
where $\Pi_E$ is a projector on an energy shell with energy $E$ (defined up to some small uncertainty 
$\Delta E$) and the normalization $V_E = \mbox{tr}\{\Pi_E\} = \exp[S_B(E)]$ is the exponential of the Boltzmann 
entropy $S_B(E)$. The state~(\ref{eq microcanonical ensemble}) is the familiar microcanonical ensemble, 
which is the central starting point of equilibrium statistical mechanics. 

Moreover, the equal-a-priori-probability postulate and the maximum entropy principle continue to be useful out of 
equilibrium. For instance, consider two systems $A$ and $B$ in thermal contact and with average energies 
$\lr{E_A}$ and $\lr{E_B}$. The maximum entropy principle then implies that the correct state to describe this 
situation is 
\begin{equation}\label{eq two canonical ensembles}
 \frac{e^{-\beta_A H_A}}{Z_A(\beta_A)} \otimes \frac{e^{-\beta_B H_B}}{Z_B(\beta_B)}.
\end{equation}
Here, the inverse temperature $\beta_{A/B}$ is chosen such that the expectation value of the Hamiltonian $H_{A/B}$
equals $\lr{E_{A/B}}$. Initial states such as~(\ref{eq two canonical ensembles}), or slight generalizations of it, 
are predominantly used throughout the literature on nonequilibrium physics for both quantum and classical systems 
and independent of the employed methods (master equations, Green's function techniques, scattering theory, 
etc.)~\cite{DeGrootMazur1984, ZwanzigBook2001, NazarovBlanterBook2009, StefanucciVanLeeuwenBook2013, SchallerBook2014, 
StrasbergBook2022}. 

Let us continue to consider the same setup, but at a different time. Due to the thermal contact, the systems will now 
have energies $\lr{E'_{A/B}}$ different from $\lr{E_{A/B}}$ in general. The maximum entropy principle then 
predicts again 
\begin{equation}\label{eq two canonical ensembles 2}
 \frac{e^{-\beta'_A H_A}}{Z_A(\beta'_A)} \otimes \frac{e^{-\beta'_B H_B}}{Z_B(\beta'_B)}
\end{equation}
with suitably adapted $\beta'_{A/B}$. But quite discomfortingly, the states~(\ref{eq two canonical ensembles})
and~(\ref{eq two canonical ensembles 2}) have different von Neumann entropies in general such that there cannot 
exist any Hamiltonian dynamics mapping state~(\ref{eq two canonical ensembles}) 
to state~(\ref{eq two canonical ensembles 2}). 

A way out of this dilemma is to use the equal-a-priori-probability postulate or the maximum entropy principle only 
once. However, the ensuing dynamics can then quickly become very complex and intractable in practical applications. 
On the other hand, it is known that the repeated use of these principles gives rise to a \emph{classical} stochastic 
process, which is \emph{Markovian} and obeys \emph{local detailed balance} (precise definitions of these notions and a 
derivation are presented below). Indeed, such a description is a common starting point of many disciplines such as 
stochastic thermodynamics~\cite{SekimotoBook2010, SeifertRPP2012, SchallerBook2014, PelitiPigolottiBook2021, 
StrasbergBook2022}, which is well confirmed experimentally~\cite{BustamanteLiphardtRitortPhysTod2005, CilibertoPRX2017}. 

The main focus of the present paper is to provide a justification from reversible microscopic dynamics of the 
repeated use of the equal-a-priori-probability postulate or the maximum entropy principle, which has been called 
the \emph{repeated randomness assumption} by van Kampen~\cite{VanKampenBook2007}. In fact, it seems that van Kampen 
has been particularly unsatisfied by it as he somewhat laconically notes at the end of his book with respect to the 
repeated randomness assumption that ``[this] statement [has] not been proved mathematically, but it is better to say 
something that is true although not proved, than to prove something that is not true'' 
(page 456 in Ref.~\cite{VanKampenBook2007}). 

Although the high complexity of the situation does not allow us to cast our results into the form of
mathematically rigorous theorems, we give plausible physical arguments together with reasonable mathematical
estimates that justify the repeated randomness assumption for \emph{slow} and \emph{coarse} observables. Thus, based
solely on \emph{one} common assumption about the observable, we are able to explain the emergence of three
seemingly distinct and usually separately studied concepts: classicality, Markovianity and local detailed balance.
Remarkably, our derivation works for pure states and avoids any ensemble averages, by using the eigenstate
thermalization hypothesis (ETH)~\cite{DeutschPRA1991, SrednickiPRE1994, SrednickiJPA1999,
RigolDunjkoOlshaniiNature2008} and typicality arguments in the form of Levy's Lemma~\cite{PopescuShortWinterNatPhys2006},
thereby providing a detailed microscopic understanding of why and when maximum entropy inference
can be applied repeatedly.

\subsection{Related literature}

Our work overlaps with so many research directions that giving an exhaustive literature overview at this place appears 
impossible. Thus, we only list the literature that we found most influential and most closely related to our work. 
We ask for the forbearance of any colleagues who might think that we have missed some important publication here; 
it is not intentional. 
 
First, we cannot take any credit for the idea to focus on slow and coarse observables, which is a central concept of 
statistical mechanics since its inception~\cite{EhrenfestEhrenfest1911, EhrenfestEhrenfestBook1959}. In fact, the 
present treatment is much inspired by an old paper from van Kampen~\cite{VanKampenPhys1954}, which well summarizes the 
underlying physical picture. 

Second, our approach follows the philosophy of pure state statistical mechanics, which is based on the idea that 
quantum mechanics alone suffices to explain statistical mechanics behaviour. In fact, the use of the 
equal-a-priori-probability postulate and the maximum entropy principle to compute equilibrium expectation values of 
observables at a \emph{single time} has by now been well justified within that approach~\cite{GemmerMichelMahlerBook2004,
DAlessioEtAlAP2016, BorgonoviEtAlPR2016, GogolinEisertRPP2016, GooldEtAlJPA2016, DeutschRPP2018, MoriEtAlJPB2018}. 

Quite naturally, research on pure state statistical mechanics has started to focus on nonequilibrium phenomena. 
For instance, it has been shown that typicality arguments remain useful even out of equilibrium (``dynamical 
typicality''~\cite{BartschGemmerPRL2009, ReimannPRE2018, XuGuoPolettiPRA2022}) and can be used to derive master 
equations~\cite{GemmerMichelEPJB2006, BreuerGemmerMichelPRE2006, GemmerBreuerEPJ2007, HahnGuhrWaltnerPRE2020}. 
Moreover, random matrix theory has been used to predict the time evolution of expectation values of 
observables~\cite{ReimannNC2016, BalzReimannPRL2017, ReimannDabelowPRL2019, DabelowReimannPRL2020, RichterEtAlPRE2020b, 
DabelowReimannJSM2021} and general results on the time-scales of thermalization have 
been found~\cite{GoldsteinHaraTasakiPRL2013, GarciaPintosEtAlPRX2017, DeOliveiraEtAlNJP2018, WilmingEtAlBook2018, 
NickelsenKastnerPRL2019, HevelingKnipschildGemmerPRX2020, SimenelGodbeyUmarPRL2020}. 

However, this research did not yet consider \emph{multi-time} processes (e.g., temporal joint probabilities or
correlation functions), apart from two exceptions mentioned below.
Also the role of the slowness of the observable and its implications for the three properties of classicality,
Markovianity and local detailed balance has not been at the focus of these previous works. Instead, these properties 
have been typically investigated within a (repeated) ensemble average approach to statistical mechanics. 

First, we comment on the emergence of classicality, which is commonly explained with
\emph{decoherence}~\cite{ZurekRMP2003, JoosEtAlBook2003, SchlosshauerPR2019}. It combines two ideas: first, all systems
are essentially open systems and, second, open systems decohere, i.e., their density matrix becomes diagonal in a
particular fixed basis (``pointer basis''). We emphasize that it is not our intention to question the
correctness of the decoherence approach. While there is fundamental criticism (see, e.g., Refs.~\cite{LeggettJP2002, BallentineFP2008, KnipschildGemmerPRA2019, BerjonOkonSudarskyPRD2021}), our results are not in contradiction with
decoherence, which is motivated by the central question: ``Which is the preferred measurement
basis?''~\cite{ZurekPRD1981} Instead, we consider a different \emph{perspective} and significantly extend the realm in 
which quantum dynamics appears classical. In particular, from the perspective of pure state statistical mechanics one 
would like to derive classical behaviour for a single wave function $|\psi\rangle$ and realistic many-body systems. 
Yet, for any observable that does not have a definite deterministic outcome when measured in state $|\psi\rangle$, 
$|\psi\rangle$ must necessarily have coherences in the eigenbasis of that observable. Global decoherence can therefore 
not happen, as a mathematical fact of linear algebra, but still one would expect that also pure states can behave 
classical in an appropriate sense. Here, by extending previous numerical studies~\cite{GemmerSteinigewegPRE2014,
SchmidtkeGemmerPRE2016}, we argue that slow and coarse observables behave classical and estimate deviations from
classical behaviour to be exponentially small in the system size. Our approach hints at a possibly deep connection
between pure state statistical mechanics, the ETH and classical behaviour, which remains unrecognized within the
conventional open quantum systems paradigm, where the bath is typically modeled as integrable and as staying
approximately in a canonical ensemble~\cite{BreuerPetruccioneBook2002, DeVegaAlonsoRMP2017}. Our approach also
provides physical substance to recent abstract studies of multi-time classicality~\cite{SmirneEtAlQST2018,
StrasbergDiazPRA2019, MilzEtAlQuantum2020, MilzEtAlPRX2020} and it might offer interesting insights for the consistent
histories approach to quantum mechanics~\cite{GriffithsJSP1984, OmnesRMP1992, Griffiths2019} and quantum
Darwinism~\cite{ZurekNP2009, ZurekEnt2022}, as recently explored by one of us~\cite{StrasbergSP2023}.

Secondly, much recent research has been devoted to understanding non-Markovianity in quantum 
systems~\cite{RivasHuelgaPlenioRPP2014, BreuerEtAlRMP2016, LiHallWisemanPR2018, MilzModiPRXQ2021}. This research mostly 
revolved around the question how to properly define and quantify non-Markovianity, but surprisingly little rigorous and 
general results are known about the question which physical properties give rise to Markovianity. For instance, it 
is known that open quantum systems are Markovian in the weak coupling \emph{limit}~\cite{DuemckeJMP1983}, which 
literally requires to scale the system-bath coupling to zero, among other questionable assumptions. Indeed, without 
that limiting procedure it has been claimed that no physical system is Markovian~\cite{FordOConnellPRL1996}.
Somewhat reconciling these two results, recent research has highlighted that typical processes are almost 
Markovian~\cite{FigueroaRomeroModiPollockQuantum2019, FigueroaRomeroPollockModiCP2021}, but with the caveat that 
``typical'' is defined with respect to an abstract mathematical measure, which is likely not typical in reality. 
Moreover, we would like to point out that an important aspect of (non-)Markovianity cannot be captured when
using ensemble averages instead of pure states. Indeed, if the system dynamics is non-Markovian, this implies that the
system reacts very sensitively to different microstates of the bath, or, conversely, if the system dynamics is
insensitive to the precise state of the bath, it must be Markovian. But by using an initial ensemble average over a
highly mixed canonical ensemble, the influence from the different microstates is washed out. To the best of our
knowledge, only recently the question of (non-)Markovianity has been studied for pure
states~\cite{FigueroaRomeroModiPollockQuantum2019, FigueroaRomeroPollockModiCP2021}. We believe, however, that the
result that almost all open quantum systems are almost Markovian is too strong. Based on our findings it seems that Markovianity is also crucially related to the observable we are probing, and cannot be deduced from the unitary dynamics alone as in Refs.~\cite{FigueroaRomeroModiPollockQuantum2019, FigueroaRomeroPollockModiCP2021}.

Thirdly, the property of local detailed balance ensures thermodynamic consistency at each time step of the 
process and it is thus build into the framework of stochastic thermodynamics~\cite{SekimotoBook2010, SeifertRPP2012, 
SchallerBook2014, PelitiPigolottiBook2021, StrasbergBook2022}. For systems that equilibrate in the macroscopic sense, 
it has been derived in its most general 
form by van Kampen based on the repeated randomness assumption~\cite{VanKampenPhys1954}. Since the notion of local 
detailed balance, which is sometimes also called ``detailed balance'' (without the attribute ``local'') or
``microreversibility'', might be less familiar to some readers, we explain it more thoroughly later on. 

We end this short literature survey with two remarks for specialists in open quantum systems theory. First, the 
repeated randomness assumption is commonly known as the \emph{Born approximation} in this field. Second, our work is
motivated by a lack of any satisfactory explanation of the \emph{repeated} randomness assumption, but sceptical voices 
might claim that Nakajima-Zwanzig projection operator techniques show that the Born approximation is only needed 
\emph{once} in the derivation of the (quantum) master equation~\cite{BreuerPetruccioneBook2002}. There are, however, 
two subtle pitfalls. First, this statement is only true for the exact Nakajima-Zwanzig 
master equation: once one applies perturbation theory the failure of the Born approximation at later times can give 
rise to additional correction terms even to lowest order in the perturbation theory~\cite{MitchisonPlenioNJP2018}. 
Second, we are here interested in processes characterized by \emph{multi-time} statistics in contrast to the
single-time statistics that are accessible with a master equation. Related recent work has also investigated the
multi-time statistics for pure state dynamics using long-time averages~\cite{DowlingEtAlQuantum2023,
DowlingEtAlSPC2023}, which we do not use here. From the perspective of open quantum system theory, our results
thus explain why and when the intuitive Born approximation is justified even though the exact unitarily time-evolved
system-bath state no longer complies with the Born approximation (for a related numerical study see
Ref.~\cite{KolovskyPRE2020}). Equivalently, if one insists to apply the Born approximation only at the initial time,
our results microscopically justify the quantum regression theorem~\cite{LiHallWisemanPR2018}.

\subsection{Outline}

Section \ref{sec preliminaries} starts by introducing an intuitive picture for our setup while establishing notation 
along the way (Sec.~\ref{sec setup}), gives a first explanation of ``slow'' observables (Sec.~\ref{sec slow}), and 
briefly reviews the main tools we are using, namely the ETH and Levy's Lemma (Sec.~\ref{sec ETH Levy}). 

Sections \ref{sec classicality}, \ref{sec Markovianity} and \ref{sec local detailed balance} contain the core results 
of this paper about classicality, Markovianity and local detailed balance, respectively. They start with a brief 
definition and discussion of the respective notion together with their derivation based on the repeated randomness 
assumption. Afterwards, we show how each of these properties arises from pure state dynamics. 

Section \ref{sec numerics} then presents numerical results for density waves in a spin chain, which confirm our main 
ideas. Since we have tested many features numerically, we decided to shift some of them to a supplemental material 
to keep the main manuscript focused. 

However, the numerical results also raise awareness about various subtleties, some of which are discussed more 
generally in Sec.~\ref{sec further discussion}. In particular, we return to the subtle notion of ``slowness'' and 
questions related to multiple observables (Sec.~\ref{sec multiple}). Moreover, Sec.~\ref{sec symmetric} discusses 
which properties of the process are not fixed by our general considerations (namely the time scales). 

Finally, Sec.~\ref{sec conclusions} presents conclusions. Furthermore, two short technical proofs are 
relegated to the Appendix. 

\section{Preliminaries}
\label{sec preliminaries}

\subsection{Setup and intuitive picture}
\label{sec setup}

We consider a time-independent isolated quantum system with Hamiltonian $H = \sum_k E_k|k\rl k|$ with ordered 
eigenenergies $E_{k+1} \ge E_k$ and eigenvectors $|k\rangle$. Owing to the time-independence, 
we can and will restrict ourselves to some microcanonical energy shell, which is small on a macroscopic scale but large 
on a microscopic scale, i.e., the dimension $D$ of the corresponding microcanonical Hilbert space $\C H$ obeys 
$D = \C O(10^N)$ with $N$ the number of particles in the system. For simplicity we assume energy to be the only 
conserved quantity, other conserved quantities (such as particle number) could be readily included in the description. 
Moreover, we set $\hbar\equiv1$ such that the time evolution of a pure state is given by 
$|\psi(t)\rangle = \sum_k e^{-iE_kt} c_k |k\rangle$ with $c_k\in\mathbb{C}$ satisfying $\sum_k |c_k|^2 = 1$. 

We are interested in the evolution of an observable $X = \sum_{x=1}^M \lambda_x\Pi_x$ with eigenvalues $\lambda_x$ and 
corresponding eigenprojectors $\Pi_x$, which divide the Hilbert space $\C H = \bigoplus_x\C H_x$ into subspaces 
$\C H_x$. Moreover, we are only interested in \emph{coarse} observables, which means that the number $M$ of different 
projectors (or potential measurement results) is much smaller than $D$. This assumption will be satisfied 
for any realistic experiment with a many-body system. Equivalently, a coarse observable is characterized by subspaces 
$\C H_x$ whose dimension is typically very large: $V_x \equiv \dim\C H_x \gg1$. We will refer to $V_x$ as a 
\emph{volume} in view of Boltzmann's entropy concept $S_B(x) \equiv \ln V_x$ ($k_B\equiv1$ throughout), which plays 
an important role later on. We further call each $x$ a \emph{macrostate}, despite the fact that it does not need to 
be macroscopically large in an intuitive sense. For instance, $x$ could label an energy eigenvalue of an open 
quantum system, which is still a coarse observable in the full system-bath space. 

\begin{figure}[t]
 \centering\includegraphics[width=0.43\textwidth,clip=true]{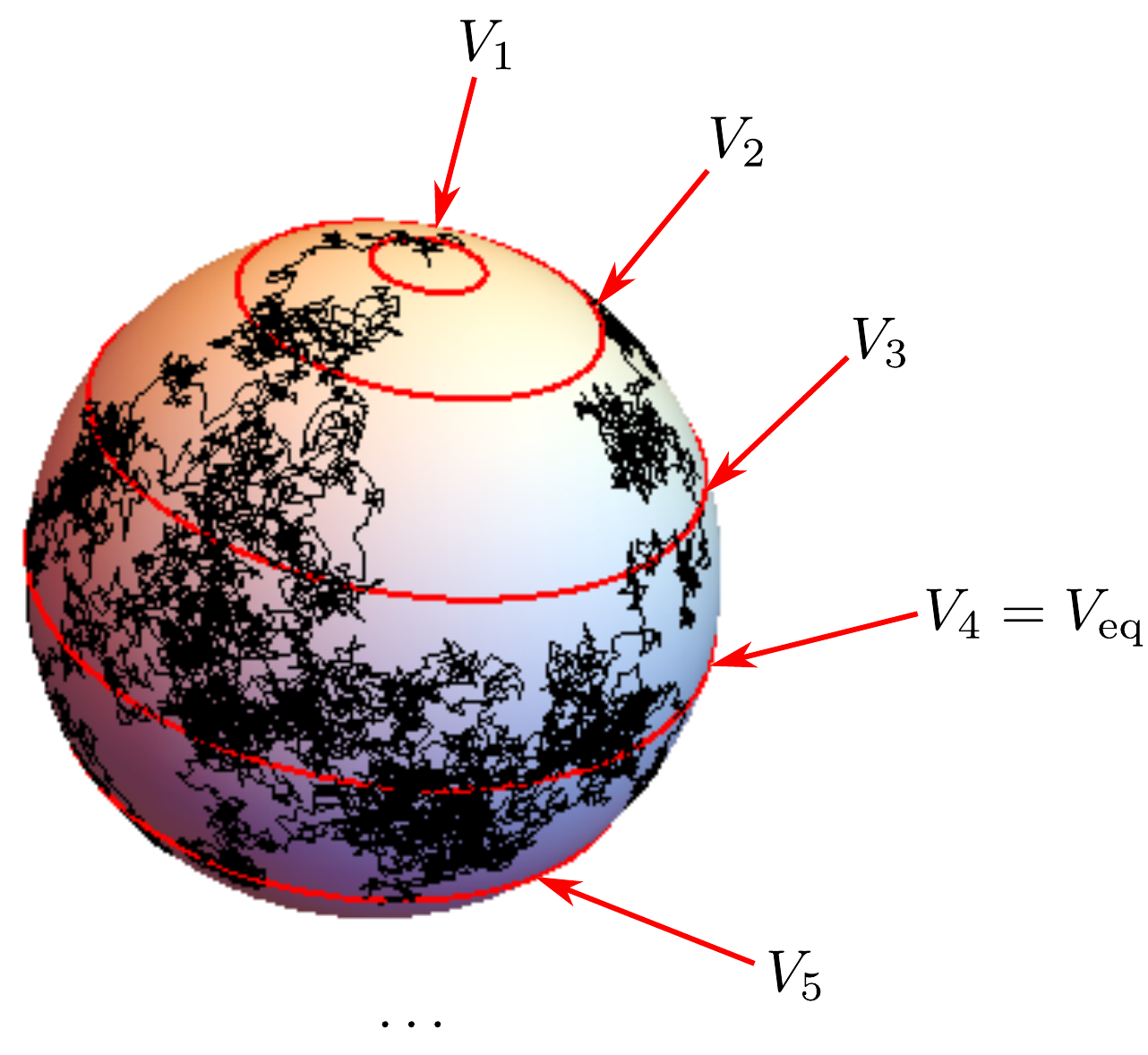}
 \label{fig rw} 
 \caption{Sketch of $\mathbb{S}^{2D-1}$ with subspaces $\C H_x$ labeled with their volumes $V_1,V_2,\dots$ (red lines). 
 The state (here initialized at the north pole) performs an effective random walk on the surface of the sphere and, 
 eventually, spends most time close to the equilibrium subspace, which has the largest volume 
 (here $V_4 = V_\text{eq}$). }
\end{figure}

We write $\Pi_x = \sum_\alpha |x_\alpha\rl x_\alpha|$, where $\alpha$ sums over the microstates $|x_\alpha\rangle$ 
spanning $\C H_x$. Decomposing the wave function in the eigenbasis of $X$ gives 
$|\psi(t)\rangle = \sum_{x,\alpha} c_{x,\alpha}(t)|x_\alpha\rangle$ with 
$c_{x,\alpha}(t) = \sum_k e^{-iE_kt} c_k \lr{x_\alpha|k}$. The normalization condition 
$\sum_{x,\alpha} |c_{x,\alpha}(t)|^2 = 1$ defines a sphere 
$\mathbb{S}^{2D-1}\subset\mathbb{C}^D\cong\mathbb{R}^{2D}$ of dimension $2D-1$ and radius 1, where the factor 
2 arises because $c_{x,\alpha}(t)$ has a real and imaginary part. Since a generic 
many-body system is non-integrable, the eigenenergies $E_k$ are incommensurate (apart from accidential 
degenercies) and the phases $e^{-iE_kt}$ vary erratically with $k$. Moreover, a typical wave function, in particular 
one prepared out of equilibrium, has many non-vanishing coefficients $c_k$~\cite{ReimannPRL2008, LindenEtAlPRE2009, 
WilmingEtAlPRL2019}. This implies that the $c_{x,\alpha}(t)$ vary in a practically unpredictable way. 
Thus, we like to picture the evolution of $|\psi(t)\rangle$ in the eigenbasis of $X$ as a \emph{random walk} on the 
high dimensional sphere $\mathbb{S}^{2D-1}$ as illustrated in Fig.~\ref{fig rw}. 

Strictly speaking, this picture is incorrect. The evolution is not truely random and, since the coefficients $c_k$ 
are constant and only the phases $e^{-iE_kt}$ vary in time, the state can only explore a $D$ dimensional 
submanifold (a hypertorus) on the sphere $\mathbb{S}^{2D-1}$. Unfortunately, our restriction of living in a 
three-dimensional world does not allow us to sketch this properly. However, what is important (and correct) for our 
purposes is that the state explores a high dimensional space in a sufficiently unbiased and random fashion. 

On the sphere $\mathbb{S}^{2D-1}$, we can picture $\C H_x$ as lower dimensional subspaces $\mathbb{S}^{2V_x-1}$ defined 
by all states $|\psi_x\rangle = \sum_\alpha c_{x,\alpha}|x_\alpha\rangle$ satisfying $\sum_\alpha |c_{x,\alpha}|^2 = 1$. 
These spaces are of measure zero (with respect to $\mathbb{S}^{2D-1}$) and therefore indicated as lines on the 
two-dimensional surface of the sphere in Fig.~\ref{fig rw}. A state drawn at random will typically overlap with many 
such volumes (which is again hard to sketch), i.e., it has coherences between different macrostates. However, many
thermodynamic variables are characterized by having one dominant subspace $x$ with $V_\text{eq} \equiv V_x \gg V_y$ 
for all $y\neq x$, which can be identified with the \emph{equilibrium} subspace (in Fig.~\ref{fig rw} the equator, 
having the longest line, corresponds to the subspace with the largest volume)~\cite{GoldsteinEtAlPRE2010}. 
Most randomly drawn states will lie very close to this equilibrium subspace. 

The present analogy suggests that the evolution of a state vector of a non-integrable system can be approximately 
viewed as a diffusion process on a high dimensional subspace of $\mathbb{S}^{2D-1}$. If we have taken into account 
all conserved quantities and if $X$ is the only relevant slowly varying observable (more on this in
Sec.~\ref{sec multiple}), then this diffusion process should be \emph{isotropic}, where the trajectory of the state
vector does not preferably select certain `narrow' regions of $\mathbb{S}^{2D-1}$ during its evolution. Based on this 
intuition, it appears plausible that the evolution of the probabilities $p_x(t) = \lr{\psi(t)|X|\psi(t)}$ should be 
describable by a \emph{classical Markov} process. It is classical because it is unlikely that the enormous number
of tiny amplitudes $c_{x,\alpha}(t)$ interfere constructively and thus give rise to a large detectable coherent effect, 
as we explain in greater detail in Sec.~\ref{sec classicality}. It is Markovian because two slightly different
initial microstates behave approximately the same from a coarse-grained point of view. Moreover, the isotropic
diffusion causes an initial nonequilibrium state, i.e., a state confined to a low entropy region of small volume,
to evolve towards larger volume regions in such a way that entropy \emph{continuously increases}, which is
the condition of local detailed balance.

The goal of this paper is to make this intuition as rigorous as possible by combining tools from the ETH and 
typicality with plausible physical assumptions.  

\subsection{Slow observables}
\label{sec slow}

Slowness is a crucial ingredient not only in our derivation but in many approaches to statistical mechanics, yet 
defining it precisely is not simple. Roughly speaking, we call an observable slow if its 
expectation value $\lr{X}(\tau)$ evolves on a characteristic time scale
\begin{equation}\label{eq slow time scale}
 \frac{1}{\delta e} \gg \tau \gg \frac{1}{\Delta E}
\end{equation}
for initial nonequilibrium states. Here, $\Delta E$ is the width of the microcanonical energy window such 
that $1/\Delta E$ corresponds to the time the system needs to evolve between two orthogonal microstates, which follows 
rigorously from the quantum speed limit~\cite{MandelstamTammJP1945, DeffnerCampbellJPA2017}. It is typically an 
extremely short time scale, impossible to resolve in most mesoscopic or macroscopic experiments since 
$\Delta E \sim \sqrt{N}$. On the other end of the spectrum, $1/\delta e$ with $\delta e \approx\Delta E/D\sim 10^{-N}$ 
the mean level spacing is an extremely long time scale known as the Heisenberg time. It corresponds to the time needed 
for a quantum system to explore the full available Hilbert space. Thus, a slow observable evolves slow compared to the 
microscopic motion of the system, but fast enough to be recognizable in an experiment as a nonequilibrium dynamics. 

Thinking further about it, we see that we can mathematically characterize a slow observable $X$ as being a narrowly 
banded matrix in an ordered energy eigenbasis. To see this, note that 
\begin{equation}\label{eq exp value}
 \lr{X}(\tau) = \sum_{k,\ell} e^{i\omega_{k\ell}\tau} c_k^* c_\ell X_{k\ell}
\end{equation}
with the transition frequency $\omega_{k\ell} = E_k - E_\ell$ and the matrix elements $X_{k\ell} = \lr{k|X|\ell}$. 
If we want to ensure that this expression varies on the time scale specified in Eq.~(\ref{eq slow time scale}) 
\emph{for all} nonequilibrium initial states (within a microcanonical energy window), we need to demand that 
$X_{k\ell}$ differs significantly from zero only for frequencies $\omega_{kl}\in[-\delta E,\delta E]$ with the width 
$\delta E$ satisfying $\delta e \ll\delta E \ll \Delta E$, i.e., $X$ is \emph{narrowly banded}.

Another perspective on slowness is offered by Heisenberg's equation of motion by defining the evolution time-scale of 
$X$ as $\tau \equiv \|X\|/|d\lr{X}/dt|$ with the operator norm $\|\cdot\|$. Then, Heisenberg's equation implies 
\begin{equation}
 \tau = \frac{\|X\|}{|\lr{[H,X]}|} \ge \frac{\|X\|}{\|[H,X]\|},
\end{equation}
where we used $|\mbox{tr}\{A\rho\}| \le \|A\|$ for any density matrix $\rho$. Now, if we demand 
\begin{equation}\label{eq slow observable norm}
 \|[H,X]\| \ll \|H\| \|X\|,
\end{equation}
one finds $\tau \gg 1/\|H\|$, which reduces to the right inequality in Eq.~(\ref{eq slow time scale}) if we 
define the (arbitrary) energy of the microcanonical energy shell to be zero. In fact, we show in 
Appendix~\ref{sec app bandedness implies small commutator} that a narrowly banded matrix implies 
Eq.~(\ref{eq slow observable norm}). Unfortunately, we were not able to show that Eq.~(\ref{eq slow observable norm}) 
implies a narrowly banded $X$, albeit we also found no counterexample. It seems likely to us that counterexamples 
require precisely tuned observables and states. For most practical purposes it thus seems reasonable to assume 
that the condition of Eq.~(\ref{eq slow observable norm}) is equivalent to a narrowly banded $X$. 

Apart from the abstract mathematical property of slowness, finding precise yet generic physical conditions for the 
existence of slow observables $X$ is not trivial. However, there are a few important cornerstones known that we list 
here. First, an obvious class is given by problems that can be cast into the form 
\begin{equation}\label{eq slow observables class 1}
 H = H_0 + \epsilon V, ~~~ [H_0,X] = 0, ~~~ [V,X] \neq 0,
\end{equation}
and where $\epsilon\ll1$ is a small perturbative parameter. In fact, this class of problems is omnipresent in the 
literature. For instance, for weakly coupled open quantum systems one has $H_0 = H_S + H_B$ with $H_S$ ($H_B$) the 
system (bath) Hamiltonian and $V$ their interaction. Then, we see that the energy of a weakly coupled open quantum 
system $X = H_S$ is a slow observable. 

More generally, it can be shown that local observables of local Hamiltonians are described by banded matrices in the 
energy eigenbasis~\cite{BeugelingMoessnerHaquePRE2015, AradKuwaharaLandauJSM2016, DeOliveiraEtAlNJP2018}. In particular, 
Refs.~\cite{AradKuwaharaLandauJSM2016, DeOliveiraEtAlNJP2018} have shown a bound of the form 
\begin{equation}
 |X_{k\ell}| \le \|X\| \exp\big(-a[|\omega_{k\ell}| - b\ln(c|\omega_{k\ell}|)]\big)
\end{equation}
for suitable constants $a$, $b$ and $c$ describing local properties of $H$ and $X$. While it is 
possible that these constants behave unfavourably in a particular application (resulting in a matrix that is not 
narrowly banded), they are importantly independent of the system size. 

Similarly, also the ETH conjectures that thermodynamically relevant observables are banded matrices characterized 
by a smooth envelope function $F(\omega_{k\ell})$ that decays for large $\omega_{k\ell}$ (see below). However, generic
results about the decay of this function are not known to the best of our knowledge.

Finally, another class of slow observables is given by extensive sums of local observables, which we like to 
illustrate with an example. Consider the 1D Ising model 
$H = \sum_{i=1}^L \sigma_z^i + \sum_{i=1}^L \sigma_x^i \sigma_x^{i+1}$ of length $L$ with periodic boundary conditions 
and $\sigma_{x,y,z}^i$ the standard Pauli matrices of spin $i$ (we ignore any prefactors because they do not change our 
point). Moreover, let the observable be the total magnetization $X = \sum_{i=1}^L \sigma_z^i$. Then, one finds that 
all operator norms in Eq.~(\ref{eq slow observable norm}) scale with $L$ and Eq.~(\ref{eq slow observable norm}) 
reduces to $L \ll L^2$, which is clearly satisfied for large $L$. This example illustrates the important point that 
observables can be slow although it is not possible to identify a perturbative parameter $\epsilon$ in the Hamiltonian, 
as it was possible for the class of observables characterized by Eq.~(\ref{eq slow observables class 1}). 

While we have focused here on presenting generic properties of slowness, they do not guarantee an \emph{isotropic} or 
\emph{unbiased} diffusion in Hilbert space as described in our intuitive picture in Sec.~\ref{sec setup}. Understanding 
this is much more subtle and closely related to the complicated problem of ergodicity. In our case, ergodicity
(exploration of the full microcanonical energy shell during the dynamics) cannot happen for reasons explained in
Sec.~\ref{sec setup}. However, what matters is a sufficiently smooth observable such that even comparably short
evolution times give representative (``typical'') averages~\cite{KhinchinBook1949}. It is a known and hard problem to
identify these observables rigorously, but we return to this question it greater detail in Sec.~\ref{sec multiple}
in context of multiple slow observables after having developed an understanding for a single observable.

\subsection{ETH and Levy's Lemma}
\label{sec ETH Levy}

In our derivation we make use of two tools, which have become widely used by now. First, the ETH conjectures that 
matrix elements in the energy eigenbasis of thermodynamically relevant observables $X$ can be written as 
\begin{equation}\label{eq ETH def}
 X_{k\ell} = \delta_{k\ell} \lr{X}_\text{mic} + \frac{1}{\sqrt{D}}F(\omega_{k\ell}) R_{k\ell}.
\end{equation}
Here, $\lr{X}_\text{mic}$ is the expectation value of $X$ with respect to the microcanonical 
ensemble~(\ref{eq microcanonical ensemble}), $F(\omega)$ is a smooth function of order one for observables with a 
second central moment (or variance in the microcanonical ensemble) of order one, 
$\mbox{tr}\{X^2\}/D - \mbox{tr}\{X\}^2/D^2 = \C O(1)$, and $R_{k\ell}$ are pseudorandom numbers of zero mean and unit 
variance. How ``random'' the $R_{k\ell}$ behave is under current investigation~\cite{FoiniKurchanPRE2019, 
ChanDeLucaChalkerPRL2019, MurthySrednickiPRL2019, RichterEtAlPRE2020, BrenesEtAlPRE2021, WangEtAlPRL2022, 
DymarskyPRL2022}. Finally, note that the ETH is a \emph{hypothesis}, but it is considered to hold for a wide
class of many-body systems in nature, see Refs.~\cite{DAlessioEtAlAP2016, DeutschRPP2018, MoriEtAlJPB2018} and
references therein for more information.

Our second tool is Levy's Lemma. To state it precisely, let $f:\mathbb{S}^{2D-1}\rightarrow\mathbb{R}$ be any 
function defined on the hypersphere of dimension $2D-1$. Moreover, let $\eta$ be the Lipschitz constant of $f$ with 
respect to the Euclidean space $\mathbb{R}^{2D}$, which is the natural embedding of $\mathbb{S}^{2D-1}$. If $f$ is 
differentiable, then $\eta = \sup|\nabla f|$. Moreover, let $\lr{f} = \mu[f(\psi)]$ denote the Haar random average of 
$f$ over the hypersphere $\mathbb{S}^{2D-1}$. Note that the Haar measure is the only measure invariant under all 
unitary transformations and therefore the natural unbiased measure on the sphere. Then, Levy's Lemma says that 
\begin{equation}
 \mu\big[|f-\lr{f}| > \epsilon \big] \le 4 \exp\left(-\frac{\epsilon^2 2D}{9\pi^3\eta^2}\right).
\end{equation}
One easily notices that even for small $\epsilon$ the right hand side quickly tends to zero for a sufficiently large 
dimension $D$. Thus, colloquially speaking, Levy's Lemma says that every ``nice'' function $f(\psi)$ on a high
dimensional hypersphere is essentially constant, i.e., it varies very little with varying $\psi$. Levy's Lemma gives 
typicality arguments a firm mathematical basis and it has been used to show that thermal equilibrium states 
are ubiquituous with respect to the Haar measure~\cite{PopescuShortWinterNatPhys2006}, among other 
applications~\cite{LindenEtAlPRE2009, FigueroaRomeroModiPollockQuantum2019, FigueroaRomeroPollockModiCP2021, 
ReimannPRL2015}. In general, it is a consequence of a phenomenon known as 
\emph{measure concentration}~\cite{TalagrandAP1996, MilmanSchechtmanBook2001}. 

\section{Classicality}
\label{sec classicality}

How to explain the emergence of classical behaviour from an underlying quantum description is an important
foundational and, nowadays, also a practical very relevant question. Clearly, the quantum-to-classical boundary is
not one-dimensional and there are \emph{many} ways to define it. For instance, one might use Bell inequalities
to find out whether a bipartite quantum state has correlations, which cannot be explained classically. This certainly
legitimate characterization, however, only probes \emph{static} quantum features of a \emph{state}. Here, instead, we
are interested in a \emph{process} and the question whether the \emph{dynamics} of $X$ reveals quantum features.
Our characterization is therefore based on the following question (see also Refs.~\cite{GemmerSteinigewegPRE2014,
SchmidtkeGemmerPRE2016, SmirneEtAlQST2018, StrasbergDiazPRA2019, MilzEtAlQuantum2020, MilzEtAlPRX2020,
StrasbergSP2023}): Can an experimenter distinguish the measurement statistics of $X$ from a classical
stochastic process?

\begin{figure}[t]
 \centering\includegraphics[width=0.35\textwidth,clip=true]{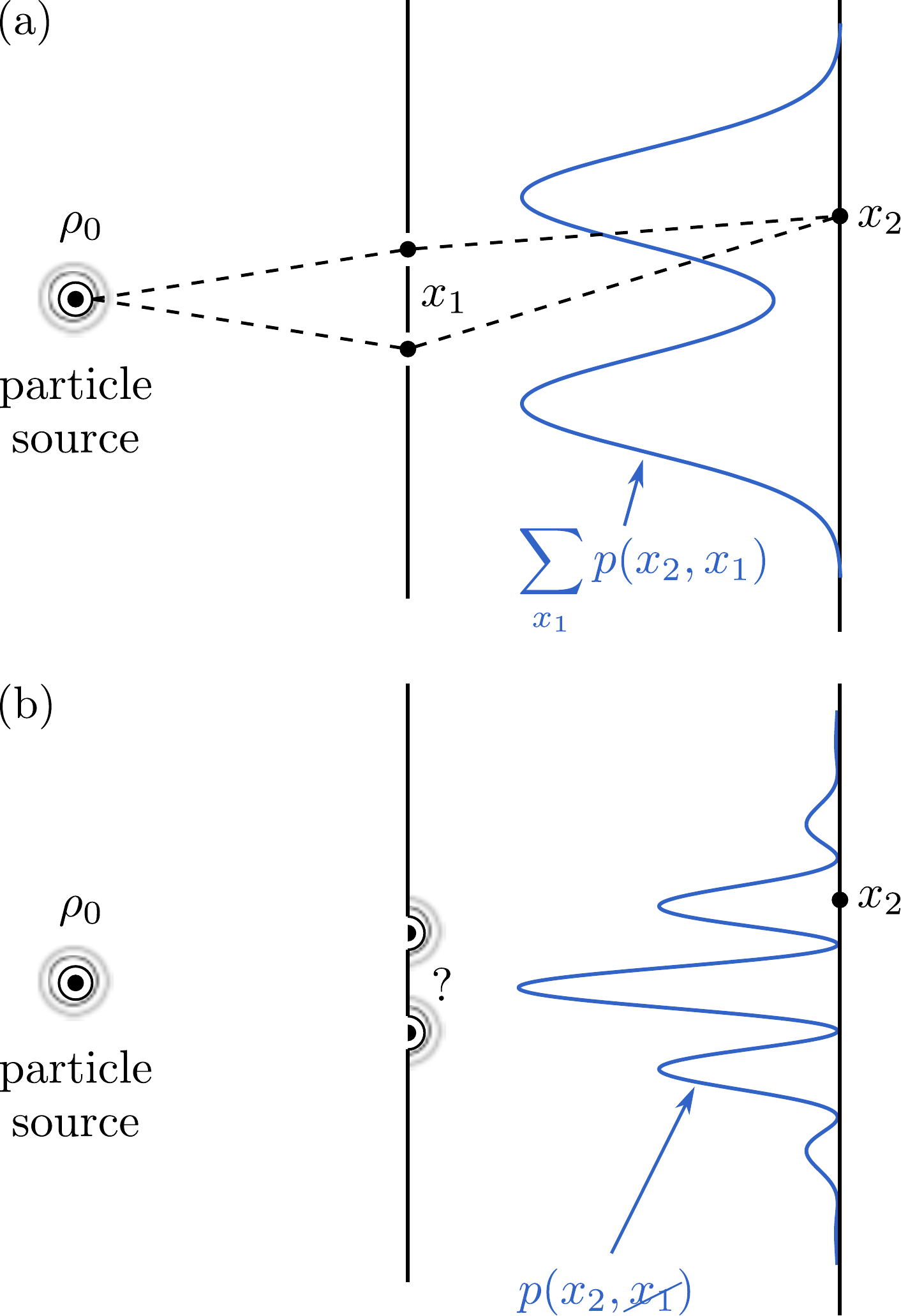}
 \label{fig double slit}
 \caption{A coherent source of particles $\rho_0$ is sent to a detection screen through a double slit. (a)~The
 particle's location at the double slit ($x_1$) and at the detection screen ($x_2$) is measured, allowing to speak
 about a definite trajectory (indicated by dashed lines) but causing a loss of wave properties. (b)~Only the
 position at the detection screen is measured, resulting in an interference pattern.
 Clearly, Eq.~(\ref{eq Kolmogorov}) is broken. }
\end{figure}

To define this rigorously, we denote the probability to obtain outcomes $x_n,\dots,x_1$ at times $t_n>\dots>t_1$ as
\begin{equation}
 \begin{split}\label{eq probability stochastic process}
  p&(x_n,\dots,x_1) = \\
  & \mbox{tr}\{\Pi_{x_n}U_n\dots U_2 \Pi_{x_1} U_1\rho(t_0) U_1^\dagger \Pi_{x_1} U_2^\dagger \dots U_n^\dagger\},
 \end{split}
\end{equation}
where $\rho(t_0)$ is some initial state and $U_k = e^{-iH(t_k-t_{k-1})}$ the unitary time evolution operator from 
$t_{k-1}$ to $t_k$. Note that the probabilities~(\ref{eq probability stochastic process}) can be experimentally
reconstructed by repeated projective measurements of the (otherwise) isolated quantum system and statistical sampling.
Next, suppose that the experimenter decides \emph{not} to measure the system at some time $t_\ell$ with $\ell<n$.
We denote the probability to obtain apart from $x_\ell$ the same outcomes $x_n,\dots,x_{\ell+1},x_{\ell-1},\dots,x_1$
by $p(x_n,\dots,\cancel{x_\ell},\dots,x_1)$, which is obtained from Eq.~(\ref{eq probability stochastic process})
by dropping the two projectors $\Pi_{x_\ell}$. Now, the defining property of a classical stochastic process
is~\cite{KolmogorovBook2018}
\begin{equation}\label{eq Kolmogorov}
 \sum_{x_\ell} p(x_n,\dots,x_\ell,\dots,x_1) = p(x_n,\dots,\cancel{x_\ell},\dots,x_1),
\end{equation}
which is also known as the Kolmogorov consistency condition or ``probability sum rule''. In words, a classical
stochastic process is characterized by the property that \emph{not measuring is equivalent to averaging over the
respective measurement outcomes}. Clearly, for a quantum process Eq.~(\ref{eq Kolmogorov}) is in general not satisfied
because quantum measurements are \emph{disturbing} and the classical example of the double slit experiment in
Fig.~\ref{fig double slit} is used to illustrate the breaking of the Kolmogorov consistency condition in the
quantum world.

We emphasize again that Eq.~(\ref{eq Kolmogorov}) is not the only way to characterize classicality, but it has
a number of desirable features. Among them are, for instance, that the Kolmogorov consistency condition implies the 
validity of all Leggett-Garg inequalities~\cite{EmaryLambertNoriRPP2014}. It is also implied by the ``consistency
condition'' imposed in the histories interpretation of quantum mechanics~\cite{GriffithsJSP1984, OmnesRMP1992, 
Griffiths2019} and it guarantees that we can apply classical reasoning to understand the physics \emph{even in absence}
of measurements. Importantly, however, testing the validity of Eq.~(\ref{eq Kolmogorov}) requires only the ability to
measure $X$ and is independent of the interpretation of quantum mechanics. Finally, note that a quantum system can 
behave classical with respect to one observable $X$, but not with respect to another observable $Y$. An extended 
discussion, in particular in relation to other approaches, is provided in Ref.~\cite{StrasbergSP2023}.

We proceed by confirming that the repeated randomness assumption implies classical measurement statistics.
To this end note that all that the experimenter knows about the state at time $t_\ell$ is some probability distribution
$p(x_\ell|x_{\ell-1},\dots,x_1)$ conditioned on earlier results $x_{\ell-1},\dots,x_1$. The state which maximizes the 
entropy given that information is 
\begin{equation}\label{eq max ent state}
 \rho_\text{max ent}(t_\ell) \equiv \sum_{x_\ell} p(x_\ell|x_{\ell-1},\dots,x_1) \frac{\Pi_{x_\ell}}{V_{x_\ell}}.
\end{equation}
Note that we would have obtained the same state by applying the equal-a-priori-probability postulate, which associates 
to every subspace $x$ the state $\Pi_x/V_x$ independent of the probability distribution. For our setup the maximum 
entropy principle and the equal-a-piori-probability postulate thus turn out to be the same. In general, this is not the 
case, although both principles still express the same basic idea: maximize ignorance given the experimentally 
available information.

Next, notice that the state~(\ref{eq max ent state}) is block diagonal with respect to the $\Pi_x$ basis.
This implies in particular that
\begin{equation}\label{eq dephasing}
 \sum_x \Pi_x\rho_\text{max ent}(t_\ell)\Pi_x = \rho_\text{max ent}(t_\ell),
\end{equation}
where the operation on the left hand side is a ``dephasing'' operation (with respect to $X$). One easily
confirms that the validity of Eq.~(\ref{eq dephasing}) implies the validity of Eq.~(\ref{eq Kolmogorov}).

It might be tempting to explain the block diagonal form of the state~(\ref{eq max ent state}) by using
decoherence. However, we are dealing with an 
isolated and not an open system here. Assuming the validity of a block diagonal state for all times either implies that 
the probabilities $p(x_t) = \mbox{tr}\{\Pi_x\rho(t)\}$ do not change in time (i.e., $X$ is a conserved quantity) or that 
the von Neumann entropy of the state is not conserved. To illustrate this, consider an initially pure state 
$\rho(t_0) = |\psi_0\rl\psi_0|$. If there are non-trivial dynamics with some probability flux, say, from $x$ to $y$, 
then this implies that $\Pi_x\rho(t)\Pi_y \neq 0$ for $x\neq y$ for at least some times $t$, i.e., the state cannot be
block diagonal. The main contribution of this section is to give a generic explanation for the emergence of classical 
measurement statistics obeying Eq.~(\ref{eq Kolmogorov}) despite the presence of a lot of coherences.

In the following, we consider three arbitrary times $t_2>t_1>t_0$ and a \emph{nonequilibrium} initial
state $\rho(t_0)$. The probability to measure $x_2$ and $x_1$ is
\begin{equation}
 p(x_2,x_1) = \mbox{tr}\{\Pi_{x_2}U_2\Pi_{x_1}U_1\rho(t_0)U_1^\dagger\Pi_{x_1}U_2^\dagger\}
\end{equation}
and the probability to \emph{only} measure $x_2$ is 
$p(x_2,\cancel{x_1}) = \mbox{tr}\{\Pi_{x_2}U_2U_1\rho(t_0)U_1^\dagger U_2^\dagger\}$. If the process is classical, 
the quantum contribution 
\begin{equation}\label{eq Q}
 Q \equiv p(x_2,\cancel{x_1}) - \sum_{x_1} p(x_2,x_1) ~ \in [-1,+1]
\end{equation}
should be zero. Below, we estimate that $Q \sim D^{-\alpha}$ with $\alpha>0$, i.e., $Q$ is exponentially small in the 
particle number and thus essentially zero for all reasonable experiments involving many-body systems. Note, however, 
that the precise value of the exponent $\alpha$ depends on the situation and is not universal.

We emphasize that smallness of $Q$ in Eq.~(\ref{eq Q}) does not imply a similarly small deviation from equality
in Eq.~(\ref{eq Kolmogorov}) in full generality, rather the extension to an \emph{arbitrary} number $n$ of time steps 
requires a separate, technically demanding argument. We recall that also the decoherence approach shows only 
decoherence of the open system density matrix, which is not sufficient to compute $n$-time correlation functions 
without additional assumptions.

The following derivation is the technically most involved part of the present paper. This comes from the fact that
we try to derive a statement valid for all slow and coarse observables satisfying the ETH, even out of equilibrium.
Since the ETH is assumed to hold for a wide class of realistic many-body systems in nature, our statement
is widely applicable. However, recalling that nonequilibrium many-body dynamics result from a complex
interplay between the initial state, the observable and the Hamiltonian (or time-evolution operator), and recalling
that there is no systematic way (e.g., in form of a perturbation theory) to take their intricate correlations
into account, it becomes evident that we must restrict our derivation to estimates and approximations. Although there
is no justification from first principles known for them, we believe them to be plausible.

Therefore, the derivation below cannot be claimed to have the status of a rigorous mathematical theorem.
Counterexamples do exist, and we also partially address them in this article. Nevertheless, the derivation below
adds considerable evidence that counterexamples are not generic. Given that, to the best of our knowledge, a
microscopic derivation of the Kolmogorov consistency condition has never been presented for an isolated system,
we find the wide scope of the derivation below certainly remarkable and, at the end, also
intuitively correct: since human senses are coarse in \emph{space and time}, this explains the emergence of a classical
world for many observables even though the true quantum state might not be diagonal (decohered) in all the eigenbases
of these observables.

\subsection{Microscopic derivation}
\label{sec classicality derivation}

How can it be that a state containing a lot of coherences gives rise to classical measurement statistics?
Roughly speaking, the idea is that the contribution of the coherences to the probabilities appearing 
in Eq.~(\ref{eq Kolmogorov}) is given by a sum of many very small and randomly oscillating terms such that the chance 
that all coherences for a coarse observable of a non-integrable many-body system align up \emph{in phase} becomes very 
small. Thus, the derivation below is essentially of statistical nature, similar to the derivation of the second law.
From that perspective it is not suprising that we have to content ourselves with some rough but reasonable estimates 
\emph{in general} (more specific models might allow, of course, for more specific conclusions, see also 
Ref.~\cite{StrasbergSP2023}). Similar to the second law, for each observable $X$ one can find precisely tuned states
and times for which classicality is \emph{violated}, yet these situations are non-generic.

\subsubsection*{Step 1: Experimentally realistic initial state dependence}

We begin with considerations about the initial state $\rho(t_0)$. In general, the further away the state is from
equilibrium the stronger it is correlated with the matrix elements of $X$. On the other 
hand, since $X$ is a coarse observable, knowing the probabilities $p(x_0) = \mbox{tr}\{\Pi_{x_0}\rho(t_0)\}$ does not 
completely determine $\rho(t_0)$, but still leaves room for some freedom. Finding the right balance between this 
freedom and the required correlations is what makes the problem delicate. Here, we solve this problem by thinking 
experimentally and by \emph{explicitly modelling the state preparation procedure}. To this end, let 
$\psi_0 = |\psi_0\rl\psi_0|$ be a pure state \emph{before} the preparation. Then, \emph{any} state preparation can be 
modeled by an \emph{instrument} $\{\C A_r\}$~\cite{KrausBook1983, BreuerPetruccioneBook2002, MilzModiPRXQ2021,
StrasbergBook2022}. Here, each $\C A_r$ is a completely positive map labeled by some (abstract) measurement outcome $r$ 
and $\C A \equiv \sum_r \C A_r$ is a completely positive and trace preserving map. This means that the state preparation 
given outcome $r$ can be written as
\begin{equation}
 \psi_0 \mapsto \rho(t_0) = \C A_r\psi_0 = \sum_{\alpha=1}^g K_\alpha(r)\psi_0 K_\alpha^\dagger(r)
\end{equation}
with operators $K_\alpha(r)$ satisfying $\sum_{\alpha=1}^g K_\alpha^\dagger(r) K_\alpha(r) \le I$, where $I$ denotes
the identity in $\C H$, and $\sum_r\sum_{\alpha=1}^g K_\alpha^\dagger(r) K_\alpha(r) = I$.
The quantum term~(\ref{eq Q}) conditioned on this preparation reads explicitly 
\begin{equation}
 Q_r(\psi_0) = \sum_{x_1\neq x'_1} \mbox{tr}\{\Pi_{x_2}U_2\Pi_{x_1}U_1(\C A_r\psi_0)U_1^\dagger\Pi_{x'_1}U_2^\dagger\}.
\end{equation}
So far, there has been no assumption.

To make progress, we now assume that the state $\psi_0$ \emph{prior} to the preparation can be randomly choosen,
i.e., it is distributed with respect to the Haar measure $\mu$. The philosophy behind this choice is that the
experimenter starts the experiment at time $t_0$ and any information about the state prior to $t_0$ is irrelevant for
the description, i.e., any possible nonequilibrium source is ``switched on'' by the preparation. Then,
since $\lr{\psi_0} \equiv \mu(\psi_0) = I/D$, we find on average
\begin{align}\label{eq mu Q}
 &\lr{Q_r} = \\ 
 &\frac{1}{D} \sum_{x_1\neq x'_1} \sum_{\alpha=1}^g 
 \mbox{tr}\{\Pi_{x_2}U_2\Pi_{x_1}U_1 K_\alpha(r) K_\alpha^\dagger(r) U_1^\dagger\Pi_{x'_1}U_2^\dagger\}. \nonumber
\end{align}
Next, Levy's Lemma implies that 
\begin{equation}
 \mu\big(|Q_r(\psi_0)-\lr{Q_r}|>\epsilon) \le 4 \exp\left(-\frac{2\epsilon^2 D}{9\pi^3\eta^2}\right).
\end{equation}
To find the Lipschitz constant $\eta$ of $Q$, we write 
\begin{align}
 &Q_r(\psi_0) = \label{eq Lipschitz Q} \\
 &\sum_{x_1} \sum_{\alpha=1}^g \lr{\psi_0|K_\alpha^\dagger(r) U_1^\dagger(1-\Pi_{x_1})U_2^\dagger \Pi_{x_2}U_2\Pi_{x_1}U_1 K_\alpha(r)|\psi_0}, \nonumber
\end{align}
and use the following five facts. First, the Lipschitz constant of a sum of Lipschitz continuous functions $f_i$ with 
Lipschitz constants $\eta_i$ is bounded by $\sum_i\eta_i$. Second, the Lipschitz constant of $\lr{\psi|A|\psi}$ is 
bounded by $2\|A\|$ for any operator $A$~\cite{PopescuShortWinterArXiv2006}. Third, $\|AB\| \le \|A\|\|B\|$ for all 
operators $A$ and $B$. Fourth, by the right polar decomposition theorem, we can write 
$K_\alpha(r) = V_\alpha(r)\sqrt{P_\alpha(r)}$ for some unitary $V_\alpha(r)$ and a positive operator 
$P_\alpha(r) = K_\alpha^\dagger(r) K_\alpha(r) \le I$. Fifth, $\|U\| = 1$ for any unitary $U$, $\|\Pi\| = 1$ for any 
projector $\Pi$ and $\|P_\alpha(r)\| \le 1$. Altogether, we then find $\eta \le 2Mg$.

Thus, Levy's Lemma shows that $\lr{Q_r} \approx Q_r(\psi_0)$ for the overwhelming majority of $\psi_0$ if
\begin{equation}
 D \gg \frac{18\pi^3M^2g^2}{\epsilon^2}.
\end{equation}
This is satisfied for a many-body system and a reasonable $\epsilon$ provided that the observable is 
\emph{coarse} such that $M\ll D$. Consequently, in the following we focus on showing that $\lr{Q}$ is small, which 
implies that $Q(\psi_0)$ is also small for the overwhelming majority of $\psi_0$. We remark that this result 
establishes already \emph{classicality at equilibrium} for any coarse observable (independent of its slowness) 
because at equilibrium there is no state preparation, i.e., $\C A = \C I$ with $\C I$ the identity map, 
such that $\lr{Q} = 0$.

We continue by specifying $\C A$ further, on which we did not put any restrictions so far. This time we use
the \emph{left} polar decomposition theorem to write $K_\alpha(r) = \sqrt{P'_\alpha(r)} V_\alpha(r)$ for some unitary 
$V_\alpha(r)$ and a positive operator $P'_\alpha(r) = K_\alpha(r) K_\alpha^\dagger(r)$, which is in general different 
from the $P_\alpha(r)$ appearing above. This yields 
\begin{equation}
 \lr{Q_r} = \frac{1}{D} \sum_{x_1\neq x'_1} \sum_{\alpha=1}^g 
 \mbox{tr}\{\Pi_{x_2}U_2\Pi_{x_1}U_1 P'_\alpha(r) U_1^\dagger\Pi_{x'_1}U_2^\dagger\}.
\end{equation}
Of course, we want that the state preparation $\C A_r$ is related to the observable $X$. It therefore appears 
reasonable to demand that $P'_\alpha(r)$ is functionally dependent on $X$ such that we can write (by Taylor
expansion) $P'_\alpha(r) = \sum_x p'_{\alpha,x,r}\Pi_x$, where the numbers $p'_{\alpha,x,r}$ are positive.
The philosophy behind this choice is related to the idea that the experimenter has no \emph{precise} control
of the microstate: they are ``only'' allowed to perform unitaries, measurements of the observable $X$ and 
post-selection. Indeed, recalling the second-law like analogy, it is clear that ``violations'' of the second law can 
be easily generated if one assumes the ability to control the velocity of every gas molecule in the air surrounding
us. Similarly, violations of classicality can be generated by a microscopic fine-tuning of the coherences in the
initial state.

Taken together, we thus arrive at the expression
\begin{equation}
 \begin{split}\label{eq Q r final}
  \lr{Q_r} =& \sum_{x_0} \sum_{\alpha=1}^g p'_{\alpha,x_0,r}\frac{V_{x_0}}{D} \\
  & \times \sum_{x_1\neq x'_1} \mbox{tr}\left\{\Pi_{x_2}U_2\Pi_{x_1}U_1 \frac{\Pi_{x_0}}{V_{x_0}} U_1^\dagger\Pi_{x'_1}U_2^\dagger\right\}.
 \end{split}
\end{equation}
Now, note that the first line equals the probability $P(r)$ to prepare the state $\C A_r\psi_0$ on average: 
\begin{equation}
 P(r) \equiv \sum_{x_0} \sum_{\alpha=1}^g p'_{\alpha,x_0,r}\frac{V_{x_0}}{D} = \mbox{tr}\{\C A_r\lr{\psi_0}\}.
\end{equation}
This probability could be small on its own and should not influence the estimate of $\lr{Q_r}$, i.e., we are interested 
in showing that $\lr{Q_r}/P(r)$ is small for all $r$, for which it is sufficient to show that the term in the second 
line of Eq.~(\ref{eq Q r final}) is small, which we denote by
\begin{equation}\label{eq q}
 q(x_2,x_0) \equiv \sum_{x_1\neq x'_1} 
 \mbox{tr}\left\{\Pi_{x_2}U_2\Pi_{x_1}U_1 \frac{\Pi_{x_0}}{V_{x_0}} U_1^\dagger\Pi_{x'_1}U_2^\dagger\right\}.
\end{equation}
Thus, as a first summary, we have reduced the task of proving the smallness of $Q$, which is a \emph{three}-point 
correlator in terms of the projectors $\Pi_x$ with \emph{unknown} correlations to the initial state $\rho(t_0)$, 
to proving the smallness of $q(x_2,x_0)$, which is a \emph{four}-point correlator \emph{without} any unknown initial 
state dependence.

\subsubsection*{Step 2: ETH for realistic projectors}

Our goal is to use an ETH ansatz of the form~(\ref{eq ETH def}) for the \emph{projectors} $\Pi_x$, i.e.,
\begin{equation}\label{eq ETH}
 (\Pi_x)_{k\ell} = \delta_{k\ell}\frac{V_x}{D} + \frac{F_x(\omega)R_{k\ell}(x)}{\sqrt{D}}
\end{equation}
for some smooth function $F_x(\omega)$ and pseudorandom coefficients $R_{k\ell}(x)$ of zero mean and unit variance. 
For an arbitrary observable $X$ with arbitrary projectors this ansatz appears questionable, but our observable $X$ is 
coarse and the sum of a few projectors only. An ETH ansatz of the form~(\ref{eq ETH}) then likely holds as it is not 
possible to generate a pseudorandom number by adding up a few non-random numbers. This point can be strengthened by 
using random matrix theory, and the validity of the ETH for coarse projectors is also a central point of Ref.~\cite{RigolSrednickiPRL2012}.

However, the ETH ansatz holds for operators whose second central moment is of order one, but since $\Pi_x$ is a
projector, this is not necessarily guaranteed. To fix this, we need to rescale $F_x(\omega)$ to ensure 
$\mbox{tr}\{\Pi_x^2\} = \mbox{tr}\{\Pi_x\} = V_x$ as required from a projector. To do so, we assume that 
$F_x(\omega) = F_x\Theta(|\omega|-\sqrt{M}\delta E)$ can be modeled by a rescaled Heaviside step function.
Indeed, in Appendix~\ref{sec app proof bandedness} we show that if $X$ has bandwidth $\delta E\ll\Delta E$ (due to its 
slowness) then $\Pi_x$ has roughly a bandwidth $\sqrt{M}\delta E \ll \Delta E$ (recall that $M$ is number of projectors 
or measurement results), which justifies the truncation of $F_x(\omega)$ for large enough $\omega$. Moreover, assuming 
a constant $F_x$ for $|\omega| \le \sqrt{M}\delta E$ is not a strong assumption for two reasons. First, it becomes clear 
from our result below that it is not crucial that the $R_{k\ell}(x)$ have \emph{exactly} unit variance, i.e., a mild 
variation of $F_x(\omega)$ with $\omega$ can be conveniently absorbed in the pseudorandom coefficients. Second, existing 
numerical studies about the off-diagonal elements in the ETH ansatz indeed indicate that $F_x(\omega)$ is often roughly 
constant up to some cutoff frequency, where it starts to quickly fall off~\cite{BeugelingMoessnerHaquePRE2015, 
ChanDeLucaChalkerPRL2019, KhaymovichHaqueMcClartyPRL2019, BrenesEtAlPRL2020, SantosPerezBernalTorresHerreraPRR2020, 
LeBlondRigolPRE2020, BrenesGooldRigol2020, RichterEtAlPRE2020, BrenesEtAlPRE2021}. Indeed, a specifically structured 
profile of the off-diagonal elements can cause anomalous behaviour~\cite{KnipschilGemmerPRE2020} to which we return 
in Sec.~\ref{sec multiple}.

After these preliminary agreements we proceed to find
\begin{align}
 \mbox{tr}\{\Pi_x^2\} &= \sum_{k,\ell} \lr{k|\Pi_x|\ell}\lr{\ell|\Pi_x|k} \\
 &= \sum_k \frac{V_x^2}{D^2} + 2\sum_k \frac{V_x}{D}\frac{F_x^2 R_{kk}}{\sqrt{D}} + 
 \sum_{k\approx\ell} \frac{F_x|R_{k\ell}(x)|^2}{D}, \nonumber
\end{align}
where we used the bandedness of $\Pi_x$, which allows us to restrict the sum over all $k$ and $\ell$ to a sum over 
all $k$ and all $\ell$ \emph{close to} $k$, which we indicate by writing $\sum_{k\approx\ell}$. This sum runs over 
$Dd$ many coefficients (instead of $D^2$), where $d$ denotes the number of states defined by the band width 
$\sqrt{M}\delta E$ of $\Pi_x$, i.e., $(\Pi_x)_{k\ell} = 0$ for all $|k-\ell|\ge d$.

Next, to evaluate the sum over $k$ we use an estimate that we will also repeatedly use below. Recall that the
$R_{k\ell}(x)$ are pseudorandom numbers of zero mean and unit variance. Summing over them can therefore be pictured
as a random walk in the complex plane with (average) unit step size and no preferred direction. Clearly, on average
such a random walk remain at the origin of the complex plane, but we are here interested in estimating the
spread of the distribution to characterize typical fluctuations. Since it is known that the standard deviation of a
random walk with unit step size equals the square root of the number of steps, we can set
$\sum_k R_{kk}(x) \approx \sqrt{D}r$ with $r$ some random number of zero mean and unit variance. Moreover, replacing 
$|R_{k\ell}(x)|^2 \approx 1$, we obtain 
\begin{equation}
 \mbox{tr}\{\Pi_x^2\} = \frac{V_x^2}{D} + 2r\frac{V_x F_x}{D} + F_x^2d,
\end{equation}
which must equal $\mbox{tr}\{\Pi_x\} = V_x$. Thus, we find the condition 
\begin{equation}
 F_x \approx \sqrt{\frac{V_x}{d}}\sqrt{1-\frac{V_x}{D}} - \frac{rV_x}{Dd},
\end{equation}
where we disregarded a term proportonal to $1/D^2$ in the square root.

To avoid a tangle of case studies below, we are interested in a worst case scenario for $F_x$. Typically,
$V_x<D$ will scale with $D$, but depending on the observable the scaling for different $x$ can be very different. 
Note, however, that always multiple $F_x$ for different $x$ enter Eq.~(\ref{eq q}). In a scenario where $M$ is small, 
all $F_x$ can be huge if we look at an observable characterized by equal volumes for each $x$: $V_x = D/M$.
Inserting this we find up to negligible corrections 
\begin{equation}\label{eq F variance}
 F_x \approx \sqrt{\frac{D}{Md}}.
\end{equation}
In the following, we consider this equal-volume-case only, which we have found to provide the worst case scenario,
but keeping track of different $F_x$ for a more refined analysis poses no conceptual challenges.
Moreover, to safe space in the notation, we write $R_{k\ell}^0 \equiv R_{k\ell}(x_0)$, etc.

\subsubsection*{Step 3: Energy level shifts}

As a matter of fact, $q(x_2,x_0)$ will contain exponential phases of the form $e^{iE_kt}$. The precise value of
them are not known (because $E_k$ is not known and also because we are interested in an estimate valid for all times)
and in addition they can be correlated with the pseudorandom coefficients $R_{k\ell}(x)$. To make our live simpler
we adapt the following line of reasoning~\cite{DabelowReimannPRL2020, RichterEtAlPRE2020b, DabelowReimannJSM2021}.

We first recall a central result of quantum chaos theory~\cite{BrodyEtAlRMP1981, HaakeBook2010,
DAlessioEtAlAP2016}, namely that the spectrum of a generic non-integrable many-body system looks immensely dense with 
a mean level spacing $\delta e$ and approximately statistically independent level spacings $E_{k+1} - E_k$ with variance 
$\delta e^2$ (note that we assume the eigenenergies $\{E_k\}$ to be ordered). Thus, we set $E_k \equiv k\delta e + c_k$ 
with $c_k$ a random correction term, which is very small (of the order of $\delta e$), and approximate in all 
time-evolution operators $\exp(i E_k t) \approx \exp(i k\delta e t)$. This appears justified for all times 
$t \ll 1/\delta e$, where $1/\delta e$ is the immensely long Heisenberg time mentioned already in Sec.~\ref{sec slow}, 
which in particular is much longer than the thermalization time.

\subsubsection*{Step 4: Estimation of $q$}

We can finally turn to the evaluation of $q(x_2,x_0)$ from Eq.~(\ref{eq q}). To make our lives as easy as
possible, it is useful to note some general properties of it. First, $q(x_2,x_0) = 0$ for either $t_1\rightarrow0$ or
$t_2\rightarrow0$, which is a consequence of the quantum Zeno effect. Second, we confirm $\sum_{x_2} q(x_2,x_0) = 0$, 
a property which has its origin in the normalization of the probabilities $p(x_2,x_1)$ and $p(x_2)$ and we will use
it to assume without loss of generality 
\begin{equation}
 x_0\neq x_2.
\end{equation}

Next, we express $q(x_2,x_0)$ in the energy eigenbasis
\begin{equation}
 \begin{split}
  q(x_2,x_0) =\frac{1}{V_{x_0}} & \sum_{x_1\neq x'_1} \sum_{k,\ell,m,n} e^{-i\omega_{k\ell}t_2} e^{-i\omega_{mn}t_1} \\
  & (\Pi_{x_2})_{\ell k} (\Pi_{x_1})_{km} (\Pi_{x_0})_{mn} (\Pi_{x'_1})_{n\ell}
 \end{split}
\end{equation}
and use the ETH ansatz from Eq.~(\ref{eq ETH}) with $F_x$ as in Eq.~(\ref{eq F variance}). Since the ETH 
ansatz~(\ref{eq ETH}) contains two terms and there are four projectors, $q(x_2,x_0)$ can be split into eight terms. 
However, if we set $(\Pi_{x_2})_{\ell k} = \delta_{k\ell} V_{x_2}/D$ or $(\Pi_{x_0})_{mn} = \delta_{mn} V_{x_0}/D$, 
it follows that $q(x_2,x_0) = 0$ because $x_1\neq x'_1$. We are thus left with estimating
\begin{equation}
 \begin{split}
  q(x_2,x_0) = \frac{1}{Dd}
  &\sum_{x_1\neq x'_1} \sum_{k\approx\ell\approx m\approx n} e^{-i\omega_{k\ell}t_2} e^{-i\omega_{mn}t_1} \\ 
  & R_{\ell k}^2 (\Pi_{x_1})_{km} R_{mn}^0 (\Pi_{x'_1})_{n\ell},
 \end{split}
\end{equation}
where $\sum_{k\approx\ell\approx m\approx n}$ denotes a sum over all quadruples $(k,\ell,m,n)$, where each index pair
is at most a distance $d$ away from each other.
We split $q(x_2,x_0) = q_1 + q_2 + q_3 + q_4$ into four terms according to the prescription
\begin{align}
 q_1: &~~~ (\Pi_{x_1})_{km} \rightarrow \frac{\delta_{km}}{M}, ~~~
 (\Pi_{x'_1})_{n\ell} \rightarrow \frac{\delta_{n\ell}}{M}, \\
 q_2: &~~~ (\Pi_{x_1})_{km} \rightarrow \frac{R_{km}^1}{\sqrt{Md}}, ~~~
 (\Pi_{x'_1})_{n\ell} \rightarrow \frac{\delta_{n\ell}}{M}, \\
 q_3: &~~~ (\Pi_{x_1})_{km} \rightarrow \frac{\delta_{km}}{M}, ~~~
 (\Pi_{x'_1})_{n\ell} \rightarrow \frac{R_{n\ell}^{1'}}{\sqrt{Md}}, \\
 q_4: &~~~ (\Pi_{x_1})_{km} \rightarrow \frac{R_{km}^1}{\sqrt{Md}}, ~~~
 (\Pi_{x'_1})_{n\ell} \rightarrow \frac{R_{n\ell}^{1'}}{\sqrt{Md}},
\end{align}
and estimate them separately in the following.

We start with $q_1$ using the slight shift in the energy levels as explained in step 3 above:
\begin{equation}
 q_1 = \frac{1}{DM^2d}
 \sum_{x_1\neq x'_1} \sum_{k\approx\ell} e^{-i(k-\ell)\delta e(t_2+t_1)} R_{\ell k}^2 R_{k\ell}^0.
\end{equation}
Since the terms no longer depend on $x_1$ and $x'_1$ we set $\sum_{x_1\neq x'_1} = M(M-1) \approx M^2$. Moreover,
we see that the time dependent phase only depends on the difference $\Delta \equiv k-\ell$ in the indices. Thus,
\begin{equation}
 q_1 = \frac{1}{Dd} \sum_{\Delta} e^{-i\Delta\delta e(t_2+t_1)} \sum_k R_{k-\Delta,k}^2 R_{k,k-\Delta}^0.
\end{equation}
where the sum over $\Delta$ is restricted to integers $|\Delta| \le d$. In complete analogy with the random walk
argument already made above we approximate
\begin{equation}
 q_1 \approx \frac{1}{\sqrt{D}d} \sum_{\Delta} e^{-i\Delta\delta e(t_2+t_1)} r(\Delta),
\end{equation}
where $r(\Delta)$ is some pseudorandom number of zero mean and unit variance depending on $\Delta$. Since we have
already used an assumption for the time dependent phases in Step 3, we like to avoid further assumptions here and
assume the worst case scenario where $e^{-i\Delta\delta e(t_2+t_1)}$ and $r(\Delta)$ are perfectly correlated (which is
clearly not realistic, but it can only weaken our final result). We thus assume that the sum over $\Delta$ scales like
$d$ (instead of $\sqrt{d}$ as expected from a random walk argument) and finally find
\begin{equation}
 q_1 \approx \frac{1}{\sqrt{D}}.
\end{equation}

We continue with $q_2$. Since the terms do not depend on $x'_1$, we set
$\sum_{x_1\neq x'_1} \approx M\sum_{x_1}$ and obtain
\begin{equation}
 \begin{split}
  q_2 \approx
  \frac{1}{DM^{1/2}d^{3/2}} &\sum_{x_1} \sum_{\Delta,\Delta'} e^{-i\Delta\delta et_2}e^{-i\Delta'\delta et_1} \\
  & \sum_\ell R_{\ell,\ell+\Delta}^2 R_{\ell+\Delta,\ell+\Delta'}^1 R_{\ell+\Delta',\ell}^0.
 \end{split}
\end{equation}
The last line is again estimated as $\sqrt{D}r(\Delta,\Delta',x_1)$ with some pseudorandom number
$r(\Delta,\Delta',x_1)$ of zero mean and unit variance. Assuming again a worst case scenario where this number is
perfectly correlated with the time dependent phases, we get the scaling
\begin{equation}
 q_2 \approx \frac{1}{DM^{1/2}d^{3/2}} Md^2\sqrt{D} = \sqrt{\frac{Md}{D}}.
\end{equation}

Next, we observe that the term $q_3$ is structurally identical to $q_2$ and gives rise to the same scaling.

Finally, we consider $q_4$:
\begin{equation}
 \begin{split}
  q_4 \approx
  \frac{1}{DMd^2} &\sum_{x_1\neq x_1'} \sum_{\Delta,\Delta'} e^{-i\Delta\delta et_2}e^{-i\Delta'\delta et_1} \\
  & \sum_{\ell\approx n} R_{\ell,\ell+\Delta}^2 R_{\ell+\Delta,n+\Delta'}^1 R_{n+\Delta',n}^0 R_{n,\ell}^{1'}.
 \end{split}
\end{equation}
Using the same estimates as above and assuming again the worst case scenario for the sums over $\Delta$ and $\Delta'$,
we arrive at
\begin{equation}
 q_4 \approx \frac{1}{DMd^2} M^2 d^2 \sqrt{Dd} = M\sqrt{\frac{d}{D}}.
\end{equation}

\subsubsection*{Summary and discussion}

The estimates we got for all terms scales with the Hilbert space dimension as $\sqrt{d/D}$ at worst, i.e., it is
given by the square root of the relative bandwidth of the projector of a slow and coarse observable relative to the
``width'' of the total Hilbert space. Since $d$ scales like $D^\beta$ with $\beta\in(0,1)$, we obtain the
scaling $D^{(\beta-1)/2} \equiv D^{-\alpha} = \C O(e^{-\alpha N})$ for some $\alpha>0$. Thus, unless $\beta$ is
extremely close to one, which will not happen for a coarse and slow observable, the quantum contribution to the
measurement statistics becomes exponentially suppressed in the particle number $N$. Note that it is not possible to
provide a universal exponent $\alpha$ as it depends on the observable $X$ and, as we will confirm numerically, also
on the particular initial state. Clearly, given the complexity of nonequilibrium many-body dynamics, it would be
suspicious if we had found a universally valid exponent $\alpha$ (albeit it is an intriguing question whether
lower/upper bounds exist).

It is also important to summarize the specific assumptions we made along the way (apart from coarseness, slowness
and the ETH). First, we assumed a realistic state preparation procedure, where the experimenter has no fine-grained
control over the microstate. Second, we assumed that the ETH holds for projectors and that the envelope function
$F_x(\omega)$ has no specifically tuned pattern. Third, we shifted the energy levels to be multiples of the mean
level spacing $\delta e$. We believe that these three assumptions are very reasonable for a slow and coarse observable
and a non-integrable many-body system. A fourth assumption we added was the random walk argument to estimate sums over
products of the pseudorandom coefficients $R_{k\ell}$. We must be self-critical here as we remarked in
Sec.~\ref{sec ETH Levy} that these coefficients are not purely random, see Refs.~\cite{FoiniKurchanPRE2019,
ChanDeLucaChalkerPRL2019, MurthySrednickiPRL2019, RichterEtAlPRE2020, BrenesEtAlPRE2021, WangEtAlPRL2022,
DymarskyPRL2022} for discussions. Since the products of the pseudorandom coefficients above always contained projectors
from different macrostates (since $x_1\neq x'_1$ and $x_0\neq x_2$), we are not aware of any way how to treat
them explicitly. In addition, we tried to compensate for deviatons from pure randomness by assuming a worst case scenario
(perfect correlations) between the time dependent phases and the remaining pseudorandom numbers. This scenario certainly 
is unnecessarily pessimistic and should alleviate the error introduced in our estimates involving $R_{k\ell}$.
Moreover, in the mean time an approach based on random matrix theory has also found that quantum contributions to
the measurement statistics should be small in general~\cite{StrasbergSP2023}. Finally, the numerical results below
and of Refs.~\cite{GemmerSteinigewegPRE2014, SchmidtkeGemmerPRE2016} further support our findings.

To conclude this section, we comment briefly on the generalization of our result to an arbitrary number $n$ of
times. This requires the computation of higher-order correlation functions and a more elaborate treatment compared to 
the present treatment then becomes necessary. Indeed, in a different and more restrictive setting general statements 
about $n$-time correlation functions have been found already~\cite{FigueroaRomeroModiPollockQuantum2019, 
FigueroaRomeroPollockModiCP2021, DowlingEtAlQuantum2023, DowlingEtAlSPC2023}. Extrapolating from these results, it
seems likely that classicality continuous to hold as long as $n\ll N$.

\section{Markovianity}
\label{sec Markovianity}

Defining Markovianity in the quantum regime is subtle and has been the focus of much 
debate~\cite{RivasHuelgaPlenioRPP2014, BreuerEtAlRMP2016, LiHallWisemanPR2018, MilzModiPRXQ2021}. Luckily, we showed 
in the previous section that the dynamics can be modeled by a classical stochastic process. Thus, we can use the 
conventional definition, which says that for all $k$ and all $x_k,\dots,x_1$ 
\begin{equation}\label{eq def Markovianity}
 p(x_k|x_{k-1},\dots,x_1) = p(x_k|x_{k-1}),
\end{equation}
where $p(a|b) \equiv p(a,b)/p(b)$ denotes a conditional probability as usual. In words, a Markov process is 
characterized by the fact that knowledge of the system state at any given time is sufficient to predict its future. 

Again, we start by showing that the repeated randomness assumption guarantees Markovian behaviour. The left hand side 
of Eq.~(\ref{eq def Markovianity}) is 
\begin{equation}
 \begin{split}
  & \frac{p(x_k,x_{k-1},\dots,x_1)}{p(x_{k-1},\dots,x_1)} = \\
  & \frac{\mbox{tr}\{\Pi_{x_k}U_k\Pi_{x_{k-1}}\rho(t_{k-1}|x_{k-2},\dots,x_1)\Pi_{x_{k-1}}U_k^\dagger\}}{\mbox{tr}\{\Pi_{x_{k-1}}\rho(t_{k-1}|x_{k-2},\dots,x_1)\}},
 \end{split}
\end{equation}
where $\rho(t_{k-1}|x_{k-2},\dots,x_1)$ denotes the exact microscopic system state at time $t_{k-1}$ conditioned on 
the previous outcomes $x_{k-2},\dots,x_1$. Now, by the repeated randomness assumption we can replace that state by 
\begin{equation}\label{eq repeated randomness Markovianity}
 \rho(t_{k-1}|x_{k-2},\dots,x_1) \mapsto \sum_{x_{k-1}} p(x_{k-1},\dots,x_1) \frac{\Pi_{x_{k-1}}}{V_{x_{k-1}}}
\end{equation}
as already done in Eq.~(\ref{eq max ent state}). This reveals 
\begin{equation}
 \begin{split}\label{eq repeated randomness Markovianity 2}
  &\frac{p(x_k,x_{k-1},\dots,x_1)}{p(x_{k-1},\dots,x_1)} \\
  &= \frac{p(x_{k-1},\dots,x_1)\mbox{tr}\{\Pi_{x_k}U_k\Pi_{x_{k-1}}U_k^\dagger\}/V_{x_{k-1}}}{p(x_{k-1},\dots,x_1)\mbox{tr}\{\Pi_{x_{k-1}}\}/V_{x_{k-1}}} \\
  &= \mbox{tr}\left\{\Pi_{x_k}U_k\frac{\Pi_{x_{k-1}}}{V_{x_{k-1}}}U_k^\dagger\right\}.
 \end{split}
\end{equation}
Once again, to apply this argument for all $t_k$, we have to make the 
replacement~(\ref{eq repeated randomness Markovianity}) repeatedly, which violates the unitarity of the process. 

Below, we use Levy's Lemma to argue that for a coarse and slow observable the dynamics are \emph{very likely} 
Markovian. Clearly, in any finite dimensional setting it is impossible to show strict Markovianity for all times and 
all initial states. 

To approach the problem, we will switch to a lighter notation. Let $\rho(t)$ be the exact microstate of the system 
at time $t$. The probability to find the system in state $x$ at time $t+\tau$ is 
\begin{equation}
 p_x(t+\tau) = \mbox{tr}\{\Pi_x U_\tau\rho(t) U_\tau^\dagger\}.
\end{equation}
Introducing the identity $I = \sum_y \Pi_y$ twice around $\rho(t)$, we find 
\begin{equation}
 \begin{split}
  p_x(t+\tau) =& \sum_y \mbox{tr}\{\Pi_x U_\tau\Pi_y\rho(t)\Pi_y U_\tau^\dagger\} \\
  &+ \sum_{y\neq y'} \mbox{tr}\{\Pi_x U_\tau\Pi_y\rho(t)\Pi_{y'} U_\tau^\dagger\}.
 \end{split}
\end{equation}
We know from the previous section that we can neglect the quantum contribution in the second line. Introducing the 
state $\rho_y(t) = \Pi_y\rho(t)\Pi_y/p_y(t)$ conditioned on outcome $y$ with $p_y(t) = \mbox{tr}\{\Pi_y\rho(t)\}$, 
we can thus write 
\begin{equation}\label{eq conditional prob}
 p_x(t+\tau) = \sum_y P_{x|y}[\rho_y(t),\tau]p_y(t).
\end{equation}
Here, $P_{x|y}[\rho_y(t),\tau] \equiv \mbox{tr}\{\Pi_x U_\tau\rho_y(t)U_\tau^\dagger\}$ is the conditional probability 
for a transition from $y$ to $x$ in time $\tau$ given that the microstate is $\rho_y(t)$. 

The reader might criticize that Eq.~(\ref{eq conditional prob}) looks already Markovian, but recall that $\rho_y(t)$ 
is the \emph{exact microstate}. In particular, this microstate could depend on any number of measurements results 
prior to time $t$. We have just suppressed this dependence for notational simplicity, but---as long as $\rho_y(t)$ is 
the exact microstate---we have \emph{not} made any assumption so far. Moreover, it is now particularly easy to see 
that the dynamics is Markovian if we can apply the equal-a-priori-probability postulate and replace $\rho_y(t)$ 
by $\Pi_y/V_y$. 

In the following section we prove in a mathematically rigorous way that $P_{x|y}[\rho_y(t),\tau]$ is almost constant 
as a function of $\rho_y(t)$ for a coarse observable, and we argue that this result is physically 
meaningful for a slow observable. Put differently, the claim is that the exact microstate $\rho_y(t)$ becomes 
irrelevant for the evaluation of $P_{x|y}[\rho_y(t),\tau]$ and thus the dynamics is Markovian. 

\subsection{Microscopic derivation}

Since any mixed state can be written as a convex linear combination of pure states and since $P_{x|y}(\rho_y,\tau)$ 
is linear in $\rho_y$, we can and will assume that $\rho_y = \psi_y \equiv |\psi_y\rl\psi_y|$ is a pure state with 
$|\psi_y\rangle\in\C H_y$. Then, to say that $P_{x|y}(\psi_y,\tau)$ is ``almost constant as a function of $\psi_y$'' 
requires us to choose a measure on $\C H_y$, which we take to be the unbiased Haar measure $\mu_y$ (note the subscript 
$y$ to indicate that we only sample randomly in $\C H_y\subset\C H$). Then, from $\mu_y(\psi_y) = \Pi_y/V_y$ we find 
\begin{equation}\label{eq P average}
 P_{x|y}(\tau) \equiv \mu_y\big[P_{x|y}(\psi_y,\tau)\big] = \frac{1}{V_y}\mbox{tr}\{\Pi_x U_\tau\Pi_yU_\tau^\dagger\}.
\end{equation}
Note that this term is identical to the last line of Eq.~(\ref{eq repeated randomness Markovianity 2}) 
in our new notation.

To bound the fluctuations in $P_{x|y}(\psi_y,\tau)$ with respect to different $\psi_y$, we use Levy's Lemma on the 
hypersphere $\mathbb{S}^{2V_y-1}$ defined by all pure states in the subspace $\C H_y$. To estimate the 
Lipschitz constant of $P_{x|y}(\psi_y,\tau)$, we note the rewriting 
\begin{equation}
 P_{x|y}(\psi_y,\tau) = \lr{\psi_y|U_\tau^\dagger\Pi_xU_\tau|\psi_y}
\end{equation}
and, using the same arguments spelled out below Eq.~(\ref{eq Lipschitz Q}), we find that the 
Lipschitz constant of $P_{x|y}(\psi_y,\tau)$ is no more than two: $\eta\le 2$. Hence, Levy's Lemma implies 
\begin{equation}\label{eq Levys lemma result}
 \mu_y\left[\left|\frac{P_{x|y}(\psi_y,\tau)}{P_{x|y}(\tau)} - 1 \right| > \epsilon \right] 
 \le 4\exp\left(-\frac{\epsilon^2P_{x|y}^2(\tau)V_y}{18\pi^3}\right).
\end{equation}

To get a feeling for this bound, we consider some numbers. First, let us assume we do not like to tolerate an error 
larger than $\epsilon = 10^{-6}$. Second, let us consider a short timescale $\tau$ such that 
$P_{x|y}(\tau) \approx 10^{-6}$. Finally, we set $V_y = 10^{\C O(N)}/M$ as a rough estimate. Then, 
\begin{equation}
 \mu_y\left[\left|\frac{P_{x|y}(\psi_y,\tau)}{P_{x|y}(\tau)} - 1 \right| > \epsilon \right] 
 \lesssim 4\exp\left(-\frac{10^{\C O(N) - 26}}{M}\right).
\end{equation}
Thus, owing to the exponential growth of the Hilbert space dimension, this term is negligible small for already 
$N \approx 50$ particles provided that the observable is \emph{coarse}, i.e., provided that $M$ is not 
unrealistically large. Also if one considers subspaces where $V_y$ is very small, which characterizes macrostates 
very far from equilibrium, the above bound becomes weak. Usually, however, $V_y$ scales exponentially with $N$. 

Everything so far is based on an exact mathematical identity. From a physical point of view, we have to 
ask when is it meaningful to assume that the precise microstate $\psi_y(t)$ can be replaced by a Haar random 
average over $\C H_y$? It is here where we use the \emph{slowness} of $X$. 

The slowness of $X$ implies that $\psi_y(t)$ has time to spread over many different microstates in $\C H_y$ before 
it `hops out' to a different macrostate $x\neq y$. Restricting the picture of Fig.~\ref{fig rw} to $\C H_y$ only, we 
like to picture $\psi_y(t)$ as performing an approximately unbiased random walk on the sphere $\mathbb{S}^{2V_y-1}$. 
Note that this picture implies that we do not expect $P_{x|y}(\psi_y,\tau) \approx P_{x|y}(\tau)$ to be true for 
too short timescales during which $\psi_y(t)$ had no time to spread over many different microstates. Moreover, it is 
important to assume that $\psi_y(t)$ explores $\C H_y$ in an (approximately) \emph{unbiased} way. For instance, 
if there is some unaccounted conserved quantity or slow observable, which restricts the motion of $\psi_y$ to some 
\emph{subspace} of $\C H_y$, it is obviously no longer allowed to sample randomly according to the Haar measure 
$\mu_y$ in \emph{all} of $\C H_y$. It is exactly this notion of unbiasedness, which is difficult to get under control 
in a mathematical precise way and we will discuss this further in Sec.~\ref{sec multiple}. 

In addition, since Levy's Lemma implies that the subset of states for which $P_{x|y}(\psi_y,\tau)$ looks 
\emph{atypical} is exponentially small, we remark that this justifies applying this reasoning \emph{repeatedly} for 
different times $t$ in Eq.~(\ref{eq conditional prob}). Indeed, while it is possible that one accidentially hits an 
atypical state at some time $t$, it is extremely unlikely to remain in a subspace of atypical states during the slow 
evolution time-scales of $X$ if $\psi_y$ diffuses approximately in an unbiased way in $\C H_y$. Thus, instead of 
\emph{conjecturing} the applicability of typicality arguments at each time step as in 
Refs.~\cite{GemmerMichelEPJB2006, BreuerGemmerMichelPRE2006, GemmerBreuerEPJ2007, HahnGuhrWaltnerPRE2020}, we argue 
that it is \emph{justified} to apply them for a generic slow and coarse observable. 

Clearly, the same point we emphasized in the last paragraph of Sec.~\ref{sec classicality derivation} also applies here. 
If the argument above is repeated too many times, one finds for sure some time interval with non-Markovian dynamics, 
which must happen in a finite dimensional quantum system. However, as long as $n\ll N$ (with $n$ the 
number of time steps) such ``accidential non-Markovianity'' is unlikely. 

To conclude, there is a strong mathematical result and plausible physical assumptions that suggest that Markovianity 
arises generically for a slow and coarse observable on a coarse (i.e., not too short) time scale and it is 
very likely to persist for many time steps. Physically, we believe that this result is best understood by introducing 
the concept of \emph{microstate independence}. If different microstates of the irrelevant degrees of 
freedom (which could encode different histories of the relevant degrees of freedom) give rise to the same transition  
probabilities, i.e., if $P_{x|y}[\psi_y(t),\tau] \approx P_{x|y}[\psi'_y(t),\tau]$ for different 
$\psi_y(t)\neq\psi'_y(t)$, then any such history dependence or ``memory'' in the irrelevant degrees of freedom
becomes \emph{irrelevant} for the future evolution of the relevant degrees of freedom. We believe this is a 
transparent physical explanation for Markovianity compared to the traditional ``loss-of-memory'' explanation, which can 
never happen in a unitarily evolving system. Importantly, we believe that this result strongly depends on the
observable, and not only on the Hamiltonian or unitary dynamics. 

\section{Local detailed balance}
\label{sec local detailed balance}

So far we have established that the dynamics can be described with overwhelming probability by a classical Markov 
process with transition probabilities [see Eqs.~(\ref{eq repeated randomness Markovianity 2}) or~(\ref{eq P average})] 
\begin{equation}\label{eq P average 2}
 P_{x|y}(\tau) = \frac{1}{V_y}\mbox{tr}\{\Pi_x U_\tau\Pi_yU_\tau^\dagger\}.
\end{equation}
While there are many processes (not only in nature) that can be described by \emph{some} Markovian transition 
probabilities $P_{x|y}(\tau)$, the specific form of $P_{x|y}(\tau)$ in Eq.~(\ref{eq P average 2}) allows us to derive 
an additional important physical property, which ensures a consistent thermodynamic description for each time step. 

To find this property, we need to introduce the anti-unitary time-reversal operator $\Theta$~\cite{StrasbergBook2022, 
HaakeBook2010}. Thus, let $\Pi_x^\Theta \equiv \Theta\Pi_x\Theta^{-1}$ be the projector on the time-reversed macrostate 
of $x$, $H^\Theta \equiv \Theta H\Theta^{-1}$ the time-reversed Hamiltonian and 
$U^\text{TR}_\tau \equiv e^{-iH^\Theta\tau}$ the time-evolution operator associated to the time-reversed process. 
Using $\Theta i = -i\Theta$ and $\mbox{tr}\{\Theta\dots\Theta^{-1}\} = \mbox{tr}\{\dots\}^*$ reveals that 
\begin{equation}\label{eq symmetry time reversal}
 \mbox{tr}\{\Pi_x U_\tau\Pi_y U_\tau^\dagger\} 
 = \mbox{tr}\{\Pi_y^\Theta U^\text{TR}_\tau\Pi_x^\Theta (U^\text{TR}_\tau)^\dagger\}.
\end{equation}
Moreover, we note that $\mbox{tr}\{\Pi_x^\Theta\} = \mbox{tr}\{\Pi_x\} = V_x$, which implies that we can conclude 
from Eq.~(\ref{eq P average 2}) that 
\begin{equation}\label{eq LDB discrete}
 P_{x|y}(\tau) = \frac{1}{V_y} \mbox{tr}\{\Pi_y^\Theta U^\text{TR}_\tau\Pi_x^\Theta (U_\tau^\text{TR})^\dagger\} 
 = \frac{V_x}{V_y} P_{y|x}^\text{TR}(\tau).
\end{equation}
Here, $P_{y|x}^\text{TR}(\tau)$ is the average conditional probability to jump from the time-reversed state 
described by $\Pi_x^\Theta$ to the state described by $\Pi_y^\Theta$ in a time step $\tau$ under the dynamics 
generated by $H^\Theta$. In many applications one has to deal with the simpler case where both, the macrostates 
and the Hamiltonian obey time-reversal symmetry, i.e., $\Pi_x^\Theta = \Pi_x$ and $H^\Theta = H$. Then, 
Eq.~(\ref{eq LDB discrete}) implies 
\begin{equation}\label{eq LDB discrete TRS}
 P_{x|y}(\tau) = \frac{V_x}{V_y} P_{y|x}(\tau).
\end{equation}
We call Eqs.~(\ref{eq LDB discrete}) and~(\ref{eq LDB discrete TRS}) the condition of \emph{local detailed balance} 
(LDB), which was previously derived using the repeated randomness assumption~\cite{VanKampenPhys1954}. 

To clarify the importance of LDB, we first rewrite the dynamics in a more familiar form. Taking $\tau$ to be small 
compared to the evolution time of $X$, but still large compared to the microscopic timescale $\Delta E^{-1}$, we 
approximate $P_{x|y}(\tau) \approx \delta_{x,y} + \tau R_{x,y}$. Here, $R_{x,y}$ is the \emph{rate matrix} obeying 
$\sum_x R_{x,y} = 0$ for all $y$ owing to probability conservation. The time-evolution on this timescale is then 
described by the differential equation 
\begin{equation}\label{eq ME}
 \frac{d}{dt} p_x(t) = \sum_y R_{x,y} p_y(t)
\end{equation}
known as a (rate, Pauli or classical) master equation. Using the Boltzmann entropy $S_B(x) = \ln V_x$, LDB becomes
\begin{equation}\label{eq LDB}
 \frac{R_{x,y}}{R_{y,x}^\text{TR}} = e^{S_B(x)-S_B(y)} ~~~ \text{or} ~~~ 
 \frac{R_{x,y}}{R_{y,x}} = e^{S_B(x)-S_B(y)},
\end{equation}
depending on the question whether time-reversal symmetry is broken or not. 

Among the implications of LDB we first note that the steady state of the dynamics reads 
\begin{equation}\label{eq equilibrium state}
 \pi_x \equiv \frac{V_x}{D}.
\end{equation}
In fact, using LDB (with or without time-reversal symmetry) it can be easily checked that 
$\sum_y R_{x,y}\pi_y = 0$. Note that Eq.~(\ref{eq equilibrium state}) describes the correct equilibrium state 
as expected from the equal-a-priori-probability postulate of statistical mechanics. The probability to be in 
macrostate $x$ is proportional to its volume $V_x$. In addition, in presence of time-reversal symmetry LDB implies 
that all net currents vanish at equilibrium, $R_{x,y}\pi_y - R_{y,x}\pi_x = 0$, but note that net currents can persist 
at equilibrium if time-reversal symmetry is broken. 

Furthermore, the thermodynamic entropy of the system is~\cite{VonNeumann1929, VonNeumannEPJH2010, VanKampenPhys1954, 
SafranekDeutschAguirrePRA2019a, StrasbergWinterPRXQ2021, SafranekEtFP2021, StrasbergBook2022} 
\begin{equation}
 S(t) \equiv \sum_x p_x(t)[-\ln p_x(t) + S_B(x)].
\end{equation}
As a consequence of LDB we find 
\begin{equation}\label{eq 2nd law}
 \frac{d}{dt} S(t) \ge 0,
\end{equation}
i.e., the thermodynamic entropy increases monotonically in time: the entropy production rate is positive. 
To derive Eq.~(\ref{eq 2nd law}), we note the useful rewriting $S(t) = \ln D - S[p_x(t)\|\pi_x]$ where 
$S[p_x\|q_x] \equiv \sum_x p_x\ln(p_x/q_x)$ is the relative entropy. The positivity of the entropy production rate 
then follows from two facts~\cite{StrasbergEspositoPRE2019}: first, $\pi_x$ is a steady state of the dynamics and, 
second, the dynamics is Markovian, which implies that relative entropy is 
contractive~\cite{StrasbergBook2022, MilzModiPRXQ2021}. LDB is therefore intimately linked to the fact that the 
dynamics \emph{tends to maximize the entropy at each time step on average}. 

Further important consequences of LDB are the emergence of the Onsager relations~\cite{VanKampenPhys1954} and a 
consistent thermodynamic framework in the presence of nonequilibrium boundary conditions~\cite{BergmannLebowitzPR1955}, 
among others, which we will not discuss here. Moreover, it might be helpful to point out that LDB is often 
expressed in a form less general than Eq.~(\ref{eq LDB}). For instance, for a small open quantum system with energies 
$\epsilon_x$ in contact with a large bath at inverse temperature $\beta$, LDB reduces to
\begin{equation}\label{eq LDB OQS}
 \frac{R_{x,y}}{R_{y,x}} = e^{\beta(\epsilon_y-\epsilon_x)},
\end{equation}
which follows from a Taylor expansion of the Boltzmann entropy and the definition of the inverse 
temperature $\beta = S'_B(E)$. Equation~(\ref{eq LDB OQS}) is used as the starting point of much current work in
classical stochastic and quantum thermodynamics~\cite{VanKampenBook2007, SekimotoBook2010, SeifertRPP2012, 
SchallerBook2014, PelitiPigolottiBook2021, StrasbergBook2022, BreuerPetruccioneBook2002}. However, Eq.~(\ref{eq LDB}) 
is the most general expression of LDB~\cite{VanKampenPhys1954, MaesNetocnyJSP2003} and it might become more important 
than Eq.~(\ref{eq LDB OQS}) in the future, for instance, to describe systems in contact with finite 
baths~\cite{RieraCampenySanperaStrasbergPRXQ2021}. 

\subsection{Microscopic derivation}
\label{sec microscopic derivation LDB}

After all the previous work, which already justified the use of Eq.~(\ref{eq P average 2}), not much remains to be 
done. In fact, all we have to ensure is that $X^\Theta = \Theta X \Theta^{-1}$ is a slow observable if $X$ is 
also slow. To show this, we first note that if $|k\rangle$ is an eigenvector of $H$ with eigenvalue $E_k$, then 
$\Theta|k\rangle \equiv |\Theta k\rangle$ is an eigenvector of $H^\Theta$ with the same eigenvalue $E_k$. Moreover, 
since $\Theta$ is anti-unitarity we have by definition $\lr{\Theta\psi|\Theta\phi} = \lr{\psi|\phi}^*$ for any two 
vectors $|\psi\rangle$ and $|\phi\rangle$. We then find 
\begin{align}
 \lr{\Theta k|\Theta X\Theta^{-1}|\Theta\ell} 
 &= \sum_{k',\ell'} \lr{\Theta k|\Theta k'}\lr{k'|X|\ell'}\lr{\Theta\ell'|\Theta\ell} \nonumber \\
 &= \sum_{k',\ell'} \delta_{kk'}\lr{k'|X|\ell'}\delta_{\ell\ell'} = X_{k\ell}.
\end{align}
Thus, $X^\Theta$ inherits the same band structure from $X$, but with respect to the eigenbasis of $H^\Theta$. 

Finally, we note a special peculiarity. All that matters to derive LDB is that there is \emph{some} anti-unitary 
operator $\Theta$, but it does not matter which $\Theta$ one chooses. Clearly, some choices are physically more 
appealing, but choosing different $\Theta$ can give rise to multiple LDB conditions, which could be advantageous for 
applications. We give an example for such different choices in the next section. 

\section{Numerics}
\label{sec numerics}

We check our ideas numerically by exact integration of the Schr\"odinger equation for an XXZ spin chain of length $L$ 
with periodic boundary condition. The dimensionless Hamiltonian is
\begin{equation}
 H = \sum_{\ell=1}^L \left(s_x^\ell s_x^{\ell+1} 
 + s_y^\ell s_y^{\ell+1} + \frac{3}{2} s_z^\ell s_z^{\ell+1} + \frac{1}{2} s_z^\ell s_z^{\ell+2}\right),
\end{equation}
where $s_\alpha^\ell$ are spin-1/2 operators at site $\ell$ and we consider the zero magnetization subspace in the 
following, which has dimension $D = \binom{L}{L/2}$ and requires $L$ to be even. 
It is known that for the present choice of parameters the Hamiltonian is non-integrable and satisfies 
the ETH~\cite{RichterEtAlPRE2020}.  

The observable that we consider is a spin density wave 
\begin{equation}\label{eq observable numerics}
 X_q = \frac{1}{\C N} \sum_{\ell=1}^L \cos\left(\frac{2\pi\ell q}{L}\right) s_z^\ell,
\end{equation}
where the wavenumber $q$ can be tuned between the longest ($q=1$) and shortest ($q=L/2$) wavelengths $\lambda = 1/q$ 
in the system. For longer wavelengths the observable becomes slower because local perturbations typically need longer 
times to induce large changes in $\lr{X_q}$ owing to the finite speed with which excitations can travel along the 
chain (Lieb-Robinson bound~\cite{NachtergaeleSimsArXiv2010}). Moreover, $\C N$ is a normalization constant which 
fixes the second central moment to one: $\mbox{tr}\{X^2_q\}/D-(\mbox{tr}\{X_q\}/D)^2 = 1$. This ensures that the 
domain of eigenvalues of $X_q$ for different $q$ is approximately the same, which makes it easier to define a common 
coarse-graining now. 

\begin{figure}[t]
 \centering\includegraphics[width=0.48\textwidth,clip=true]{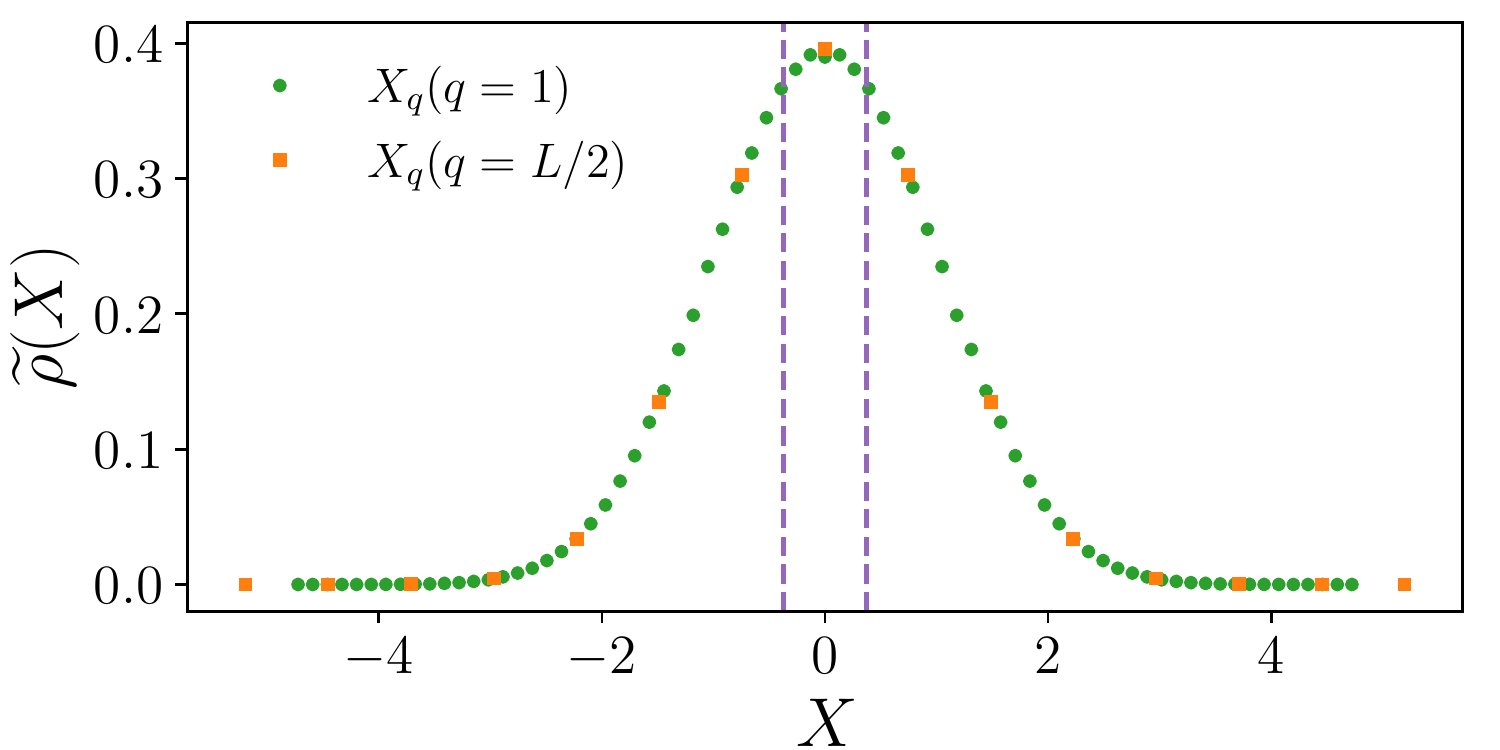}
 \label{fig-dos} 
 \caption{Eigenvalue distribution of the density wave operator $X_q$ for $q=1$ and $q=L/2$ for chain length $L = 28$. 
 The dashed lines indicate the coarse-graining window for $x=0$. }
\end{figure}

To define it, we note that all eigenvectors of $X_q$ can be conveniently labeled by 
$|\bb z\rangle = |z_1,\dots,z_L\rangle$ with $|z_\ell\rangle$ denoting a local eigenstate of $s_z^{(\ell)}$ with 
$z_i\in\{\pm1\}$. We further observe that to each eigenvector $|\bb z\rangle$ with eigenvalue $\lambda(\bb z)$ there 
is another eigenvector $|{-\bb z}\rangle$, obtained by flipping all $z_i$ to $-z_i$, with eigenvalue 
$\lambda(-\bb z) = -\lambda(\bb z)$. Thus, the eigenvalues come in pairs symmetrically distributed around zero, as 
shown in Fig.~\ref{fig-dos}. Due to the symmetry of the spectrum, it is convenient to label the coarse-grained 
eigenspaces as $x\in\{\dots,-1,0,+1,\dots\}$ with projectors $\Pi_x = \sum_{\bb z\in I_x} |\bb z\rl \bb z|$, where 
\begin{equation}\label{eq coarse graining interval}
 I_x = \big\{\bb z\big|\lambda(\bb z) \in \big[(x-1/2)\delta X,(x+1/2)\delta X\big]\big\}.
\end{equation}
Here, $\delta X$ denotes the coarse-graining width, as indicated in Fig.~\ref{fig-dos}. Unless otherwise mentioned, 
we set $\delta X = 0.74$. 

As an initial state we choose in the following 
\begin{equation}\label{eq initial state}
 |\psi\rangle \sim e^{-\kappa X_q/2}|\psi_{R}\rangle,
\end{equation}
where $|\psi_R\rangle$ is a random state with coefficients drawn from a zero-mean Gaussian distribution. Thus, 
$|\psi_R\rangle$ mimics an infinite temperature state, which makes optimal use of the available Hilbert space dimension. 
Moreover, $e^{-\kappa X_q/2}$ prepares the system out of equilibrium, where $\kappa$ is a perturbation strength. 
Below, we choose a moderate perturbation $\kappa=0.1$, but we have also checked results for different initial states 
corresponding to different temperatures and different perturbations and observed similar behaviour. 
These results are relegated to the supplemental material~\cite{SM2}. 

\begin{figure}[t]
 \centering\includegraphics[width=0.48\textwidth,clip=true]{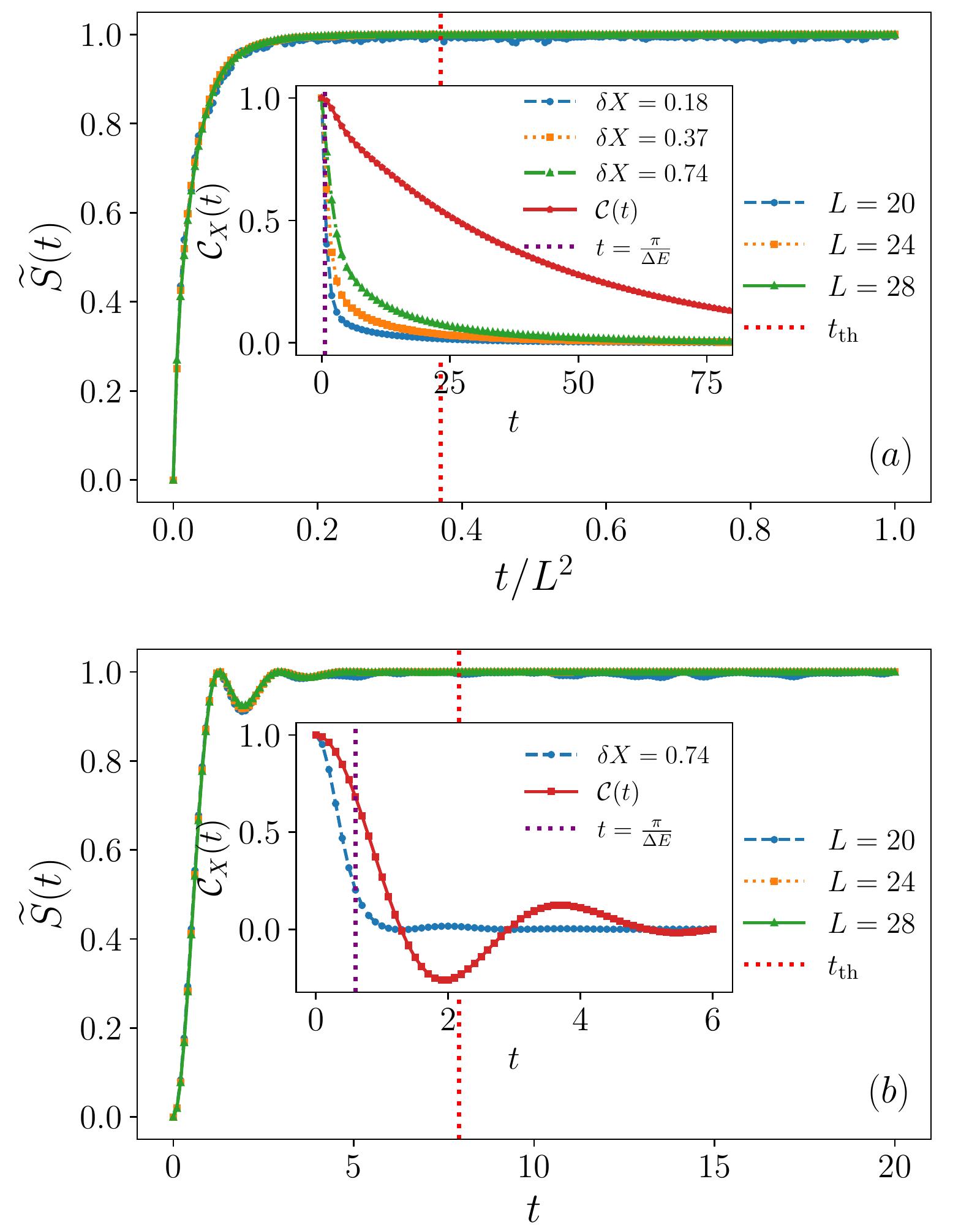}
 \label{fig obs ent} 
 \caption{Time evolution of thermodynamic entropy for the longest (a) and shortest wave length (b) for different 
 chain lengths $L$ as indicated in the plot. The time axis is rescaled in (a) because the relaxation time scales 
 like $L^2$ for slow observables whereas it is system size independent for fast observables. 
 Insets: Time evolution of correlation functions for $L=28$ as explained 
 in the main text. }
\end{figure}

From what we said below Eq.~(\ref{eq observable numerics}) we expect that our theory works well for $q=1$ but not for 
$q=L/2$. A first indicator for this is shown in Fig.~\ref{fig obs ent}, where we plot for better comparison the 
rescaled thermodynamic entropy 
\begin{equation}\label{eq S rescaled}
 \tilde S(t) \equiv \frac{S(t) - S(0)}{S(\infty) - S(0)}
\end{equation}
with $S(\infty) = \ln D$. We see that it increases monotonously for $q=1$, as predicted by Eq.~(\ref{eq 2nd law}), 
whereas it clearly violates Eq.~(\ref{eq 2nd law}) for $q=L/2$. This violation also does not seem to become 
smaller for larger system size. Moreover, the vertical red dashed line in the figures indicates the 
\emph{thermalization time} $t_\text{th}$ for comparison. It is defined to be the time by which the rescaled expectation 
value $\langle\tilde X_q\rangle$ for the case $L=28$, defined similar to Eq.~(\ref{eq S rescaled}), decayed to 1 
percent and stayed below this threshold afterwards. Since $t_\text{th}$ fluctuates in each realization, 
we averaged it over 10 different initial states. 

To further investigate the slowness of $X_q$, the insets of Fig.~\ref{fig obs ent} show the behaviour of the 
correlation function $\C C_X = \mbox{tr}\{\Pi_0(t)\Pi_0\}/\mbox{tr}\{\Pi_0\}$ as a function of time for $q=1$ and 
$q=L/2$ as well as for different coarse-graining sizes $\delta X$. Moreover, we also plot the correlation function 
$\C C = \mbox{tr}\{X(t)X\}/\mbox{tr}\{X^2\}$ and the dotted purple vertical line indicates the microscopic evolution 
time scale $\pi/\Delta E$ (since we do not use a microcanonical energy window, $\Delta E$ here denotes the standard 
deviation of the energy spectrum). The insets thus demonstrate again that $X_q$ is slow for $q=1$ but not for $q=L/2$. 
Moreover, we confirm that projectors are slower for coarser coarse-grainings as derived in 
Appendix~\ref{sec app proof bandedness}. 

\begin{figure}[t]
 \centering\includegraphics[width=0.48\textwidth,clip=true]{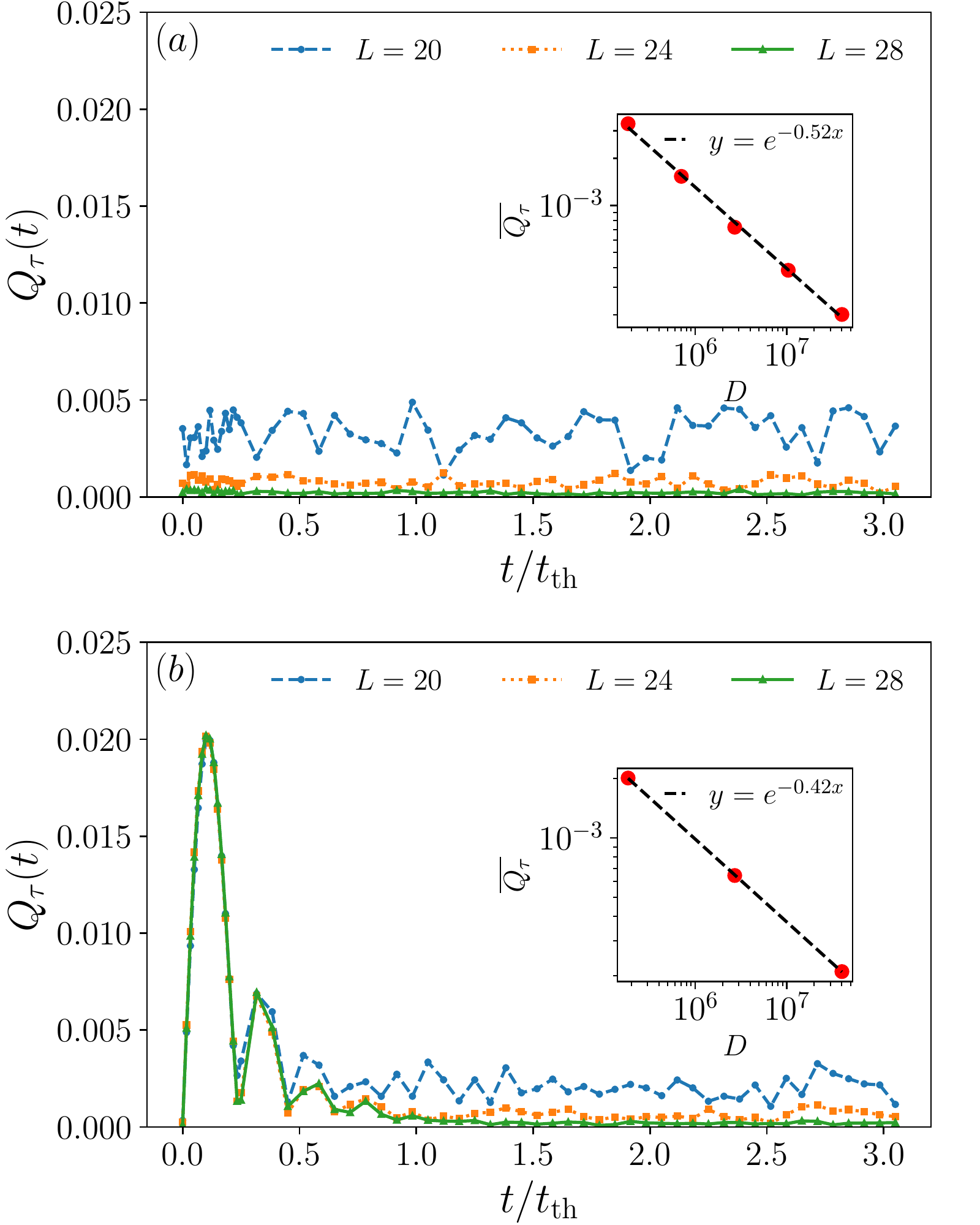}
 \label{fig quantum term} 
 \caption{Time evolution of the quantum term for $q=1$ (a) and $q=L/2$ (b). The insets show a log-log plot of a 
 suitable time-average as a function of the dimension $D$. }
\end{figure}

Next, we consider in Fig.~\ref{fig quantum term} the influence of the quantum part $Q$ on the dynamics as discussed 
in Sec.~\ref{sec classicality}. Specifically, we consider the quantity 
\begin{equation}
 Q_\tau(t) = \sum_x\left|\sum_{y\neq z}\text{tr}\{\Pi_{x}U_\tau\Pi_{y}\rho(t)\Pi_{z}U_\tau^\dagger\}\right|
\end{equation}
for $\tau = t_\text{th}/30$, which is a time scale at which nonequilibrium phenomena happen. Now, in 
Sec.~\ref{sec classicality} we have argued that the term inside the absolute value should be very small for slow 
observables and estimated that it scales like $D^{-\alpha}$ with unknown $\alpha$. The smallness of $Q_\tau(t)$ for a 
slow observable becomes immediately obvious from Fig.~\ref{fig quantum term}, whereas it is an order of magnitude larger 
for the fast case for times up to $t_\text{th}/2$. However, we also see that $Q_\tau(t)$ fluctuates for all times. 
To better check the scaling and to smooth out fluctuations, we consider the time average 
\begin{equation}
 \overline{Q_\tau} = \frac{1}{t_f-t_\text{th}}\int^{t_f}_{t_\text{th}} dt Q_\tau(t),
\end{equation}
where the integral is taken in the time interval during which $Q_\tau$ has approximately reached a steady value. 
The insets of Fig.~\ref{fig quantum term} reveal that the scaling exponent is $\alpha \approx 0.56$ for the slow case. 
Interestingly, also the fast case obeys a scaling law for times $t\ge t_\text{th}$ with a smaller exponent 
$\alpha \approx 0.42$. This indicates that classicality could be a universal feature for any coarse observable of a 
nonintegrable many-body system for not too small times. Moreover, as in Fig.~\ref{fig obs ent} we see that for the fast 
case the initial violation of classicality does not seem to become smaller for larger system sizes. This challenges the 
universal validity of the penultimate paragraph of Sec.~\ref{sec slow}, where we stated that sums of local observables 
tend to become slow for $L\rightarrow\infty$. Although we are numerically far away from the 
$L\rightarrow\infty$ case we have no direct explanation for that behaviour. 

\begin{figure}[t]
 \centering\includegraphics[width=0.48\textwidth,clip=true]{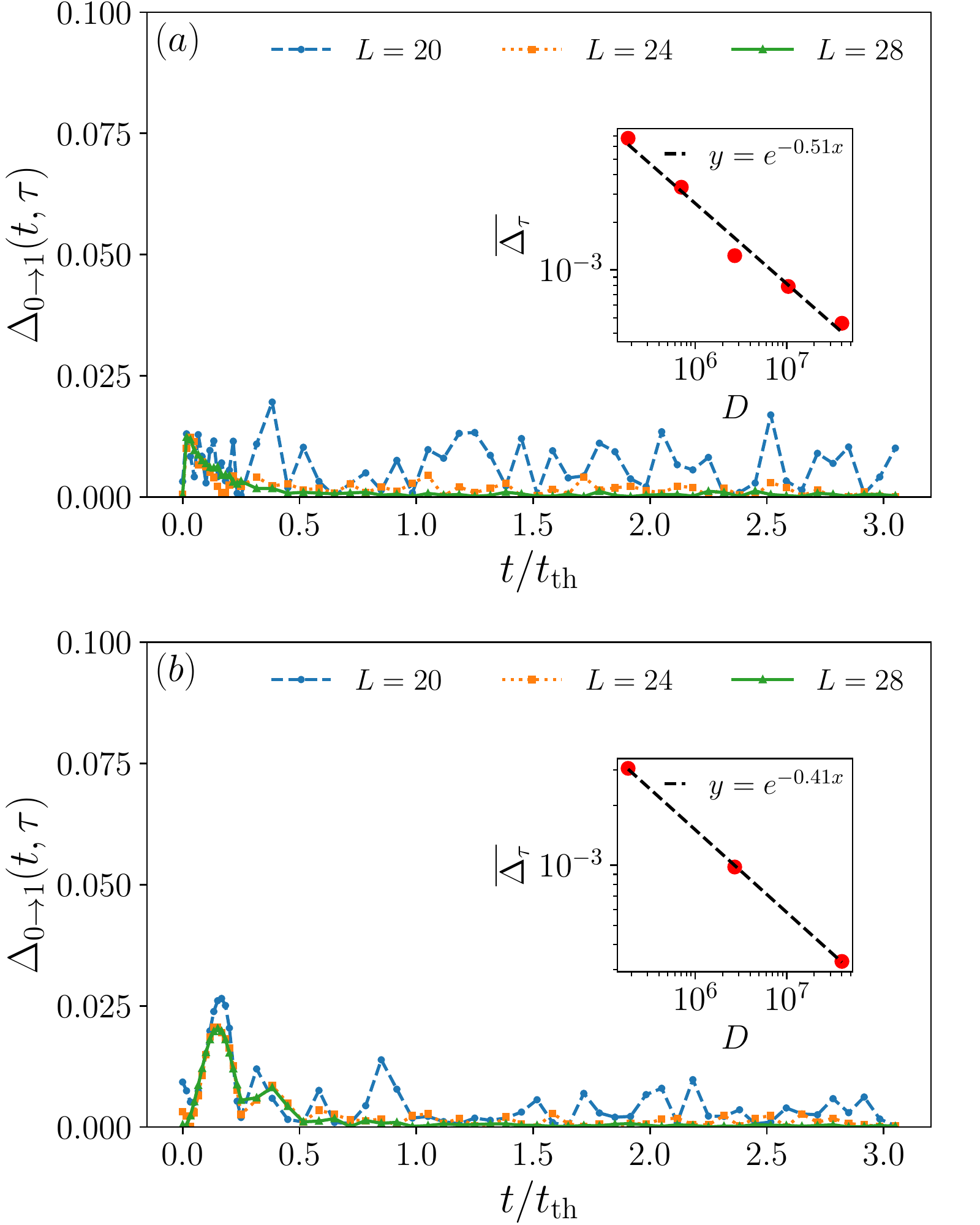}
 \label{fig LDB} 
 \caption{Check of LDB for $\Theta = K_z$ for $q=1$ (a) and $q=L/2$ (b). The insets show a log-log plot of a suitable 
 time average as a function of the dimension $D$. Note that we only check the LDB condition in case (b) if 
 $L=4k$ (with $k\in\mathbb{N}$) because it turns out that the $x=0$ subspace is empty if $L=4k+2$ (recall that we 
 restricted the dynamics to the zero magnetization subspace). }
\end{figure}

Finally, we turn to the condition of LDB and consider the quantity 
\begin{equation}\label{eq LDB check}
 \Delta_{0\rightarrow1} (t,\tau) 
 = \left|\frac{{V}_{1}R_{1,0}(t,\tau)}{{V}_2 R^\text{TR}_{0,1}(t,\tau)}-1\right|.
\end{equation}
Here, $R_{x,y}(t,\tau) = \mbox{tr}\{\Pi_x U_\tau\rho_y(t)U_\tau^\dagger\}$ is the rate to jump from $y$ to $x$. 
To define $R^\text{TR}_{y,x}(t,\tau)$ we need to introduce a time-reversal operator $\Theta$. As discussed in 
Sec.~\ref{sec microscopic derivation LDB}, multiple choices are conceivable. 

We first choose $\Theta = K_z$, where $K_z$ denotes complex conjugation in the local $s_z$ basis. For that 
choice we find $\Theta(s_x,s_y,s_z)\Theta^{-1} = (s_x,-s_y,s_z)$ and easily confirm that both the Hamiltonian and 
observable are symmetric and hence $R^\text{TR}_{y,x}(t,\tau) = R_{y,x}(t,\tau)$. In Fig.~\ref{fig LDB} we plot 
$\Delta_{0\rightarrow1} (t,\tau)$ for the slow and fast case with $\tau = t_\text{th}/30$ again. Furthermore, to check the 
scaling, the insets show again the time-average 
\begin{equation}
 \overline{\Delta_{\tau}} = \frac{1}{t_{f}-t_\text{th}}\int_{t_\text{th}}^{t_{f}}\Delta_{0\rightarrow1}(t,\tau).
\end{equation}
It is evident that LDB is well satisfied for the slow observables at all times. Also for the fast observable LDB is 
well satisfied for most times, only initially some slight violations do not vanish even with increasing system size. 
We attribute this behaviour to the particular choice of initial state in Eq.~(\ref{eq initial state}), which is very 
smoothly spread out over all microstates and thus looks quite ``typical'' even for the fast observable. The situation 
changes for different initial states as studied in the supplemental material~\cite{SM2}. 

\begin{figure}[t]
 \centering\includegraphics[width=0.48\textwidth,clip=true]{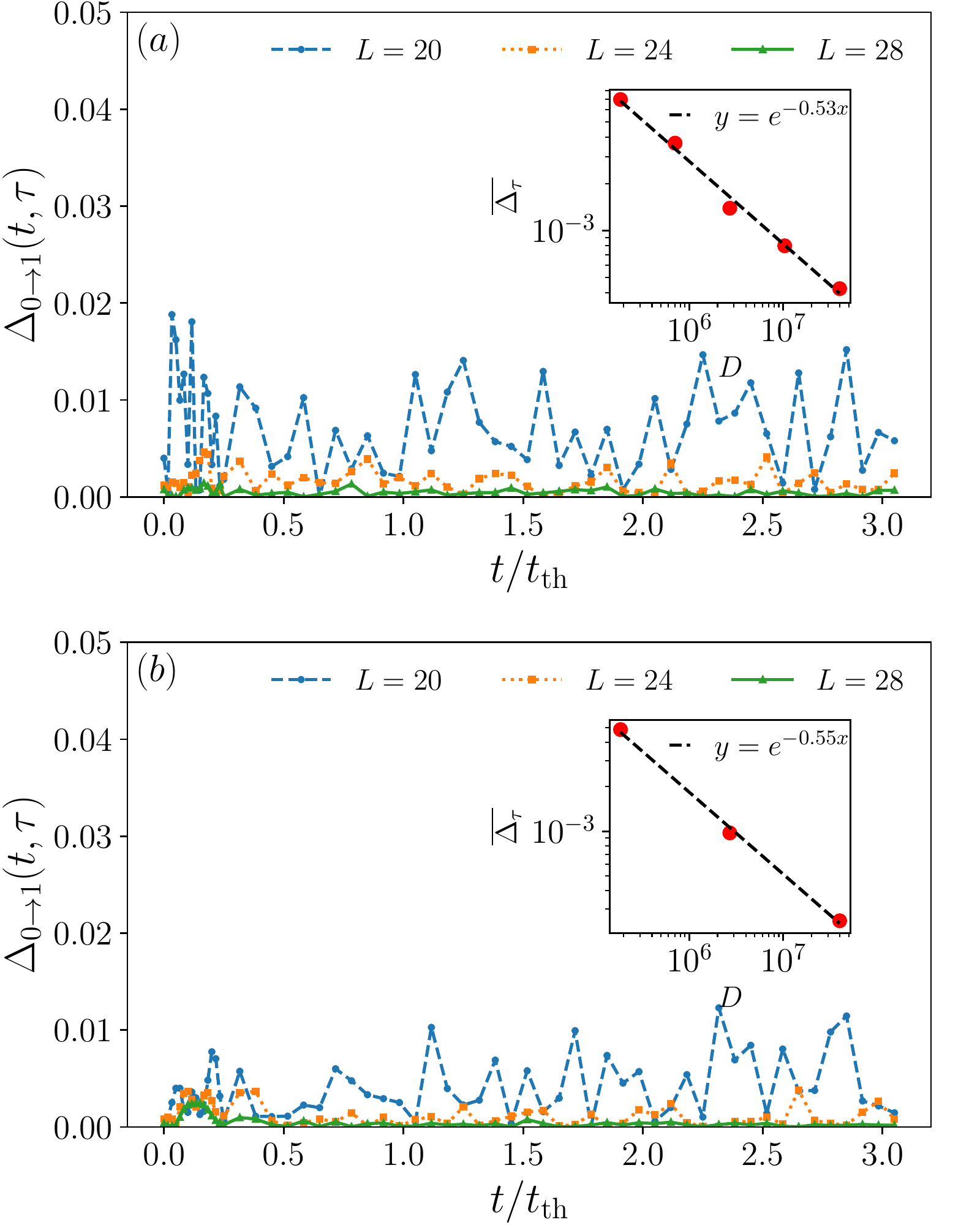}
 \label{fig LDB 2} 
 \caption{The same as Fig.~\ref{fig LDB} but with $\Theta$ defined in Eq.~(\ref{eq time reversal 2}). }
\end{figure}

As a second choice we consider 
\begin{equation}\label{eq time reversal 2}
 \Theta = \exp\left[\frac{i\pi}{2}(\sigma_y^{(1)}+\dots+\sigma_y^{(L)})/2\right] K_z
\end{equation}
for which we find $\Theta(s_x,s_y,s_z)\Theta^{-1} = -(s_x,s_y,s_z)$. Since angular momentum is odd under 
time-reversal in classical mechanics, this could be considered the ``conventional'' choice. In this case we find 
$\Theta X_q\Theta^{-1} = -X_q$ whereas $H$ is still symmetric under time-reversal. From what we said about the 
eigenvalues and eigenvectors above Eq.~(\ref{eq coarse graining interval}), we infer that 
$\Theta \Pi_x\Theta^{-1} = \Pi_{-x}$ and hence $R^\text{TR}_{y,x}(t,\tau) = R_{-y,-x}(t,\tau)$. This LDB condition is 
\emph{different} from the previous one and we check its validity in Fig.~\ref{fig LDB 2}. The conclusions are, 
however, the same as the plots look very similar to Fig.~\ref{fig LDB}. 

To conclude, we clearly see the emergence of classicality and LDB for the pure state dynamics of a slow observable 
of a non-integrable many body system. Instead, for the fast observable classicality does not hold for transient times, 
whereas LDB is quite well satisfied due to the smoothness of the initial state. These results, together with the 
additional results presented in the supplemental material~\cite{SM2}, firmly support our main ideas. 

\section{Further Discussion}
\label{sec further discussion}

\subsection{Multiple observables}
\label{sec multiple}

While we argued that slowness is a necessary condition for classicality, Markovianity and LDB, we have also 
collected some evidence that it is not sufficient. In particular, it was important that the state vector 
explores the available Hilbert space in a seemingly ``unbiased'' fashion. Here, we further discuss the subtlety of 
slowness, mostly from the perspective of multiple observables. Certainly, more research is required in the future. 

To begin with, we emphasize once more the importance to take all strictly conserved quantities into account. In most 
applications this will be energy and particle number, but extensions to other non-commuting are desirabe
too~\cite{MurthyEtAlArXiv2022}. In any case, if one misses one of those conserved quantities, it is clear that
one can not assume the state to spread equally over the subspace $\C H_x$ corresponding to a macrostate $x$. 

Next, let $X, Y, Z, \dots$ be slow observables that are not conserved. If these observables 
mutually commute, our results carry over immediately by replacing $\C H_x$ with $\C H_{x,y,z,\dots}$ provided the 
dimension of these subspaces remains large enough. Examples of this kind include, e.g., the local energy and particle 
number of an open system. 

The situation is more complicated if the observables do not commute. To examine this situation, let us first 
assume that their mutual commutator is small, i.e., $\|[X,Y]\| \ll \|X\|\|Y\|$. Now, if there are only \emph{two} 
such slow observables, then it is possible to construct approximations $X'$ and $Y'$ to $X$ and $Y$ that satisfy 
$[X',Y'] = 0$~\cite{LinFIC1995, HastingsCMP2009} and our approch can be applied to $X'$ and $Y'$. 
We believe this covers a large class of relevant situations in statistical mechanics, but it is interesting 
to ask what happens beyond. 

If there are more than two observables that approximately commute, von Neumann thought that it is still possible to 
approximate them by commuting observables~\cite{VonNeumann1929, VonNeumannEPJH2010}. Also van Kampen assumed that the 
issue of non-commutativity can be overcome by coarse-graining, i.e., that the quantum uncertainty is drowned by the 
experimental measurement error~\cite{VanKampenPhys1954}. We now know that three or more approximately commuting 
observables can \emph{not} be approximated by commuting observables \emph{in general}~\cite{ChoiPAMS1988}. However, 
observables of macroscopic systems that are sums of local observables, and thus of particular relevance to statistical 
mechanics, can be approximated by commuting observables~\cite{OgataJFA2013}. An explicit construction for the 
subspaces of the corresponding macrostates was given in Ref.~\cite{HalpernEtAlNatComm2016}, which could be used to 
extend the present theory. 

\begin{figure}[t]
 \centering\includegraphics[width=0.48\textwidth,clip=true]{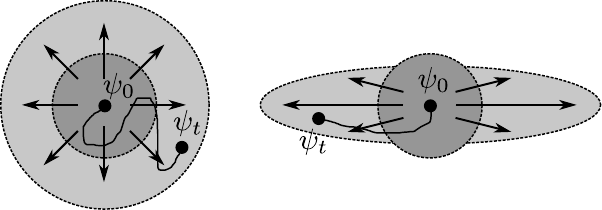}
 \label{fig diffusion} 
 \caption{Two local regions of the sphere in Fig.~\ref{fig rw} with respect to the effective observable mentioned in 
 the text and exemplary trajectories $\psi_0\mapsto\psi_t$. Our theory is applicable whenever a small $\epsilon$-ball 
 of initial states around $\psi_0$ (dark shaded region) approximately explores the sphere in an isotropic fashion on a 
 time-scale $t$ obeying $\tau\gg t\gg\Delta E^{-1}$ as shown on the left. In contrast, on the right there is a strong
 preference to move in a horizontal direction, indicating that there is another slow observable associated to slow
 motion in the vertical direction that need be accounted for. }
\end{figure}

What remains is the case of multiple slow observables that are ``strongly'' non-commuting, although we are not aware of 
any example in statistical mechanics that clearly demonstrates the necessity to consider that case. However, it is a 
legitimate point of view to claim that, even if an experimenter attempts to measure multiple observables (whether 
commuting or not), the resulting transformation on the system is mathematically always described by one set of 
projectors (or more generally a set of positive operator-valued measures) belonging to \emph{one effective} observable. 
While it might be hard to infer that effective observable, it only matters that it is slow. Our theory will work for 
this effective observable whenever the dynamics in this basis is approximately isotropic on the sphere introduced in 
Sec.~\ref{sec setup} because then the state vector explores the space in an unbiased way from a coarse-grained point 
of view, which allows to use typicality arguments. This intuition is sketched in Fig.~\ref{fig diffusion}. 

Unfortunately, it is unknown to us whether there are generic arguments that explain whether an observable gives rise to 
a behaviour as illustrated on the left or on the right of Fig.~\ref{fig diffusion}. Some intuition on this question 
can be gained from Ref.~\cite{KnipschilGemmerPRE2020}, where ``strange relaxation dynamics'' were generated by 
choosing an envelope function $F(\omega)$ in the ETH ansatz, which is very different from a step function (as assumed 
in Sec.~\ref{sec classicality derivation}), e.g., a step function \emph{modulated by a cosine}. As long as this 
observable remains narrowly banded, it would still qualify as slow. However, the unusual modulation with a cosine 
function causes non-Markovian dynamics~\cite{KnipschilGemmerPRE2020}. From the perspective of the present paper, 
we would say that the cosine modulation preferably selects certain energy coherences over others such that the 
dynamics of the microstate no longer appears isotropic or unbiased. 

\subsection{Symmetric part of the rates}
\label{sec symmetric}

This paper argued in great generality that the dynamics of a slow and coarse observable (modulo the difficulties 
mentioned in the previous section) is given by a classical Markov process describable by a master equation with rates 
$R_{x,y}(\psi_y)$, which are almost constant as a function of the microstate $\psi_y$. This implied that their 
asymmetric part obeys LDB and one possible parametrization of the rates is therefore 
$R_{x,y}(\psi_y) = S_{xy} \sqrt{V_x/V_y}$, where the symmetric part $S_{xy} = S_{yx}$ remained unspecified. 

It is not surprising that it turns out that our methods employed so far are \emph{too general} to draw any decisive 
conclusions about the symmetric part $S_{xy}$. It is strongly model-dependent and contains all the informaton about the 
nature of the system, its precise interactions, whether it is dominated by fast or slow transitions, etc. The symmetric 
part therefore can only be inferred by specifying more details about the model, and additional approximation schemes such 
as perturbation theory are likely necessary to proceed analytically. Since the numerous models and techniques are well 
covered in the existing literature, we do not follow down this path here. 

However, at least one property of $S_{xy}$ can be inferred rather easily and we briefly sketch how to do this here. 
This property concerns the \emph{topology} formed by the network of macrostates 
$x$. In fact, we can view a master equation description as a graph, where the vertices are formed by the states $x$, 
and two states $x$ and $y$ are connected by an edge whenever $S_{xy} \neq 0$. Thus, while we are not able to fix or 
estimate any finite value of $S_{xy}$, we can at least decide whether $S_{xy} \neq 0$ or not, which already provides 
valuable information about the concrete physical behaviour of the system. 

To do so, we follow Ref.~\cite{HamazakiPRXQ2022} and directly cast the von Neumann equation 
$\partial_t\rho(t) = -i[H,\rho(t)]$ as a continuity equation for the probabilities $p_x(t)$. Namely, we find 
\begin{equation}
 \frac{d}{dt}p_x(t) = \sum_{y(\neq x)} J_{x,y},
\end{equation}
where $J_{x,y} \equiv -i\mbox{tr}\{H_{xy}\rho_{yx}(t) - \rho_{xy}(t)H_{yx}\}$ is a probability current from $y$ to $x$ 
(with $H_{xy} \equiv \Pi_x H\Pi_y$ and $\rho_{xy} = \Pi_x\rho\Pi_y$). Thus, we see that $S_{x,y}$, and hence $R_{x,y}$, 
is non-zero only if $J_{x,y} \neq 0$, which implies that $H_{xy}$ must not be the null operator. Now, while it is 
hard to compute the transition rates $R_{xy}$ or probabilities $P_{x|y}(\tau)$ directly, it is typically easy to 
know whether $H_{xy} = 0$ or not. 

For instance, consider two weakly coupled subsystems $A$ and $B$ exchanging particles (say, electrons) with each other. 
Most microscopic interaction Hamiltonians will only contain terms proportional to $c_A c_B^\dagger + c_B c_A^\dagger$, 
where $c_{A/B}^\dagger$ ($c_{A/B}$) in the creation (annihilation) operator of an electron in $A/B$. Thus, we see that 
a coarse-graining of the particle number operator of $A$ with states $n\in\mathbb{N}$ describing $n$ electrons in $A$ 
will give rise to a master equation in form of a birth-and-death process, where only neighbouring probabilities are 
connected, i.e., $R_{n,n'} = 0$ if $|n-n'|>1$. If the subsystems are superconductors, then the interaction Hamiltonian 
also contains a term porportional to $(c_A)^2 (c_B^\dagger)^2 + (c_B)^2 (c_A^\dagger)^2$, which implies that also 
$R_{n,n+2} \neq 0$. In this way, we can infer the structure of the graph describing the dynamics of $p_x(t)$. 

For the future it seems worth to explore methods to quantiatively estimate the symmetric part $S_{xy}$ of the rates 
without using perturbation theory, for instance, by using tools from quantum speed limits~\cite{MandelstamTammJP1945, 
DeffnerCampbellJPA2017, HamazakiPRXQ2022} or recent results on thermalization times~\cite{GoldsteinHaraTasakiPRL2013, 
GarciaPintosEtAlPRX2017, DeOliveiraEtAlNJP2018, WilmingEtAlBook2018, NickelsenKastnerPRL2019, 
HevelingKnipschildGemmerPRX2020, SimenelGodbeyUmarPRL2020}.

\section{Conclusions}
\label{sec conclusions}

It is a fact that many natural processes are well approximated by classical Markov processes obeying local detailed 
balance. While this is a curse (or welcome challenge) for many researchers, for instance, those who are interested in 
building a large scale quantum computer or finding quantum effects in thermodynamics, it is also a blessing for many 
other researchers because it greatly simplifies their life and, in fact, it does not sound too speculative that 
stability, predictability and some reduced (but not too low) complexity are important ingredients for the existence of 
intelligent life itself. 

It is also a fact that the repeated randomness assumption---i.e., the repeated use of the equal-a-priori-probability 
postulate or the maximum entropy principle---explains the emergence of classicality, Markovianity and local detailed 
balance. Yet, this assumption is at odds with unitary quantum mechanics (or phase space 
preserving classical mechanics), and no satisfactory microscopic explanation has been put forward so far. 

Based on the intuitive (and already previously used) picture of a slow and coarse observable, we provided various 
estimates, mathematical theorems and numerical simulations that justify the repeated randomness assumption for the 
dynamics of a pure state of an isolated nonintegrable many-body system. Importantly, our approach is not in conflict 
with unitary quantum mechanics. 

We also find it noteworthy that our approach did not make use of common concepts and approximations. 
For instance, we did not use Nakajima-Zwanzig projection operator techniques, repeated interaction schemes, Born 
approximations, perturbation theory, various sorts of Markov assumptions, secular approximations, among others, which 
are often hard to control and justify microscopically. 

Nevertheless, the price to pay was to accept an intuitive but technically subtle notion of slowness, in particular
with respect to the question whether the microscopic state diffuses in an unbiased or isotropic way from a coarse 
perspective. We believe, however, that it was worth to pay this price for the unifying perspective we have got on 
problems that are studied in many different branches of statistical mechanics. 

Conceptually, our research offers a shift in perspective for the investigation of how classicality arises from
quantum mechanics. Central to our approach is chaos in an isolated many-body system and a definition of classicality
based on multi-time probabilities (Kolmogorov consistency), instead of focusing on the dynamics of an open quantum
system as done in the decoherence approach~\cite{ZurekRMP2003, JoosEtAlBook2003, SchlosshauerPR2019}.
Importantly, our approach is not in conflict with environmentally induced decoherence, yet we believe we cannot
agree with the statement of Zeh, one of the pioneers of the decoherence approach, that ``all attempts to describe
macroscopic objects quantum mechanically as being isolated [...] were thus doomed to fail''~\cite{ZehBook2007}.
It is an interesting future prospect to connect these approaches to semiclassical methods based on a stationary
phase approximation of the Feynman path integral~\cite{BerryInBook2001, BraunBook2001, GarbaczewskiOlkiewiczBook2002,
PadmanabhanBook2015}. Interestingly, the semiclassical limit for Hamiltonians with a
classically chaotic limit is problematic due to a rapid smearing out of the wave function over phase space, which is
usually addressed by using environmentally induced decoherence~\cite{ZurekPazPRL1994, CalzettaCQG2012}. How this
fits together with our picture is at present unclear to us.

Furthermore, we provided a systematic justification of the validity of the Markov approximation, which does not
rely on an initial ensemble average that has the potential to wash out already many non-Markovian effects. Moreover, 
we stressed the important role played by the observable (and not only the Hamiltonian) for the question of 
(non-)Markovianity.

Among the more technical insights, we want to highlight our idea to account for correlations between initial 
nonequilibrium states and the pseudorandom coefficients in the ETH ansatz (Sec.~\ref{sec classicality derivation}) 
and the observation that multiple local detailed balance relations can exist for the same setup 
(Secs.~\ref{sec microscopic derivation LDB} and~\ref{sec numerics}). 

It is further worth to comment on how breaking of time-reversal symmetry emerges in our framework given that our 
approach does not use any of the common mechanisms to break it: there are no special initial or repeated ensemble 
averages and both, the ETH and typicality, are arguments that are time-symmetric. In fact, our approach leaves room 
for a time-symmetric picture because we have ``only'' shown that it is overwhelmingly more likely to evolve into the 
direction of an increasing entropy gradient. Applying a time-reversal operator to a microstate along this dynamics 
would indeed provide an example for one of those \emph{atypical} nonequilibrium states for which the master equation 
does not apply. However, since entropy is proportional to the volume of the macrostates and since these volumes grow 
very quickly the closer we are to equilibrium, it is very unlikely to accidentially hit such an atypical state. 
Of course, this general picture complies well with Boltzmann's intuition about the 
second law~\cite{LebowitzPhysToday1993}. 

Our results also suggest that the repeated randomness assumption quickly breaks down for observables that are neither 
slow nor coarse. In that regime, it seems that not many universal features remain, an exception being the laws of 
thermodynamics including fluctuation theorems (see Refs.~\cite{SeifertRPP2012, SchallerBook2014, PelitiPigolottiBook2021, 
StrasbergBook2022} and references therein). This makes the (thermo)dynamic description of such processes very rich 
in variety, but also extremely hard to describe with common principles. 

Finally, our work leaves much room for future research, for instance, related to classicality, decoherence and 
interpretations of quantum mechanics, or related to the subtle notion of slowness, possible refinements thereof and 
systematic correction terms for observables that are a little faster and finer than the present observables but not 
too fast and fine. 

\subsection*{Acknowledgements}

Discussions with Josh Deutsch, Mark Srednicki and Nicole Yunger Halpern are gratefully acknowledged. PS further
acknowledges collaboration and discussion with John Goold, Mark Mitchison and Kavan Modi on a related project (unpublished).
PS and AW acknowledge financial support from the European Commission QuantERA grant ExTRaQT (Spanish MICINN project 
PCI2022-132965), by the Spanish MINECO (project PID2019-107609GB-I00) with the support of FEDER funds, the Generalitat 
de Catalunya (project 2017-SGR-1127), by the Spanish MCIN with funding from European Union NextGenerationEU 
(PRTR-C17.I1) and the Generalitat de Catalunya. 
PS is financially supported by ``la Caixa'' Foundation (ID 100010434, fellowship code LCF/BQ/PR21/11840014). 
AW is furthermore supported by the Alexander von Humboldt Foundation, as well as the Institute of Advanced Study of the 
Technical University Munich.
JW and JG are supported by the Deutsche Forschungsgemeinschaft (DFG) within the Research Unit FOR 2692 under Grant 
No.~397107022 (GE 1657/3-2). 


\bibliography{/home/philipp/Documents/references/books,/home/philipp/Documents/references/open_systems,/home/philipp/Documents/references/thermo,/home/philipp/Documents/references/info_thermo,/home/philipp/Documents/references/general_QM,/home/philipp/Documents/references/math_phys,/home/philipp/Documents/references/equilibration,/home/philipp/Documents/references/time}

\appendix
\section{Bandedness implies small commutator}
\label{sec app bandedness implies small commutator}

For any operator $A$ the operator norm is defined as 
$\|A\| = \sup_{\lr{\psi|\psi}=1} \sqrt{\lr{\psi|A^\dagger A|\psi}}$ and equals the maximum eigenvalue (in absolute 
value) for normal operators satisfying $AA^\dagger = A^\dagger A$. For simplicity we can will assume that the 
groundstate energy is zero, $E_0=0$, and that the largest eigenvalue of $H$ satisfies $E_D = 1$. This implies $\|H\|=1$. 

A straightforward calculation reveals that 
\begin{equation}
 \begin{split}\label{eq operator norm commutator}
  &\lr{\psi|[H,X]^\dagger[H,X]|\psi} \\
  &= \sum_{k,\ell,m} c_k^* c_m X_{k\ell} X_{\ell m}(E_m-E_\ell)(E_k-E_\ell).
 \end{split}
\end{equation}
Now, if $X$ is banded, the sum over $k$ and $m$ can be restricted to `nearby' values around $\ell$ and we can bound 
the energy differences by the band width $\delta E$: 
\begin{equation}
 \begin{split}
  \text{Eq.~(\ref{eq operator norm commutator})}
  &\le \delta E^2 \sum_{k,\ell,m} c_k^* c_m X_{k\ell} X_{\ell m} \\
  &= \delta E^2 \lr{\psi|X^\dagger X|\psi}.
 \end{split}
\end{equation}
Our assumptions $\|H\| = 1$ and that $X$ is narrowly banded imply $\delta E\ll1$. Since the previous equation holds 
for any state $|\psi\rangle$, we conclude 
\begin{equation}
 \|[H,X]\| \le \delta E\|X\| \ll \|X\|.
\end{equation}

\section{Proof of bandedness of projectors}
\label{sec app proof bandedness}

We consider an arbitrary observable $A$ and its spectral decomposition $A = \sum_{a=1}^M \lambda_a \Pi_a$. We like to 
construct a function $f_a$ that satisfies $f_a(A) = \Pi_a$. It can be shown that this is the case if $f_a$ 
satisfies $f_a(\lambda_b) = \delta_{ab}$ by looking at $f_a(A)|\psi\rangle$ for an arbitrary $|\psi\rangle$ expanded 
in the eigenbasis of $A$. Moreover, it is clear that the function $f_a(x) = c\prod_{b(\neq a)} (x-\lambda_b)$ satisfies 
$f_a(\lambda_b) = \delta_{a,b}$ if the constant $c$ is chosen such that $f_a(\lambda_a) = 1$. Thus, we see that $f_a$ 
is a polynomial of degree $M-1$. 

This implies that, if $A$ is banded, then so is $\Pi_a$, although it is not as narrowly banded. Suppose the $A_{k\ell}$ 
are distributed around the diagonal according to a Gaussian with variance $\sigma^2$, then the elements 
$(\Pi_a)_{k\ell}$ are distributed around the diagonal with variance $(M-1)\sigma^2$, i.e., the standard 
deviation is approximately $\sqrt{M}\sigma$.

\section{Supplemental Material}

We present further numerical results about density waves in the XXZ spin chain, in particular for different initial states and different coarse-graining widths. We consistently find that classicality and local detailed balance (LDB) holds for slow observables, though their scaling $D^{-\alpha}$ with the dimension can be different from the $\alpha = 0.5$ case in the main text. In contrast, pronounced violations are observed for fast observables. Moreover, we find the counterintuitive result that LDB can sometimes be better satisfied for finer (instead of coarser) observables for short times, and explain it with the special structure of the observable. Finally, we also briefly study the different setup of energy exchanges between two spin chains and confirm our main conclusions also there.

\subsection{Results for other initial states}\label{sect-2}

\subsubsection{Results for initial states far from equilibrium}

In this section, we  consider an initial state as in Eq.~(79) but for $\kappa=0.3$ (instead of $\kappa=0.1$).

\begin{figure}[t]
	\centering\includegraphics[width=0.45\textwidth,clip=true]{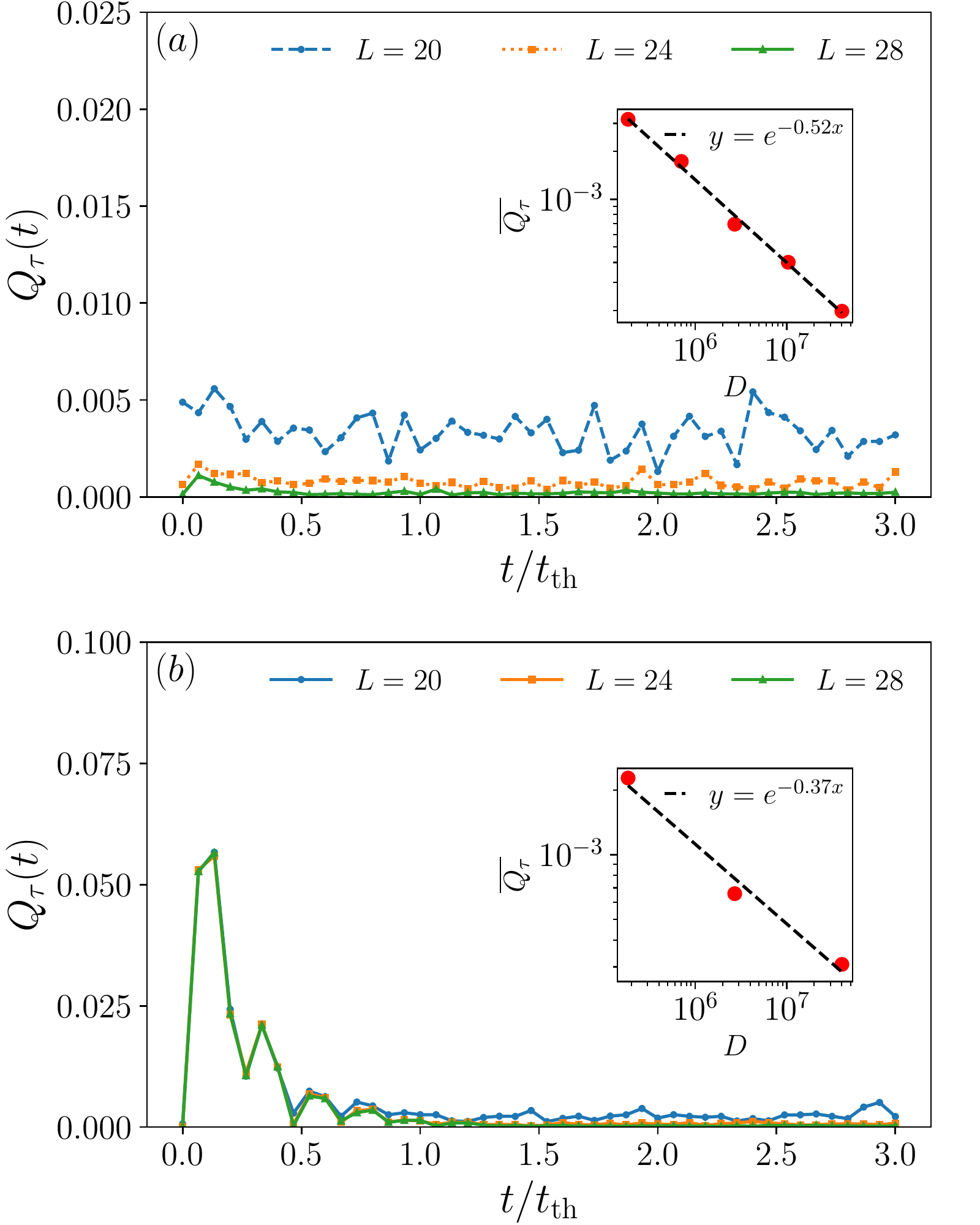}
	\label{fig-CL-0.3}
	\caption{Time evolution of the quantum term $q=1$ (a) and $q=L/2$ (b) for coarse-graining width $\delta X=2.0$ for far-from-equilibrium initial states with $\kappa = 0.3$.
		The insets show a log-log plot of a suitable time-average as a function of the dimension $D$ of the total Hilbert space. }
\end{figure}

The influence of the quantum part $Q_\tau(t)$ are shown in Fig.~\ref{fig-CL-0.3}.
Similar to the main text, the quantum part for the slow observable is small and scales like $D^{-\alpha}$ with
$\alpha = 0.52$. In contrast, for the fast observable the quantum part is much larger for times up to $t_\text{th}/2$
and roughly three times larger than in Fig.~4, suggesting that classicality does not hold far from equilibrium for
fast observables. However, for times $t\ge t_\text{th}/2$ a scaling with $\alpha =0.37$ is again confirmed.

\begin{figure}[t]
	\includegraphics[width=1\columnwidth]{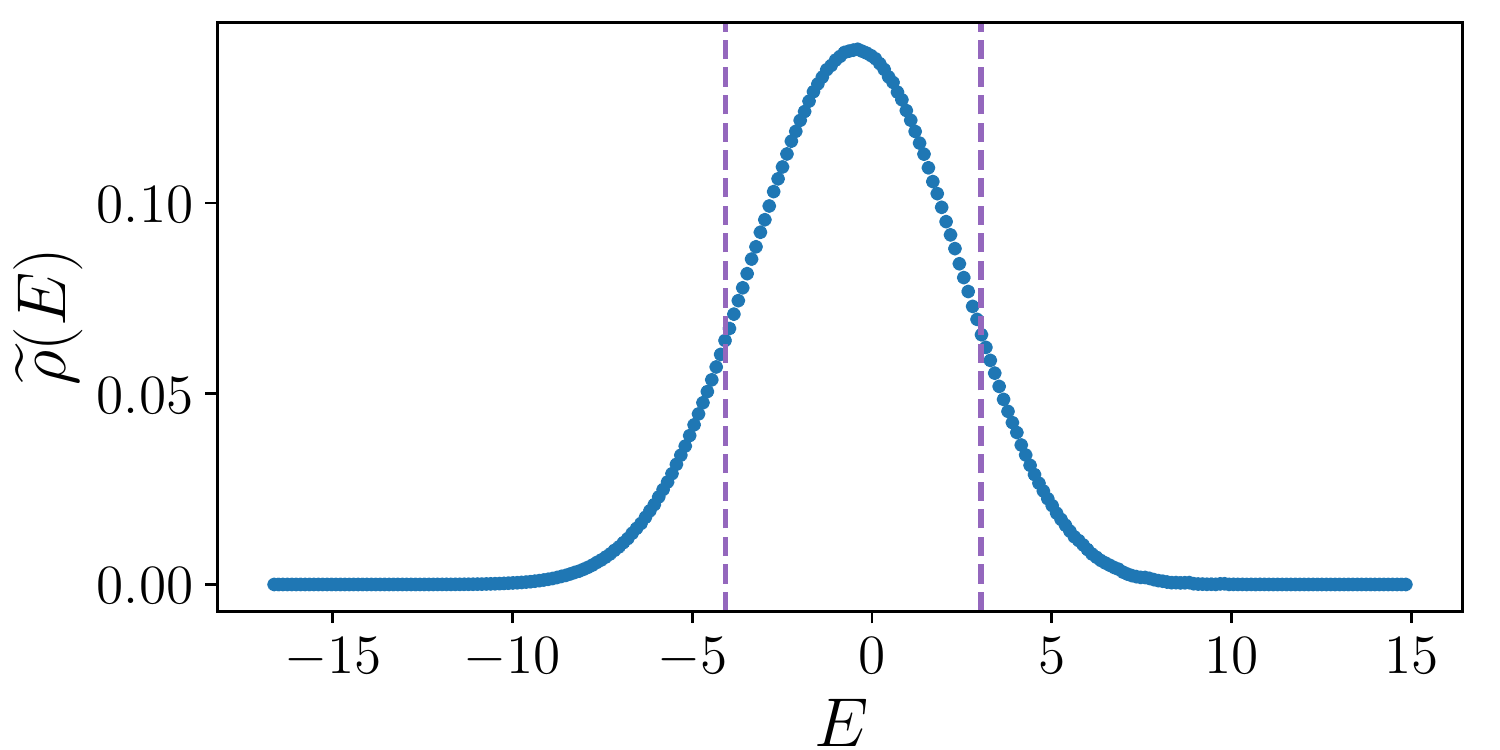}

	\caption{Rescaled density of state $\widetilde{{\rho}}(E) = \rho(E)/D$ for the XXZ model for $L=28$. The dashed line indicate the microcanonical window considered in Sect.\ref{sect-C} . }

	\label{DOS}
\end{figure}

Next, we check the condition of LDB, which involves a technical subtlety. Recall that LDB was derived for the dynamics
contained inside a \emph{microcanonical energy shell}. In Sec.~VI of the main text, we already relaxed that condition
as we considered the entire Hilbert space. This was possible because the coefficients of a slightly perturbed initial
state with $\kappa=0.1$ are well distributed across the entire Hilbert space, whose spectrum closely resembles a
Gaussian distribution with zero mean, thus effectively corresponding to an infinite temperature state
(see Fig.~\ref{DOS} for the spectrum). However, for states far from equilibrium as considered here the coefficients in the
energy eigenbasis are far from being equally distributed throughout the whole Hilbert space, so effectively, we are
restricted to a different Hilbert space. Thus, instead of the dimension of the whole subspace ($V_x$), we need to
consider the effective volume
\begin{equation}
 \widetilde{V}_x = \pi_x D,
\end{equation}
which is dependent on the equilibrium state probabilities $\pi_x$. In our numerical simulations, the effective dimension
$\widetilde{V}_x$ is calculated as
\begin{equation}
\widetilde{V}_{x}=\frac{1}{t_{f}-t_\text{th}}\int_{t_\text{th}}^{t_{f}}\text{tr}(\rho(t)\Pi_{x})dt.
\end{equation}
Since all initial states considered in this section are far from equilibrium, the effective dimension will be used
throughout this section to check the condition of LDB.

\begin{figure}[t]
 \centering\includegraphics[width=0.45\textwidth,clip=true]{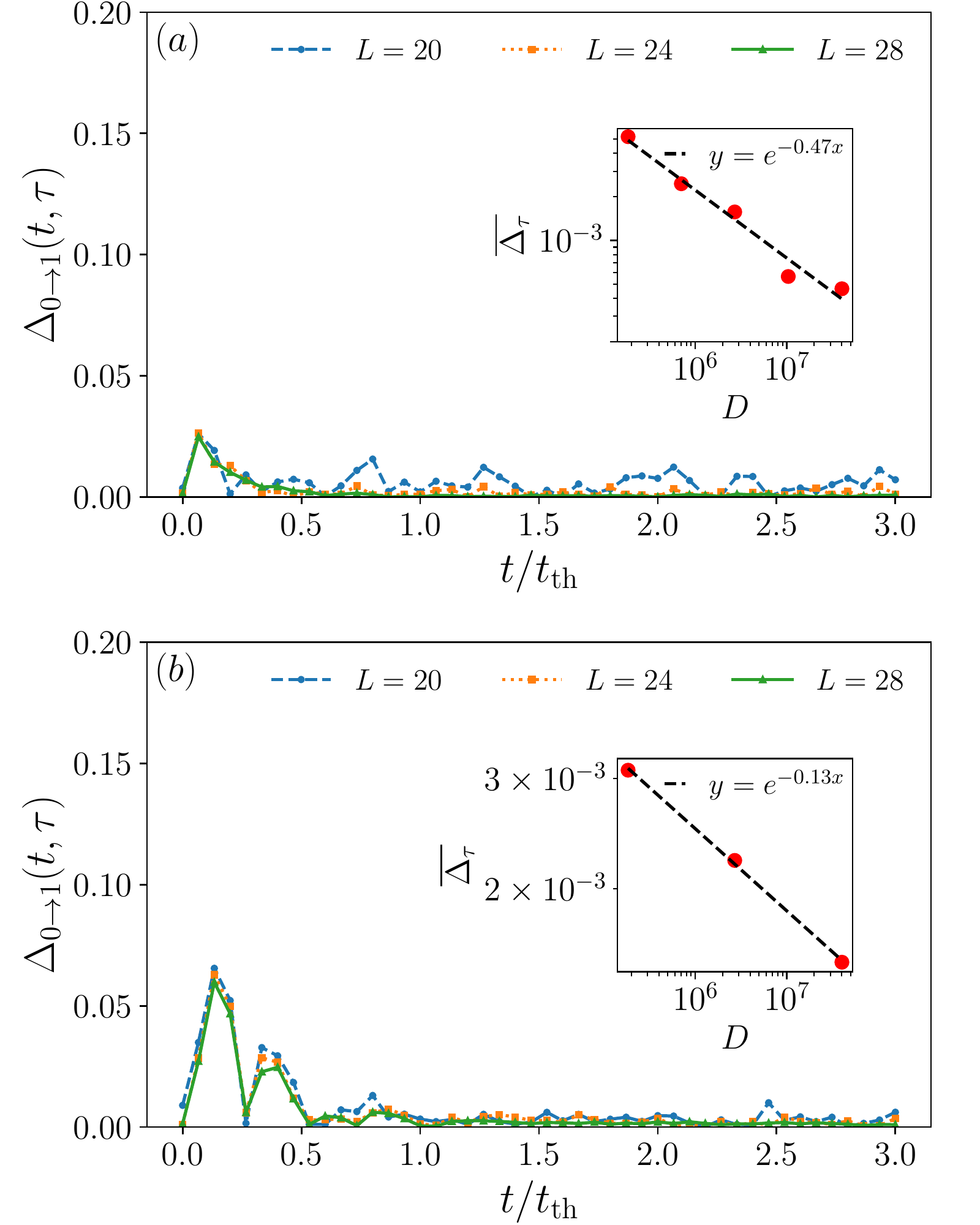}
 \label{fig-LDB-0.3}
 \caption{Check of LDB for $\Theta = K_z$ for $q=1$ (a) and $q=L/2$ (b) for coarse-graining width $\delta X=0.74$ for far-from-equilibrium initial states with $\kappa = 0.3$. The insets show a log-log plot of time average as a function of the dimension $D$ of the
 total Hilbert space. }
\end{figure}

Similar to the results for the near-equilibrium initial state, one can see from Fig.~\ref{fig-LDB-0.3} that LDB is
well satisfied for slow observables also for the far-from-equilibrium intial state, and obeys a scaling law with
$\alpha = 0.47$. In contrast, initial violations up to $t_\text{th}/2$ (roughly twice as large as in Fig.~5) are clearly
visible for the fast observable, and the scaling is notably worse with $\alpha = 0.13$.

\subsubsection{Results for initial states distributed in the two largest subspace}

In this section, we consider a different kind of initial state. It is initially distributed only in the two subspace
$I_0$ and $I_1$ defined in  Eq.~(78) and can be written as
\begin{equation}\label{eq-istate}
 |\psi\rangle=\sqrt{p_{0}}\Pi_0|\psi_{R}^{0}\rangle+\sqrt{p_{1}}\Pi_1|\psi_{R}^{1}\rangle,
\end{equation}
where $|\psi^{1,2}_R\rangle$ are random states and $p_0 = \frac{1+\delta p}{2},\ p_1 = \frac{1-\delta p}{2}$.

\begin{figure}[t]
 \centering\includegraphics[width=0.45\textwidth,clip=true]{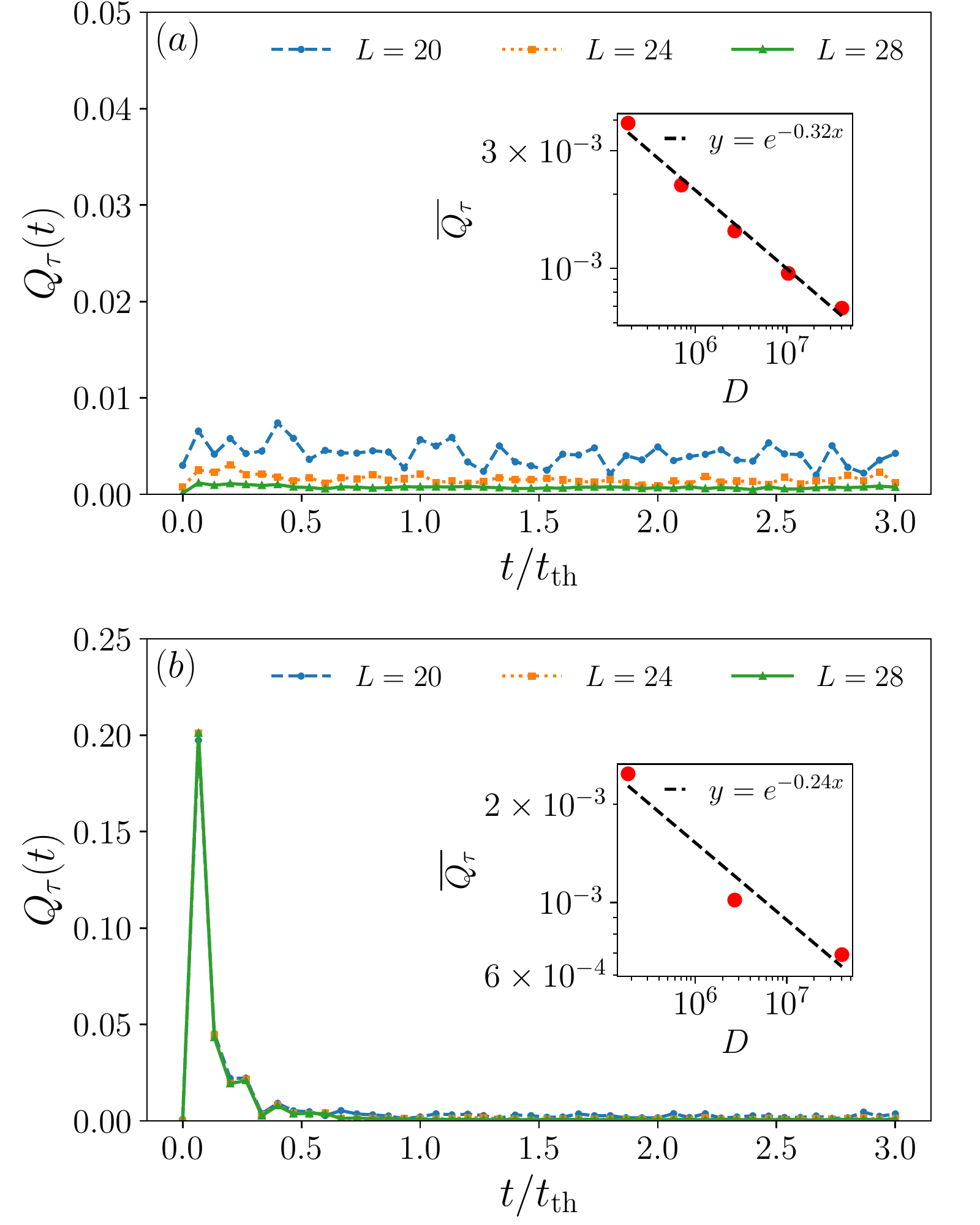}
 \label{fig-CL-Mix}
\caption{Time evolution of the quantum term $q=1$ (a) and $q=L/2$ (b) for coarse-graining width $\delta X=0.74$ for initial states in Eq.~(\ref{eq-istate}) for $\delta p = 0.2$.
 The insets show a log-log plot of a suitable time-average as a function of the dimension $D$ of total Hilbert space. }
\end{figure}

Starting with the investigation of classicality again, one can see from Fig.~\ref{fig-CL-Mix} that---in unison with our
main claim---for slow observables the quantum part $Q_\tau(t)$ is very small. It is interesting to note, however, that
the scaling with $\alpha = 0.32$ is \emph{worse} compared to the previous examples, and the detailed reasons for this
behaviour remain to be understood. Moreover, we observe \emph{very strong violations} of classicality for the fast
observable up to times $t_\text{th}/2$. Afterwards, a similar scaling law with $\alpha = 0.26$ is observed.

\begin{figure}[t]
 \centering\includegraphics[width=0.45\textwidth,clip=true]{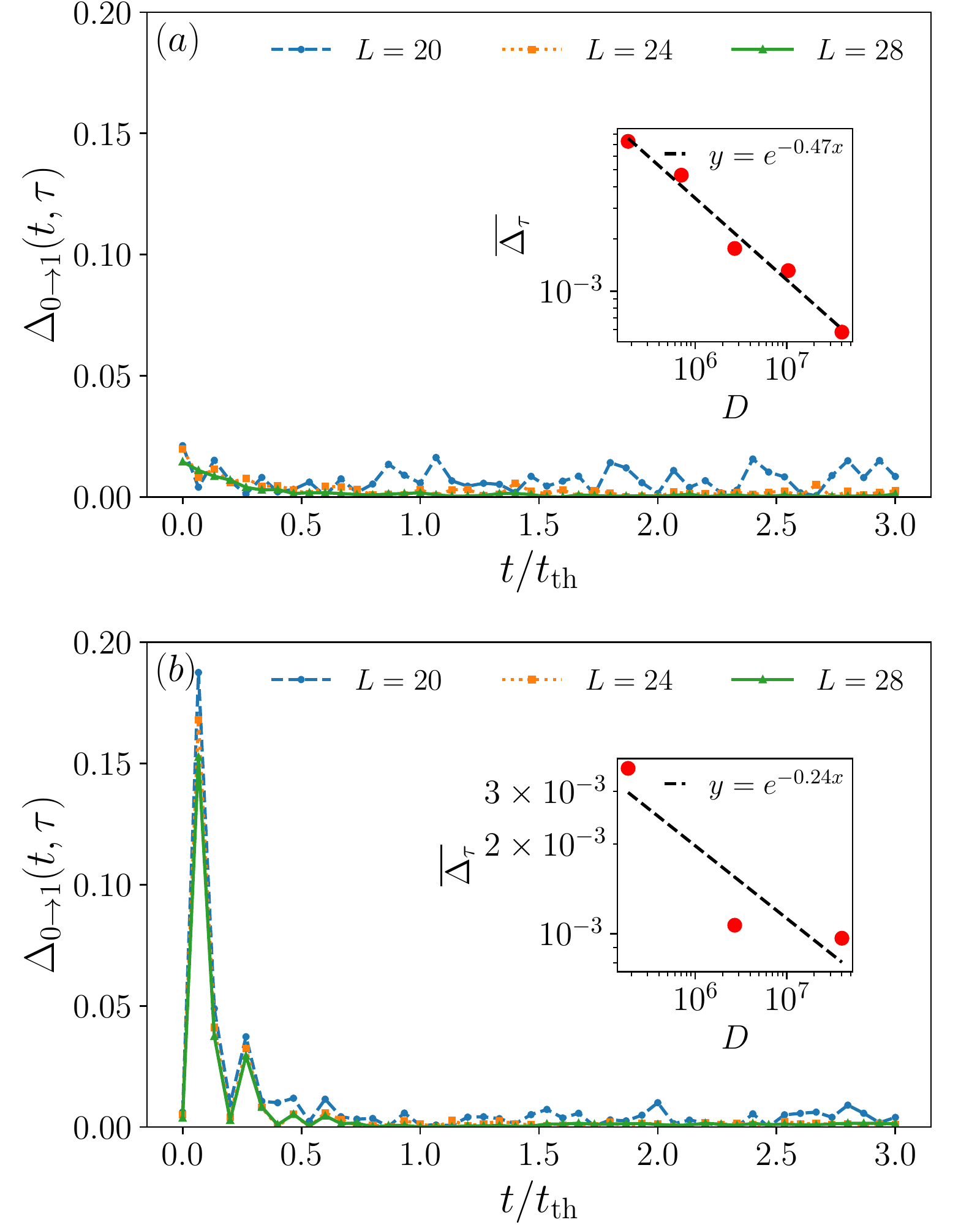}
 \label{fig-LDB-Mix}
 \caption{Check of LDB for $\Theta = K_z$ for $q=1$ (a) and $q=L/2$ (b) for coarse-graining width $\delta X=0.74$ initial states in Eq.~(\ref{eq-istate}) for $\delta p = 0.2$. The insets show a log-log plot of time average as a function of the dimension $D$ of total Hilbert space. }
\end{figure}

Next, we check the condition of LDB in Fig.~\ref{fig-LDB-Mix}. The general behavior of it for both the slow and fast
observable case is similar to what we found before. However, for the fast case strong violations of LDB are found at
early times indicating that slowness is crucial for LDB to hold out of equilibrium, which was less visible in Fig.~5
and~6. The probable reason for that is that the initial state considered in the main text is close to a fully mixed
state with respect to \emph{every} coarse-grained subspace. As a result, the state may still stay close to typical
states during the time evolution, even for the fast observable. Instead, the state in Eq.~(\ref{eq-istate}) considered
here is only fully mixed in two subspaces, so during transient times it is more likely to evolve into atypical states
while ``exploring'' new subspaces. Finally, after some transient time, LDB is also well satisfied for the fast case,
but with a clearly different scaling ($\alpha = 0.24$) compared to the slow case ($\alpha=0.47$).

\begin{figure}[t]
 \centering\includegraphics[width=0.45\textwidth,clip=true]{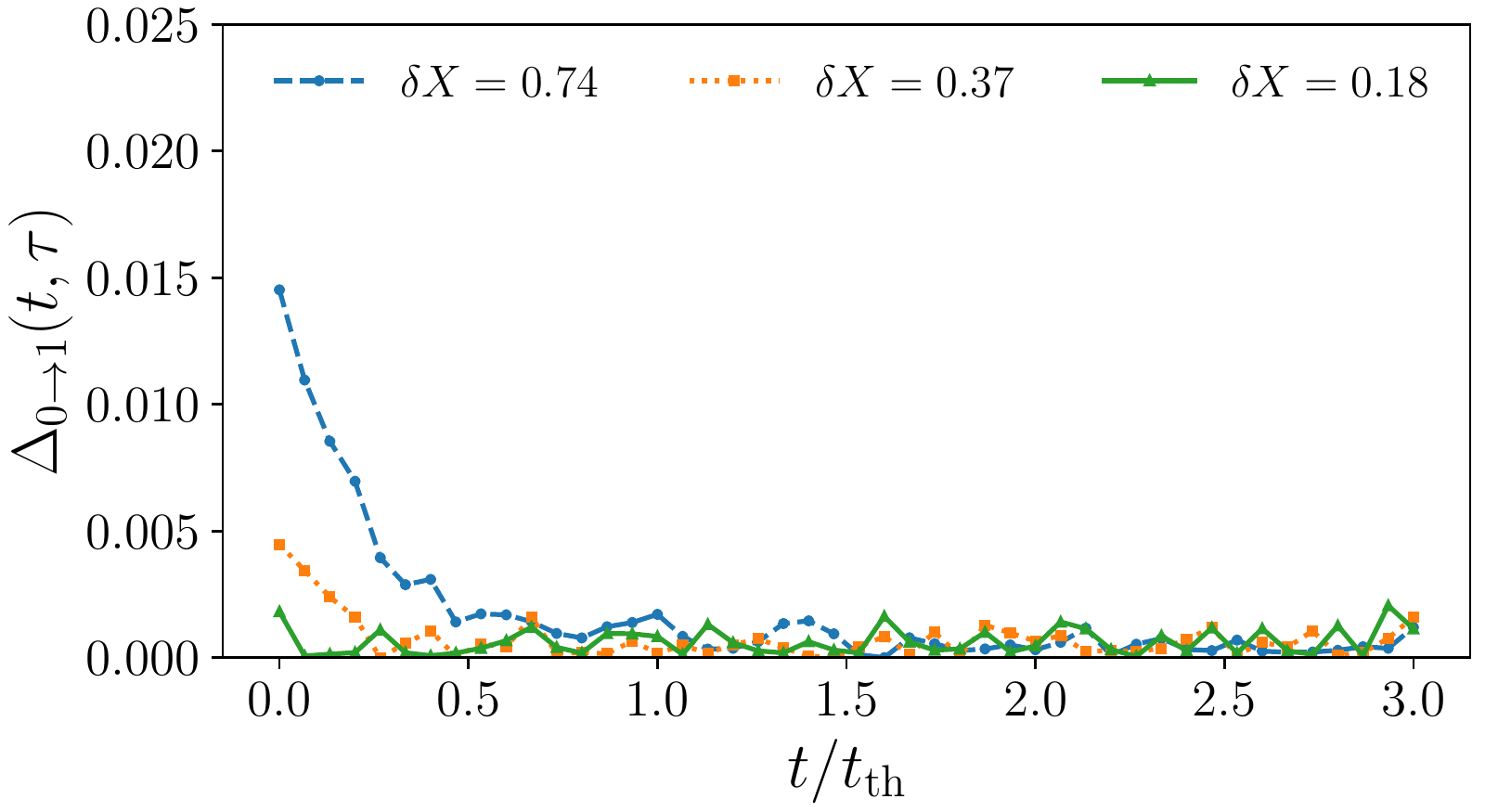}
 \label{fig-Slow-Mix-CG}
 \caption{Check of LDB for $\Theta = K_z$ for $q=1$ for different coarse-graining width $\delta X = 0.74,0.37,0.18$ for initial states in Eq.(\ref{eq-istate}) for $\delta p = 0.2$, with the system size  $N = 28$.}
\end{figure}

In addition to what we have studied previously, we now also consider the influence of the coarse-graining \emph{width}
$\delta X$ on LDB for slow observables. To this end, besides the previously considered choice of $\delta X = 0.74$,
we also check LDB for finer coarse-grainings with $\delta X = 0.37$ and $\delta X = 0.18$. Figure~\ref{fig-Slow-Mix-CG}
displays the results for the largest system size with $L=28$. Remarkably, we observe the counterintuitive results
that \emph{LDB is better satisfied for finer coarse-grainings for short times}. This is in contrast to what we should
expect based on a naive application of Levy's Lemma. The reason for that can be understood by considering the
\emph{interconnection of microstates} $|\textbf{z}\rangle$ belonging to the same coarse-grained space $I_x$. The dynamics
on these states can be viewed as a network, where an edge connects two microstates $|\textbf{z}\rangle$ and
$|\textbf{z}'\rangle$ if the Hamiltonian creates an overlap between the two:
$\langle\textbf{z}|H|\textbf{z}'\rangle \neq 0$. Now, since the Hamiltonian $H$ in Eq.~(76) contains only two-body
interactions, many pairs of microstates in a given coarse-grained space $I_x$ are not directly connected by a single
edge. This implies that there is not only a slow time scale associated to the evolution of the observable $X$
(associated to changes between subspaces $I_x$), but there is also a relatively slow time scale associated to changes
\emph{between microstates in the same subspace}. This implies that, if a state starts to explore a new subspace, it
takes some time until it looks typical (i.e., evenly smeared out over different microstates) in that subspace. But it
is precisely this ``smearing out'' over different microstates within a given subspace that justifies the Haar random
average in the application of Levy's Lemma, and this smearing out \emph{takes longer for larger subspaces}, i.e., for
coarser coarse-grainings. This is exactly what is observed in Fig.~\ref{fig-Slow-Mix-CG}, which further shows how subtle
the notion of a coarse and slow observable is as the interconnection of the network of microstates plays a central role.
Moreover, note that this behaviour can not be observed for the initial state considered in the main text because this
state is already well smeared out over all microstates.

\subsubsection{Results for initial states restricted to energy window}\label{sect-C}

In this section, we consider yet another initial state which is restricted to a microcanonical energy window as sketched
in Fig.~\ref{DOS}. It has the following form
\begin{equation}
 |\psi\rangle\sim\Pi_{E,\Delta E}e^{-\kappa {X_q/2}}|\psi_{R}\rangle,
\end{equation}
where $|\psi_R\rangle$ is a random state drawn from a Gaussian distribution, and
\begin{equation}
\Pi_{E,\Delta E}=\sum_{E_{k}\in[E-\frac{\Delta E}{2},E+\frac{\Delta E}{2}]}|E_{k}\rangle\langle E_{k}|,
\end{equation}
with $E_k$ and $|k\rangle$ being the eigenvalue and eigenstate of the Hamiltonian.
Here the center of the energy window is chosen according to the canonical (inverse) temperature $\beta$ (here we only
consider the infinite temperature case $\beta = 0$) by the following equation
\begin{equation}
 E=\frac{\text{tr}(e^{-\beta H}H)}{\text{tr}(e^{-\beta H})}.
\end{equation}

\begin{figure}[t]
 \centering\includegraphics[width=0.45\textwidth,clip=true]{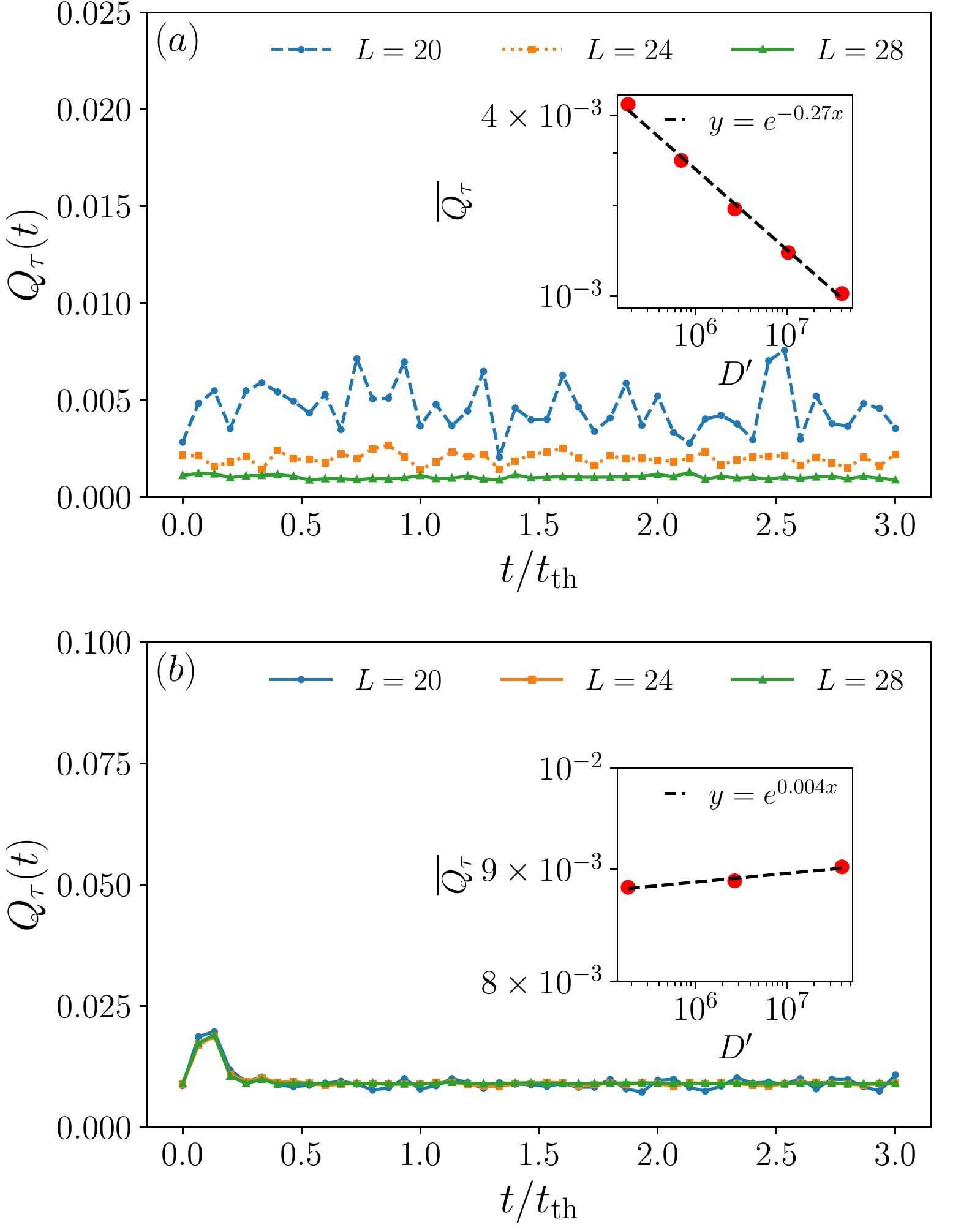}
 \label{fig-Q-mic}
 \caption{Time evolution of the quantum term for $q=1$ (a) and $q=L/2$ (b) for coarse-graining width $\delta X=0.74$ for initial state in a microcanonical energy window corresponding to infinite temperature $\beta = 0$. The energy width $\Delta E$ is chosen to be $\Delta E = 3\sqrt{L/20}$. The insets show a log-log plot of a suitable time-average as a
 function of the dimension $D^\prime$ of the microcanonical energy shell. }
\end{figure}

One can see from Fig.~\ref{fig-Q-mic} that for the slow observable the quantum part $Q_{\tau}(t)$ is small for all
times, but obeys a weaker scaling law with $\alpha = 0.27$ (similar to Fig.~\ref{fig-CL-Mix}) compared to the
$\alpha \approx 0.5$ case of the main text. Moreover, violations for the fast observable are much stronger than for
the slow observable, and no decay to zero is visible for long times. Nevertheless, we note that the violation of
classicality is small compared to the maximum possible value of one.

\begin{figure}[t]
 \centering\includegraphics[width=0.45\textwidth,clip=true]{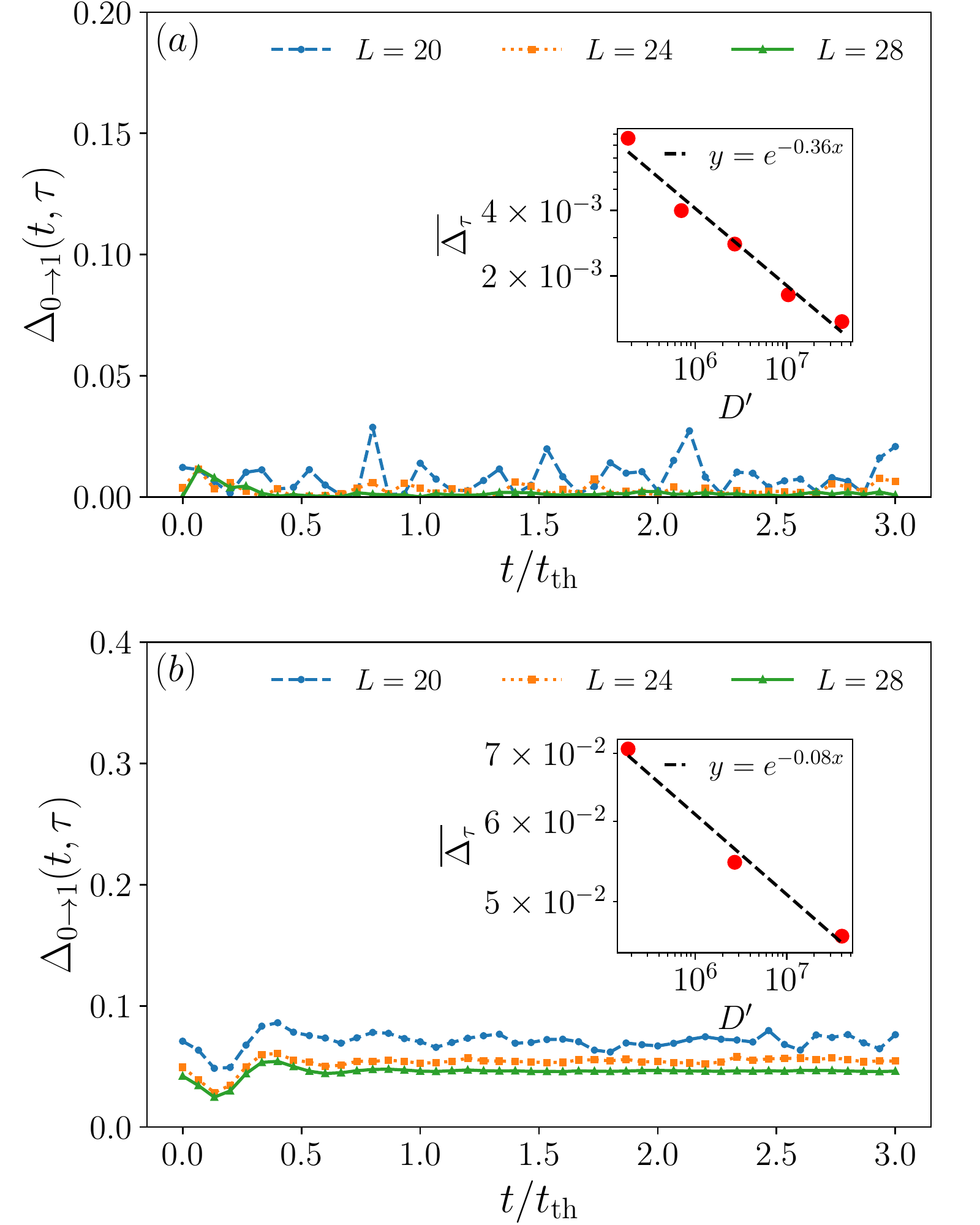}
 \label{fig-LDB-mic}
 \caption{Check of LDB for $\Theta = K_z$ for $q=1$ (a) and $q=L/2$ (b) for coarse-graining width $\delta X=0.74$.
 The insets show a log-log plot of a suitable time average as a function of the dimension $D^\prime$ of the
 microcanonical energy shell. }
\end{figure}

In Fig.~\ref{fig-LDB-mic} we check the condition of LDB. Similar to the results shown in Fig.~\ref{fig-Q-mic}, LDB works
well for the slow observable for almost all times except for small deviations until $t_\text{th}/4$. In contrast, for
the fast observable LDB is still not satisfied even after the thermalziation time. The possible reason for both
the persistent violation of classciality and LDB even after the thermalization time may be that---different from the
slow observable where one has $[\Pi_{E,\Delta E},\Pi_{x}] \approx 0$---for fast observable the commutator
$[\Pi_{E,\Delta E},\Pi_{x}]$ remains quite large. As a result, the state may still be highly atypical even after the
thermalization time.

\subsection{Numerical results in coupled tilted field Ising model for canonical initial state}

Besides the extensively studied XXZ model, we also briefly consider a different model, namely two coupled Ising models
with tilted field. The Hamiltonian reads
\begin{equation}
 H = H_A + H_B + \lambda H^I_{A}\otimes H^I_{B},
\end{equation}
where
\begin{equation}
 H_A = H_B = h_{x}\sum_{\ell=1}^{n}\sigma_{x}^{\ell}+h_{z}\sum_{\ell=1}^{n}\sigma_{z}^{\ell}+g_{z}\sum_{\ell=1}^{n}\sigma_{z}^{\ell}\sigma_{z}^{\ell+1},
\end{equation}
and
\begin{equation}
 H^I_A=H^I_B = \sigma^{n}_z.
\end{equation}
Here $n=L/2$ is the length of the Ising chain with periodic boundaries,
and we choose $h_x=1.0,\ h_z = 0.5, \ g_z = 1.0$.  The eigenvalue and eigenstate of the two subsystems are denoted by
\begin{equation}
 H_A|e_\alpha\rangle = e_\alpha|e_\alpha\rangle, \quad H_B|e_\beta\rangle = e_\beta|e_\beta\rangle.
\end{equation}
The initial state we consider here is a product state of the canonical states of the two subsystems, obtained by makeing use of typicality,
\begin{equation}\label{eq-is2C}
 |\psi\rangle=K(e^{-\beta_{A}H_{A}/2}|\psi_{R}^{A}\rangle)\otimes(e^{-\beta_{B}H_{B}/2}|\psi_{R}^{B}\rangle),
\end{equation}
where $K$ is a normalization constant and $|\psi^{A,B}_R\rangle$ are Gaussian random states.

\begin{figure}[b]
	\includegraphics[width=1\columnwidth]{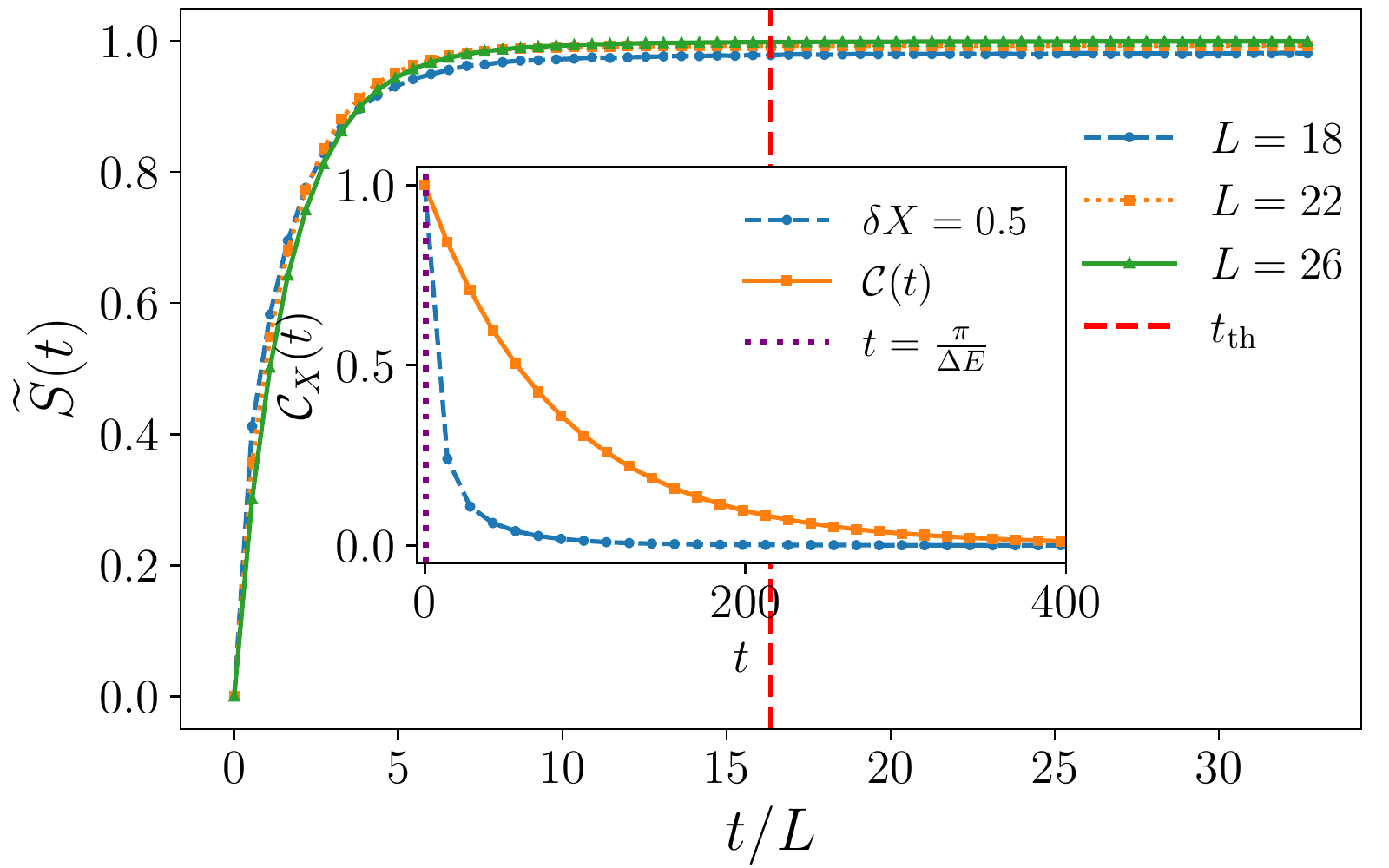}
	\caption{Time evolution of rescaled thermodynamic entropy $\widetilde{S}(t)$ for the energy difference operator for coupled Ising model as a function of $t/L$, for the initial state introduced in Eq. (\ref{eq-is2C}) for $\beta_A = 0.1,\quad \beta_B = -0.1$ and $\lambda = 0.5$. The time axis is rescaled as  the relaxation time scales like $L$ for this
	observable. Insets: Time evolution of correlation functions for $L=26$ as explained in the main text.}
	\label{2C-S}
\end{figure}

Here we consider the energy difference operator
\begin{equation}
 X = \frac{1}{\cal N}(H_A-H_B)
\end{equation}
where $\cal N$ is a normalization constant which ﬁxes the second central moment of the operator to one. The projector of the coarse-grained eigenspaces can be written as
\begin{equation}
 \Pi_x \equiv\sum_{e_{\alpha}-e_{\beta}\in[E-\frac{\delta X}{2},E+\frac{\delta X}{2}]}P_{\alpha\beta}
\end{equation}
where
\begin{equation}
 P_{\alpha\beta}=|e_{\alpha}\rangle\langle e_{\alpha}|\otimes|e_{\beta}\rangle\langle e_{\beta}|,
\end{equation}
and $\delta X$ is the coarse-graining width (which is chosen as $\delta X=0.5$ in our numerical simulation).

First we show the result for the rescaled thermodynamics entropy $\widetilde{S}(t)$ defined in Eq.~(80). It can be seen from Fig.~\ref{2C-S} that $\widetilde{S}(t)$ increases monotonically, which is expected to be the case as the energy difference operator we consider here is a slow observable according to the inset of Fig.~\ref{2C-S}.

\begin{figure}[t]
	\includegraphics[width=1\columnwidth]{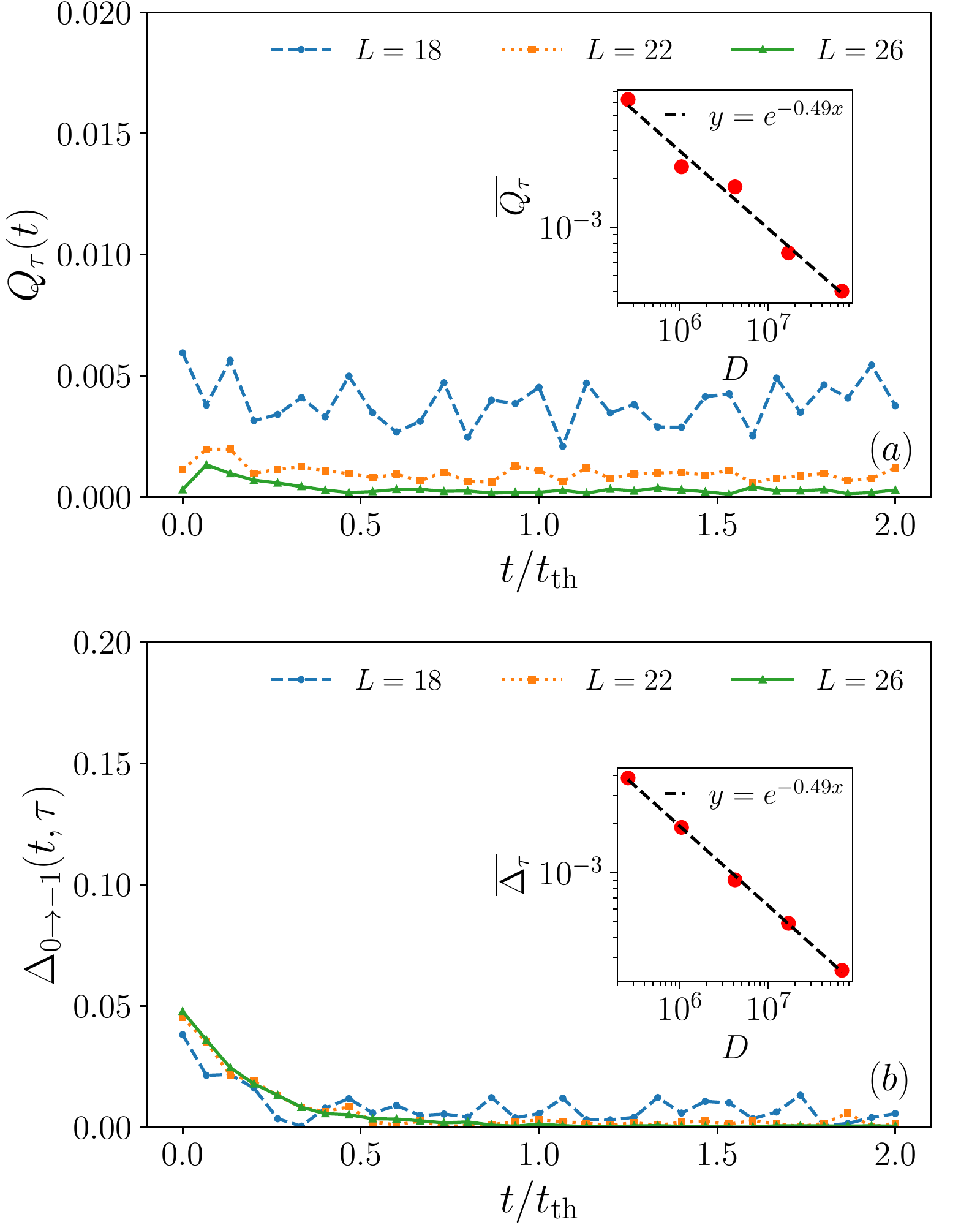}
	\caption{(a) $Q_\tau(t)$ versus $t/t_\text{th}$ and (b) $\Delta_{0\rightarrow -1}(t,\tau)$ versus $t/t_\text{th}$  in coupled Ising model for $\lambda = 0.5$ and $\delta X = 0.5$, for canonical initial state at  temperature ($\beta_A = 0.1 ,\quad \beta_B=-0.1$). The insets show a log-log plot of time average as a function of the dimension $D$ of the total Hilbert space. }
	\label{2C-DB}
\end{figure}

Next, we study the influence of the quantum part $Q_{\tau}(t)$ in Fig.~\ref{2C-DB}(a). One can see that similar to the
results we have in the XXZ model the quantum term is small for all times considered here, and its time average scales
with $\alpha = 0.49$ (close to the ideal scaling $\alpha =0.5$ estimated in Sec.~III and in unison with the numerical
results of the main text and of Fig.~\ref{fig-CL-0.3}). In Fig.~\ref{2C-DB}(b) the condition of LDB is checked, and one
can see that after a slight deviation for times up to $t_\text{th}/4$ it stays close to zero, and it obeys again a
scaling law with $\alpha = 0.49$.

\end{document}